\documentclass[12pt]{article}
\usepackage{a4wide,graphicx,amsmath,amssymb,hyperref}
\usepackage[numbers,sort&compress]{natbib}
\usepackage{rotating}

\def\GeV{{\rm GeV}}
\def\TeV{{\rm TeV}}

\def\lapproxeq{\lower .7ex\hbox{$\;\stackrel{\textstyle                                                    
<}{\sim}\;$}}                                                    
\def\gapproxeq{\lower .7ex\hbox{$\;\stackrel{\textstyle                                                    
>}{\sim}\;$}} 
\begin{document}

\titlepage

\begin{flushright}
LCTS/2012-27\\
Cavendish-HEP-2012/16\\
IPPP/12/78\\
ZU-TH 24/12\\
\end{flushright}

\vspace*{0.1cm}

\begin{center}

{\Large \bf  Extended Parameterisations for MSTW PDFs and their \\[1ex]
effect on Lepton Charge Asymmetry from $W$ Decays}

  \vspace*{0.3cm}
  \textsc{A.D. Martin$^a$, A.J.Th.M. Mathijssen$^{b,}\footnote{Work of 
A.J.Th.M. Mathijssen mainly done while at University College London.}$, W.J. Stirling$^c$, \\[1ex]
R.S. Thorne$^d$, B.J.A. Watt$^d$ and G. Watt$^e$} \\
  
  \vspace*{0.2cm}$^a$ Institute for Particle Physics Phenomenology, University of Durham, DH1 3LE, UK \\
  $^b$ Rudolf Peierls Centre for Theoretical Physics, 1 Keble Road,
Oxford, OX1 3NP, UK\\
  $^c$ Cavendish Laboratory, University of Cambridge, CB3 0HE, UK \\
  $^d$ Department of Physics and Astronomy, University College London, WC1E 6BT, UK \\
  $^e$ Institut f\"ur Theoretische Physik, Universit\"at Z\"urich, CH-8057
Z\"urich, Switzerland
\end{center}

\vspace*{0.1cm}

\begin{abstract}
We investigate the effect of extending the standard MSTW parameterisation of input parton distribution functions (PDFs) 
using Chebyshev polynomials, rather than the usual expressions which involve a 
factor of the form $(1+\epsilon x^{0.5} + \gamma x)$. We find evidence 
that four powers in the polynomial are generally sufficient for 
high precision. Applying this to valence and sea quarks, the gluon already 
being sufficiently flexible and needing only two powers, 
we find an improvement in the global fit, but a 
significant change only in the small-$x$ valence up-quark PDF, $u_V$. We investigate the 
effect of also extending, and making more flexible, the `nuclear' correction to deuteron structure 
functions. We show that the extended `Chebyshev' parameterisation results in 
an improved stability in the deuteron corrections that are required for the best fit to the `global' data.
The resulting PDFs have a significantly, but not dramatically, altered valence down-quark distribution, $d_V$. It is shown that, for the extended set of MSTW PDFs, their  
uncertainties can be obtained using 23, rather than the usual 20, orthogonal `uncertainty' eigenvectors. This is true both without and with extended 
deuteron corrections. Since the dominant effect is on the valence quarks, we
present a detailed study of the dependence of the valence--sea separation 
on the predictions for the decay lepton charge asymmetry which results from $W^\pm$ production at the LHC, illustrating the PDFs and the $x$ range 
probed for different experimental scenarios. We show that the modified MSTW
PDFs make significantly improved predictions for these data at the LHC, 
particularly for high values of the $p_T$ cut of the decay lepton. However, this is a special 
case, since the asymmetry is extremely sensitive to valence--sea details, and in particular to the combination $u_V-d_V$ of valence PDFs for $x \sim M_W/\sqrt{s}$ at low
lepton rapidities. We show that the 
predictions for a wide variety of total cross sections are very similar to those obtained using the MSTW2008 PDFs, with changes being much smaller than the PDF uncertainties.
\end{abstract}

\newpage

\section{Introduction}

In the determination of the parton distribution functions (PDFs) of the proton from fitting 
to the available deep-inelastic and related hard-scattering data, a long-standing question 
is the extent to which the limitations of a fixed form of the input parameterisations 
affect the best fit and the uncertainty of the resulting PDFs.  It is certainly the case 
that various groups performing `global' PDF analyses have had to introduce new parameters 
to facilitate a good-quality fit to some new data, either because they probe a new PDF 
combination or a new kinematic range, or simply because new data are much more precise than 
previous measurements. This has resulted in most groups 
performing fits to a wide variety of data sets 
\cite{Martin:2009iq,Lai:2010vv,Alekhin:2012ig,JimenezDelgado:2008hf} 
using about 4-6 free parameters for each type of 
parton.\footnote{The HERAPDF fit \cite{Aaron:2009aa} uses fewer free parameters 
in their study. However, in that analysis the effect of adding extra parameters 
is included as part of the additional ``parameterisation'' uncertainty.} 

The NNPDF group \cite{Ball:2012cx} circumvents this 
issue by using effectively an extremely large and flexible parameterisation, 
but in order to avoid fitting all the fluctuations in data, they must split 
data into training and validation sets and have algorithms which determine the 
methods of both convergence and `stopping'. This means they do not have an 
easily  identifiable `best fit' and it is very difficult to compare the 
sources of their PDF uncertainty to those for other groups. Indeed, there has been  
clear sensitivity to their convergence and `stopping' algorithms, though this
has been quite small since the set in \cite{Ball:2010de}. It is 
hypothesised that the lack of parameter flexibility is part of the reason 
for the need for a `tolerance', or the use of `$\Delta \chi^2 > 1$', to obtain uncertainties 
in the MSTW (and CTEQ) fits. But studies so far suggest that while this is probably a 
component, it is not all, or even the dominant reason for this need for inflation of 
$\Delta \chi^2$ \cite{Pumplin:2009bb,Watt:2012tq} (see, for example, a discussion of relative 
uncertainties between sets in \cite{DeRoeck:2011na}).

So far, there has been surprisingly little investigation of the change of the form of input 
parameterisations on the uncertainties of PDF sets based on a best fit and 
expansion about this central PDF set. There are studies by Pumplin 
\cite{Pumplin:2009bb}, by Glazov, Moch and Radescu \cite{Glazov:2010bw}, and 
one sentence in the MSTW conference proceedings \cite{Thorne:2010kj}. 

In this article we investigate 
the effect of extending the MSTW parameterisation of the input PDFs by changing the interpolating
polynomial $(1+\epsilon x^{0.5}+\gamma x)$, which was introduced 
for separate up and down valence and for sea quarks in \cite{Martin:1994kn}, 
to a term including up to an $n^{\rm th}$-order Chebyshev 
polynomial. First, we investigate the most appropriate order of polynomial to 
use such that sufficient flexibility is achieved, but not so much that one is in 
danger of fitting fluctuations in the data. To study this, general functions of a suitable shape are
generated, and pseudo-data are fitted. 
We conclude that about a $4^{\rm th}$-order polynomial should 
generally be adequate. We then try using this type of polynomial in fits to real data, 
first for the two valence distributions, but also additionally for the sea distribution. We see a significant improvement in the $\chi^2$ for the 
best global fit at both stages, but the only significant change in the PDFs    
is for the $u_V$ distribution for $x<0.03$ at high $Q^2\sim 10^4~\GeV^2$, or slightly higher $x$ at low $Q^2$. 

At present, and for at least the short-term future, we will have to continue to include the existing deep-inelastic scattering and Drell--Yan data from deuteron targets to achieve a determination of the PDFs of the different quark flavours, particularly at moderate and large values of $x$.  It is therefore important to repeat our previous study of improving the nonperturbative corrections to the deuteron structure 
functions \cite{Thorne:2010kj} from the default \cite{Badelek:1994qg}
first used in \cite{Martin:1994kk}, but now using the extended Chebyshev parametric forms of the input PDFs. The results are rather more successful than when using our standard 
PDF parameterisation, with the deuteron correction being rather similar to that 
expected from various models, with little variation when different assumptions 
are made. The change in deuteron corrections is found to change the $d_V$ distribution to a fairly significant extent. 

For the Chebyshev input parameterisations, without and with the additional 
freedom in deuteron corrections, we demonstrate that a suitable set of PDF 
uncertainty eigenvectors can be found, using 23 orthogonal directions in 
parameter space, rather than the 20 used in the standard MSTW2008 PDF set.   
Since it is the valence quark PDFs that are affected, we 
examine the detailed dependence of quark decomposition on the lepton  charge asymmetry (which results from $W^{\pm}$ production).
We show that the precise combination, and the $x$ range, 
of the PDFs probed is very dependent on the lepton rapidity and on the lepton $p_T$ cut applied to the data.

The lepton charge asymmetry is particularly sensitive to the small-$x$ valence PDFs, and 
this sensitivity is seen to increase rather dramatically as the $p_T$ cut is 
raised. The predictions obtained from the new PDFs (using Chebyshev input parameterisations) are compared with both the 
Tevatron lepton asymmetry data \cite{Abazov:2008qv} that were not used in the MSTW2008 fit, 
and the recent LHC data, namely the ATLAS
lepton rapidity data \cite{Aad:2011dm} and the CMS lepton asymmetry data \cite{Chatrchyan:2012xt}. In all cases the 
default MSTW2008 PDFs are not optimum for these lepton asymmetry data sets, whereas the new `Chebyshev' PDFs
give much improved predictions even though they are obtained from a fit to {\it exactly} 
the same global data set as used to determine the MSTW2008 PDFs. 

Although the lepton charge asymmetry, which has extreme 
sensitivity to the low $x$ valence--sea decomposition, is better fit by the change in the 
PDFs, we check that the predicted values of the $W^{\pm},~Z$, Higgs, etc. total cross sections are essentially {\it unaltered}. To be precise, they are only changed by amounts far smaller than those due to the PDF uncertainty.

\section{Parameterising Input PDFs with Chebyshev Polynomials}

  \begin{figure}[htb!]
  \centering
\includegraphics[width=1.00\textwidth,clip]{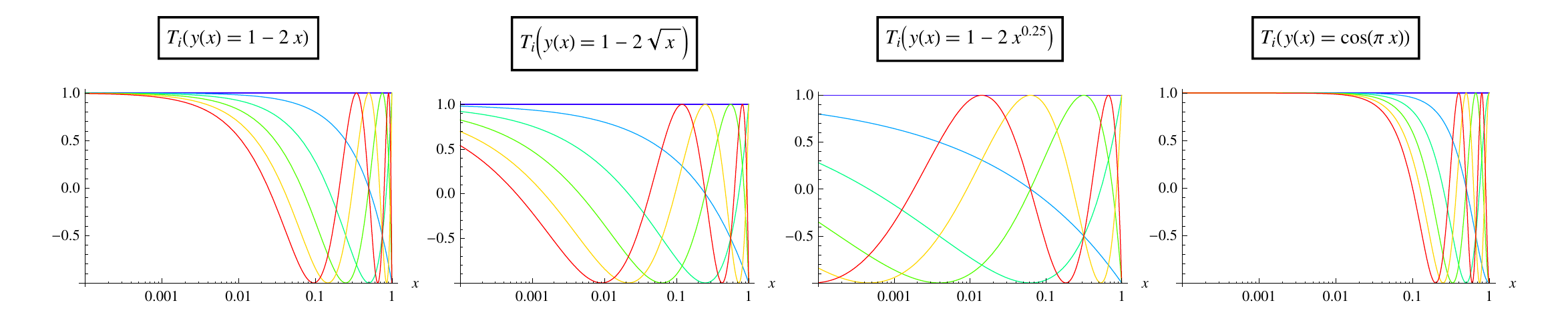}
  \caption{Behaviour of Chebyshev polynomials $T_i[y(x)]$ 
of order $i=0$ to 5 as a function of $x$ for 
different arguments for the expansion variable. The order of the polynomial 
increases as the structure extends to smaller $x$ values. The order of the polynomial also increases across the visible spectrum (i.e.~dark blue to red).}
  \label{fig:Fig19}
  \end{figure}
In a recent previous study \cite{Watt:2012tq}, an investigation of the uncertainty of our PDFs using Monte Carlo generated data replicas was performed, as opposed to the use of perturbations about the best fit as was done in the MSTW2008 analysis. Little 
change was seen when the full 28 MSTW PDF parameters were left free compared to 
the 20 used in eigenvector generation. To be precise, the uncertainty using 
$\Delta \chi^2=1$ was compared using the two approaches and significant 
difference was only seen for the $u_V$ distribution for $x<0.03$
at the input scale $Q_0^2=1~\GeV^2$, where the 
uncertainty band expands a little, and for $d_V$ in some $x$ regions. It was 
particularly reassuring that there is little change in the uncertainty on the 
gluon distribution 
despite the number of free parameters being extended from 4 to 7. Since it is 
difficult to apply our previously used `dynamical tolerance' technique~\cite{Martin:2009iq} for the uncertainty 
determination to this Monte Carlo method, and since there was little change in the results, 
it was concluded that the eigenvector approach was justified and would continue to be used in our PDF analyses.\footnote{It was, however, shown how an arbitrary number of Monte Carlo sets of PDFs could 
be generated starting with the eigenvector definition.} Nevertheless, there 
was some evidence that an extended parameterisation might lead to some 
differences in the PDFs of the valence quarks. Hence, we start by investigating this hypothesis.

For valence and sea quarks the default MSTW parameterisation for the input at 
$Q_0^2=1~{\rm GeV}^2$ was taken to be 
\begin{equation}
xf(x,Q_0^2) = A (1-x)^{\eta}x^{\delta}(1+\epsilon x^{0.5}+\gamma x).
\label{eq:MSTWparam}
\end{equation}
The $(1-x)$ power, $\eta$, allows a smooth interpolation to zero as $x \to 1$ and is
inspired by number counting rules. The single small-$x$ power, $\delta$, is 
inspired by the behaviour predicted by Regge theory at small $x$. 
We found long ago that, first at NNLO \cite{Martin:2000gq}, and also 
with improved data at NLO \cite{Martin:2001es}, 
that two terms with different small $x$ powers were needed for the gluon distribution 
to give the best fit. For the gluon the parameterisation is 
\begin{equation}
xg(x,Q_0^2) = A_g (1-x)^{\eta_g}x^{\delta_g}(1+\epsilon_g x^{0.5}+\gamma_g x)
+A_{g'}(1-x)^{\eta_{g'}}x^{\delta_{g'}}.
\label{eq:MSTWgluon}
\end{equation}
The input parameterisations for some other distributions, $\bar d - \bar u$ 
and $s - \bar s$, take slightly different forms, but these are not very 
precisely determined, and we will not consider changes to these in this article. Similarly, as previously, 
$s + \bar s$ is taken to be the same as the sea parameterisation 
except for the normalisation and $(1-x)$ power, which are left free. 
The polynomials, interpolating between the high-$x$ and 
low-$x$ limits, have no real motivation other than the separation of half-integer powers 
being again inspired by Regge theory, and the two free parameters seeming to be 
sufficient to obtain an optimum fit. An investigation of introducing
either an extra parameter of the form $ax^2$ or $ax^{0.25}$ into the valence quark
parameterisation was reported very briefly in \cite{Thorne:2010kj} since neither had a significant 
effect on the fit quality -- at best they gave $\Delta \chi^2=-4$. 
However, the introduction of an $ax^2$ term did change the small-$x$ $u_V$ distribution 
a little outside its uncertainty, and hence, as with the Monte Carlo study, suggests 
the uncertainty  on this PDF, in the range $x < 0.03$, is underestimated.

  \begin{figure}[htb!]
  \centering
\includegraphics[width=0.48\textwidth]{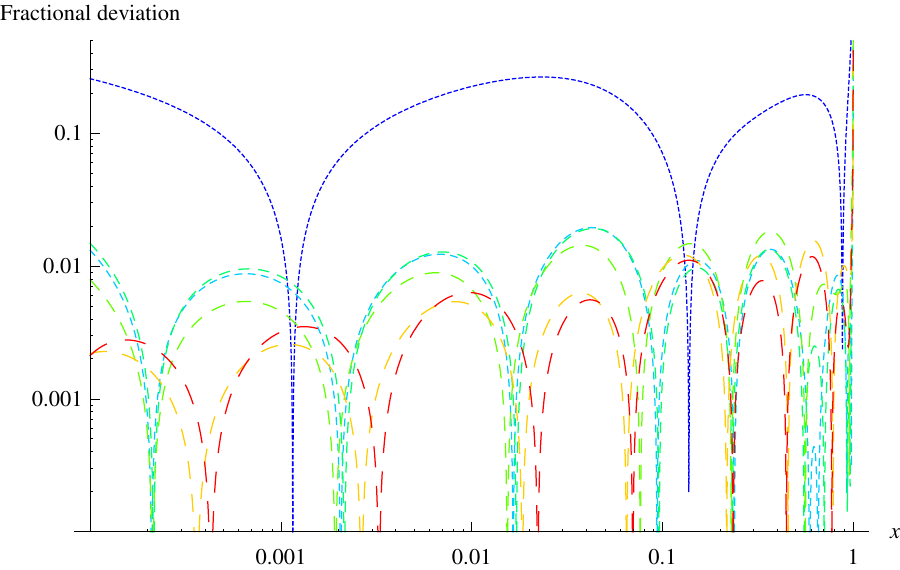}
\includegraphics[width=0.48\textwidth]{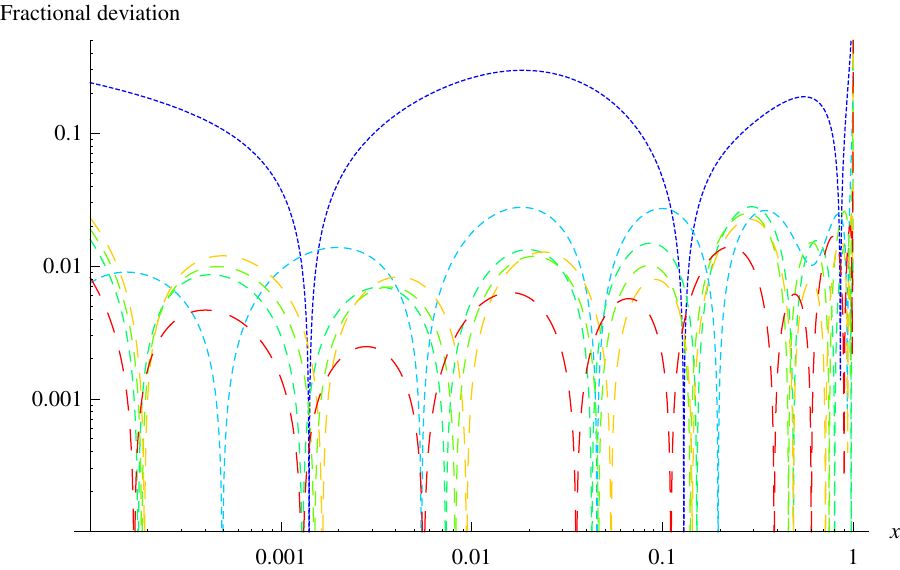}
  \caption{Two examples of the fractional deviation 
between fitted function and true function for 
fits with increasing highest order of Chebyshev polynomials for 
valence-like distributions. The dash length decreases as the highest order of the 
polynomial increases. The order of the polynomial also increases across the visible spectrum (i.e.~dark blue to red).}
  \label{fig:Fig33pdf}
  \end{figure}

Here we undertake a much more systematic study. As a basis for the interpolating 
polynomial we decide to use Chebyshev polynomials (though we looked at,
and will mention briefly, other possibilities). So we write
\begin{equation}
xf(x,Q_0^2) = A(1-x)^{\eta}x^{\delta}\left(1+\sum_{i=1}^n a_i T_{i}(y(x))\right),
\label{eq:generalparam}
\end{equation}
where $y$ is a function of $x$ to be specified.
We keep the same form of the $(1-x)$ and $x$ powers in the high- and 
low-$x$ limits. One of the main motivations for the choice of Chebyshev 
polynomials is that, not only the end points of the polynomials 
at $y = \pm 1$ have magnitude 1, but each maximum and minimum 
between the endpoints does also. Other choices, such as Legendre 
polynomials, have maxima and minima with smaller magnitudes so 
they have smaller variations in magnitude away from the endpoints.
There is still a choice to make regarding the argument $y$ of the 
polynomial. We need $y=1$ at the lower limit of $x$, i.e.~$x=0$, and 
$y=-1$ at the other limit $x=1$, but there are many choices which 
could satisfy this. In practice the PDFs are measured between a range
of roughly $0.0001<x<1$, so we want a choice such that the polynomials
vary throughout the whole of this range. The form of the first few 
polynomials is shown for various choices in  Fig.~\ref{fig:Fig19}.  
Clearly $y=1-2x$ is too concentrated at high $x$ and $y=1-2x^{0.25}$ 
extends to too low $x$. An alternative of $y=\cos(\pi x)$
is very concentrated at high $x$. We choose $y=1-2\sqrt{x}$
as a convenient definition. 
This is the same choice as in the study reported in \cite{Pumplin:2009bb}. 
It is slightly different from the choice in  \cite{Glazov:2010bw} 
which used logarithmic dependence rather than 
powers of $x$, but the results are similar. A polynomial in  $y \equiv 1-2\sqrt{x}$
also has the feature that it is equivalent to a polynomial in $\sqrt{x}$, 
the same as the default MSTW parameterisation, though for a $n^{\rm th}$-order 
Chebyshev polynomial the maximum power of $x$ is $x^{n/2}$. The half integer 
separation of terms is consistent with the Regge physics motivation of the 
MSTW parameterisation.

Pumplin \cite{Pumplin:2009bb} has explained clearly why a parameterisation like 
(\ref{eq:generalparam}) is advantageous. Most previous parameterisations, including MSTW, 
have been based on  interpolating functions like those in (\ref{eq:MSTWparam}) with a small 
number of parameters, $\epsilon,\gamma,...$. If the number of parameters is increased to 
allow more flexibility, the resulting fit becomes unstable, with parameters taking large 
values and with strong cancellations between the corresponding terms. On the other hand, 
the parameters, $a_i$, in the Chebyshev form  (\ref{eq:generalparam})  are much more 
convenient, and well-behaved, for fitting to the data. The requirement of smoothness in 
the input PDFs forces the values of the parameters $a_i$ to be reasonably small at large 
order $i$. The Chebyshev polynomials of increasingly large order, $n$, model the behaviour 
of the input distribution at an increasingly fine scale in $x$. However, it is still an 
open question as to how many parameters are needed to model a parton distribution to 
sufficient accuracy without also starting to fit fluctuations. So far  the standard
technique is to impose some artificial restriction such as a requirement on smoothness
of the function. 

In order to test how many parameters are indeed needed for a sufficiently good fit 
we generate pseudo-data for a valence quark input PDF, say $u_V$, scattered around a function
with the general shape of a valence quark distribution obtained from a very 
large order polynomial $f(x)$ with smoothness constraints applied in 
order to stop it developing kinks. The function is constrained to give an 
integrated total of two valence quarks and the fits are all constrained 
in the same manner. The 1000 pseudo-data points are 
distributed evenly in $\ln(1/x)$ with their percentage error held constant
at $3\%$, i.e.~${\rm error}_i = 0.03f(x_i)$, and with random scatter about the 
exact function according to the size of the uncertainty. 
We then find the best fit for this pseudo-data using a parameterisation 
of the form (\ref{eq:generalparam}) with increasing highest order of the Chebyshev 
polynomials.  
This procedure was repeated for a variety of different choices of starting 
function. Two typical results are shown in Fig.~\ref{fig:Fig33pdf} which shows  
the percent deviation of the fit function from the full function for two different starting 
functions. 
The highest order of the polynomial increases across the visible spectrum (i.e.~dark 
blue to red). Just using one term in the polynomial, i.e.~$1+a_1T_1(y)$, 
can give deviations of 
about $10\%$ over a wide range, but with $2$ terms in the polynomial this reduces to 
mainly a $\leq 2\%$ deviation (in most cases -- sometimes 2 polynomials can still
give significantly larger deviations $\sim 5-10\%$). 
For $4$ terms there is generally $\leq 1\%$ deviation 
except at very high $x$. This does not improve very significantly with further 
terms added. This accuracy should be compared to the uncertainty in the MSTW2008 input
PDFs. For valence quarks the 1 sigma uncertainty is at best just lower than $2\%$ for 
$u_V(x,Q_0^2)$ near $x=0.1$. For $d_V(x,Q_0^2)$ it is nearly always $> 3\%$.     

  \begin{figure}[htb!]
  \centering
  \includegraphics[width=0.32\textwidth,clip]{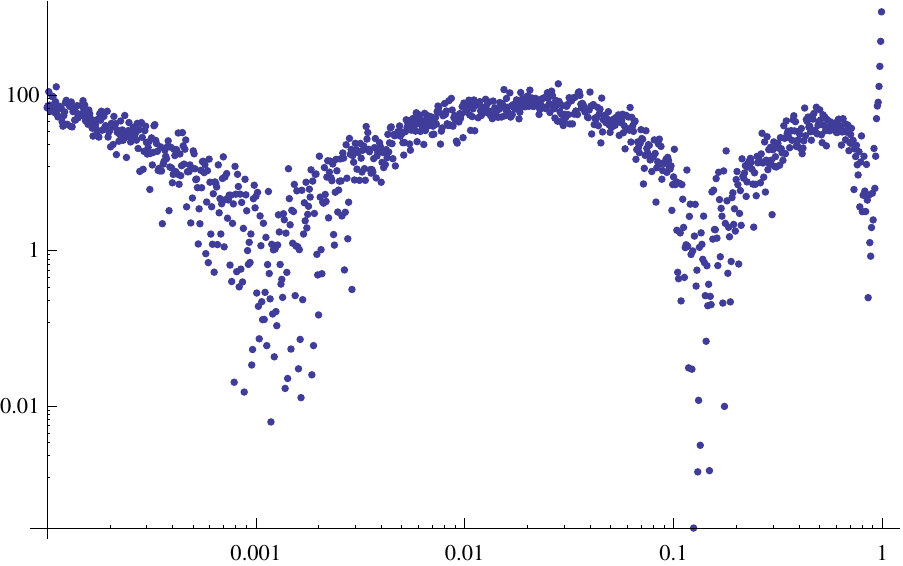}
  \includegraphics[width=0.32\textwidth,clip]{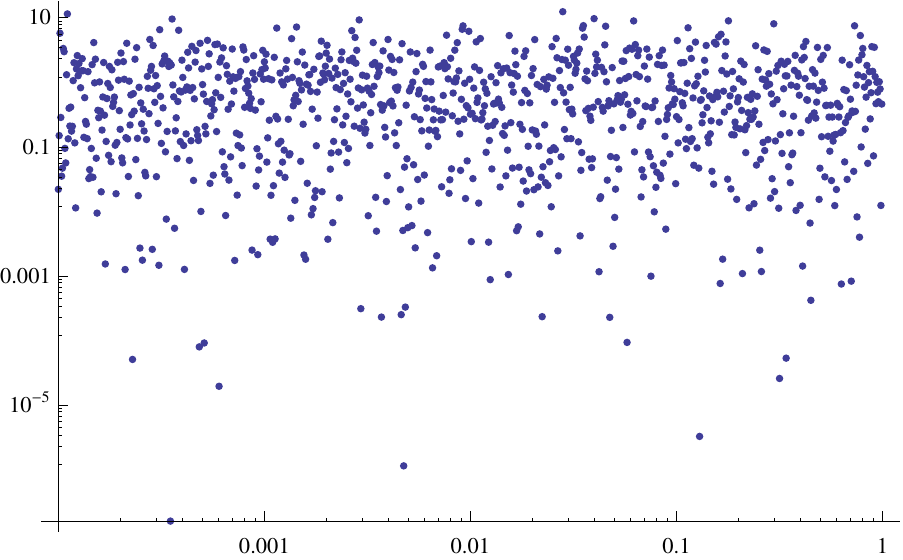}
  \includegraphics[width=0.32\textwidth,clip]{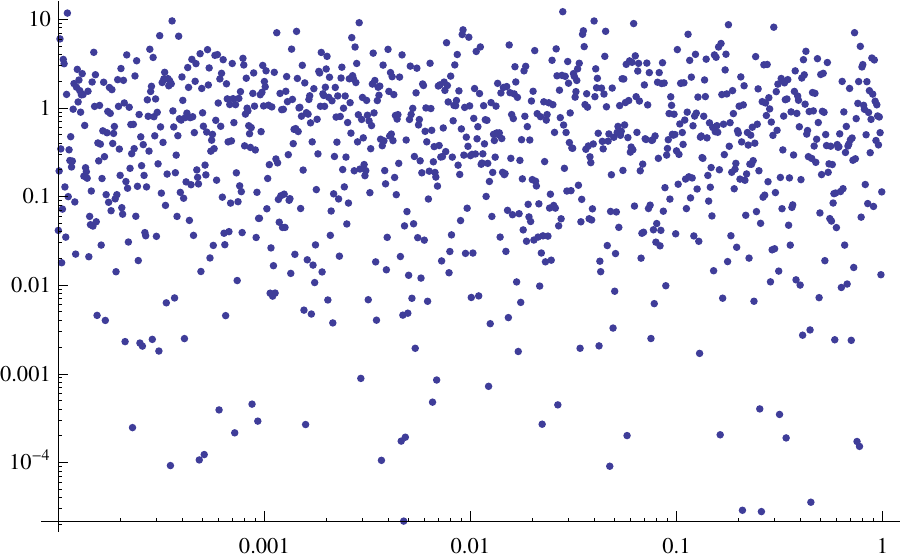}\\
  \includegraphics[width=0.32\textwidth,clip]{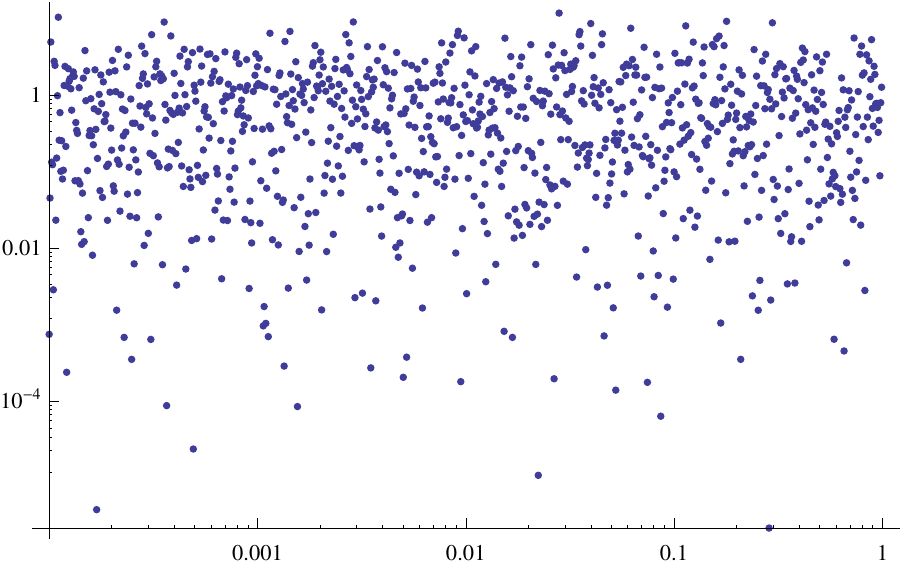}
  \includegraphics[width=0.32\textwidth,clip]{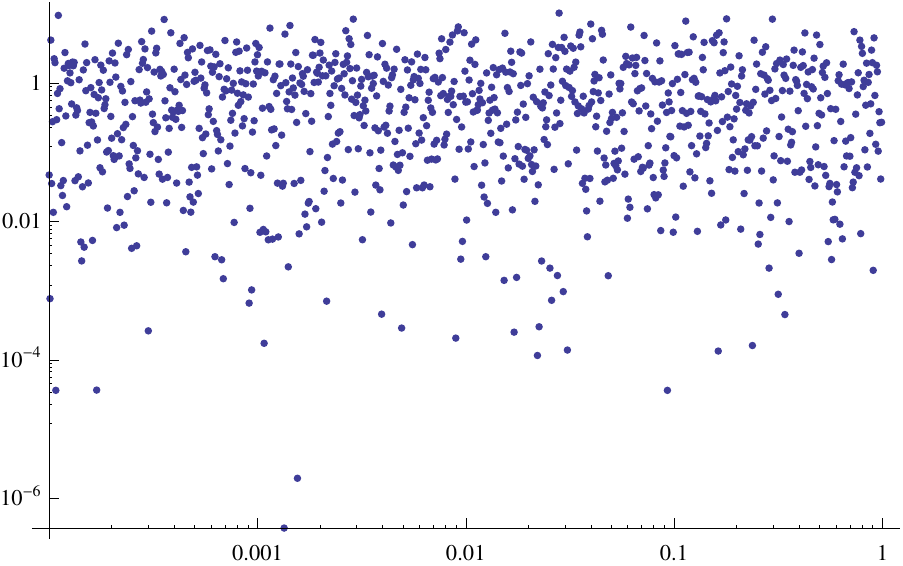}
  \includegraphics[width=0.32\textwidth,clip]{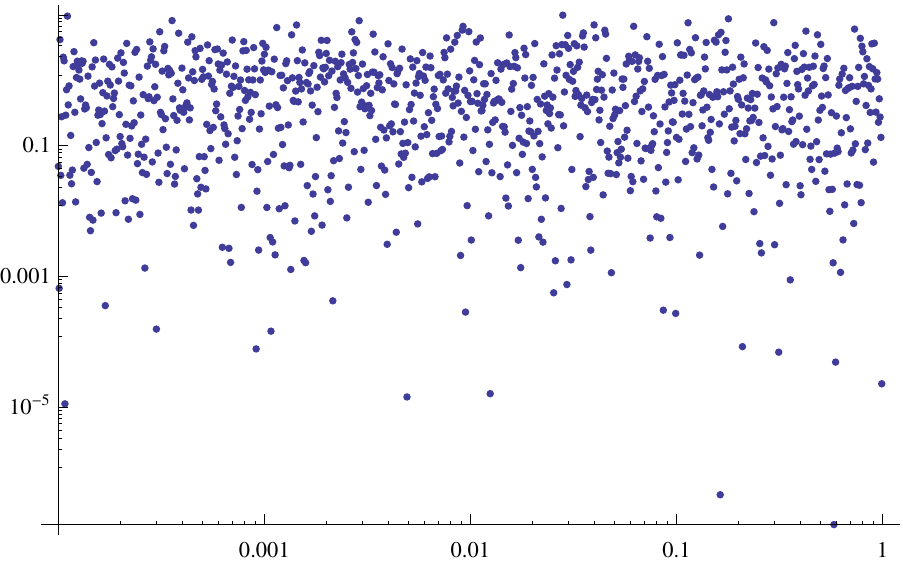}
  \caption{Distribution in $\chi^2$ values versus $x$ for 
fits to a valence-like quark distribution with increasing highest order of 
Chebyshev polynomials, going from $1^{\rm st}$ (top left) from left to right
for the first row, then from left to right for the second row until the 
$6^{\rm th}$ order. Note that the vertical axis is not the same in all 
plots as the number of points with very large $\chi^2$ decreases with 
the highest order of the polynomial. In this example, for two terms onwards there is a 
fairly random distribution. }
  \label{fig:Fig34pdf}
  \end{figure}

In Fig.~\ref{fig:Fig34pdf} we see the $\chi^2$ distribution for 
increasing highest order of Chebyshev polynomials. This reflects the deviation
from the original function, i.e.~for one term in the polynomial there are 
many points with high $\chi^2$ but the distribution becomes roughly 
as expected even with just two terms. There is no obvious structure as
a function of $x$ in 
the figure and extremely 
few $\chi^2$ values, if any, take values greater than 10. 
The total $\chi^2$ improves dramatically when going from 1 to 2 
terms in the polynomial. After this it decreases by a few units 
with  each additional polynomial up to $6^{\rm th}$ order, though this is difficult to appreciate from the plots. In some cases  
after $n$ terms in the polynomial we start fitting noise, i.e.~the $\chi^2$ becomes 
lower than that for the true function. The number of polynomials required for this varies,
but in the most extreme case happened with just 2. Somewhere between 4--6 is more common, 
but this feature is not always present when using 6 terms. 
Fits were also performed using Legendre polynomials. Since a term 
including up to the $n^{\rm th}$
order Legendre polynomial is just a re-expression of a term including up 
to the $n^{\rm th}$-order Chebyshev
polynomial, 
the fit quality is the same within numerical accuracy. However, the 
coefficients are more correlated in the case of the Legendre polynomials, presumably 
because each term has less variation with $x$ in this case. We have also 
checked that variation of the size of the error on the pseudo-data leads to 
no significant difference in the results, other than the values of the $\chi^2$,
which becomes higher for poor fits as the error decreases and {\it vice 
versa}. For $3\%$ uncertainty once the typical deviation is within a percent 
or so the $\chi^2$ becomes close to one per point.   

  \begin{figure}[htb!]
  \centering
\includegraphics[width=0.48\textwidth,clip]{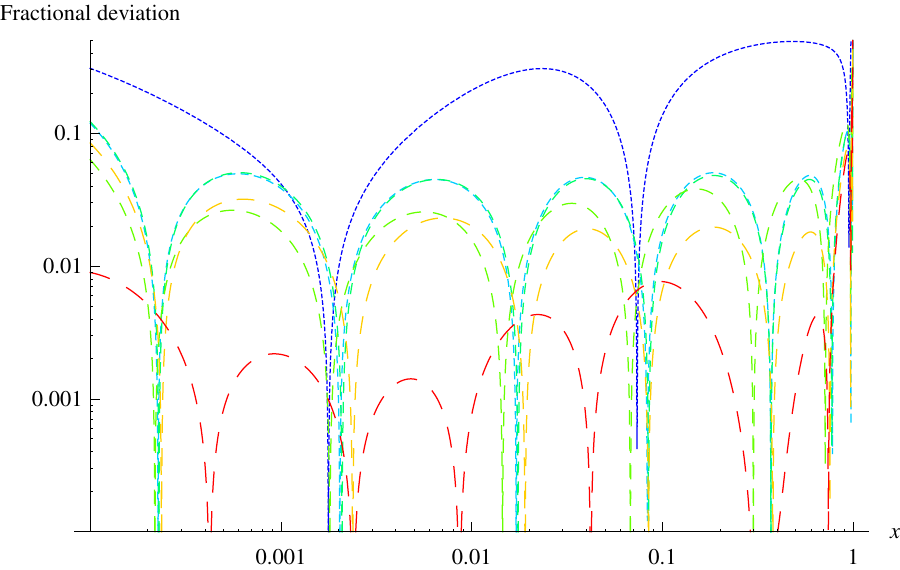}
\includegraphics[width=0.48\textwidth,clip]{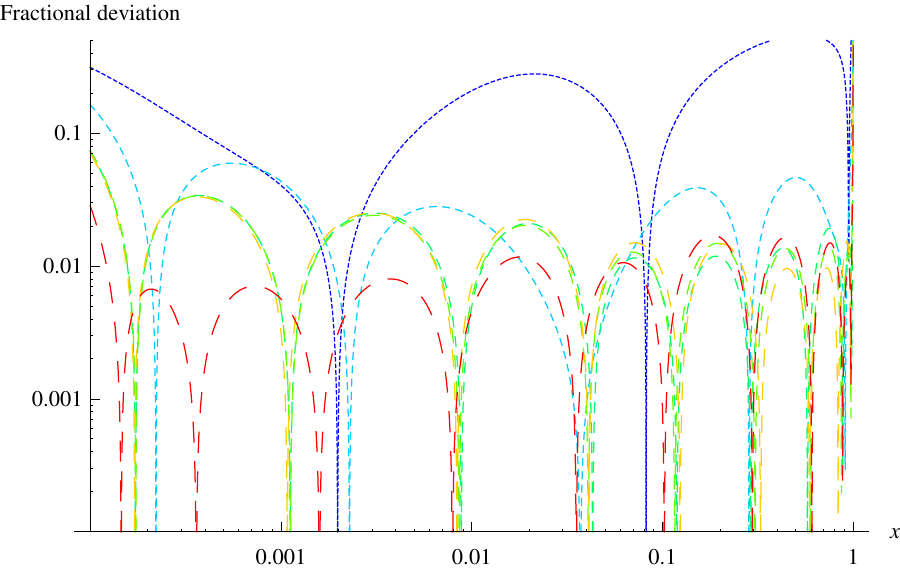}
  \caption{Two examples of the fractional deviation 
between fitted function and true function for 
fits with increasing highest order of Chebyshev Polynomials for 
sea-like distributions. The dash length decreases as the highest order of the 
polynomial increases. The order of the polynomial also increases across the visible spectrum (i.e.~dark blue to red).}
  \label{fig:SeaFig34}
  \end{figure}

\begin{figure}[htb!]
  \centering
  \includegraphics[width=0.32\textwidth,clip]{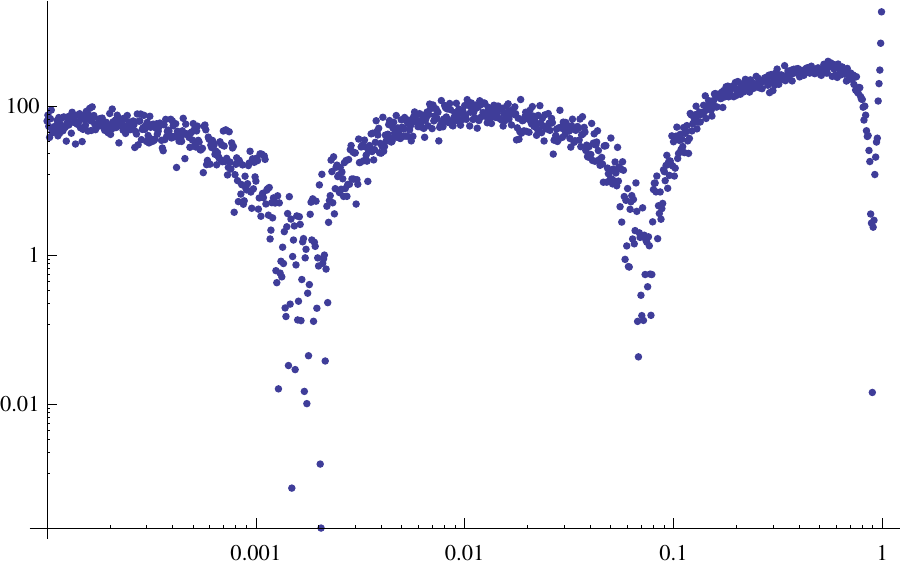}
  \includegraphics[width=0.32\textwidth,clip]{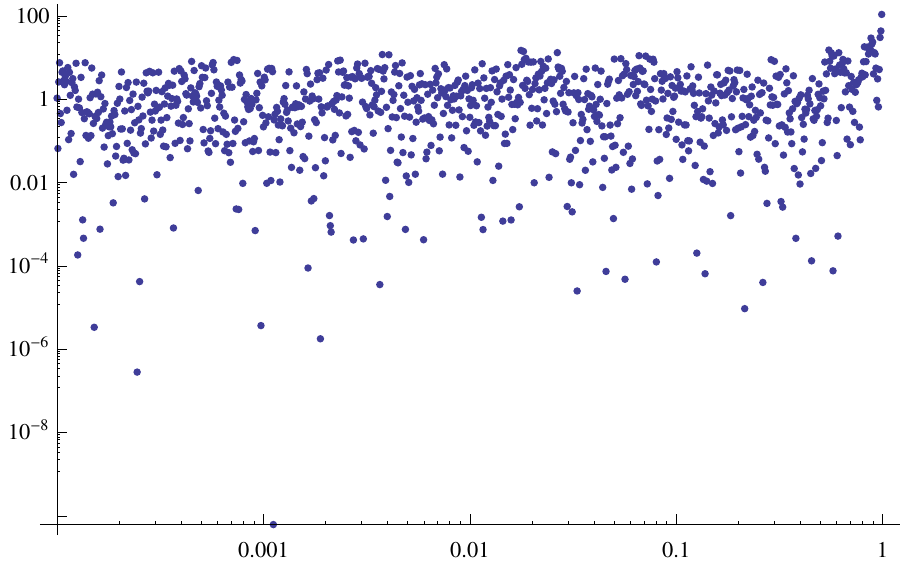}
  \includegraphics[width=0.32\textwidth,clip]{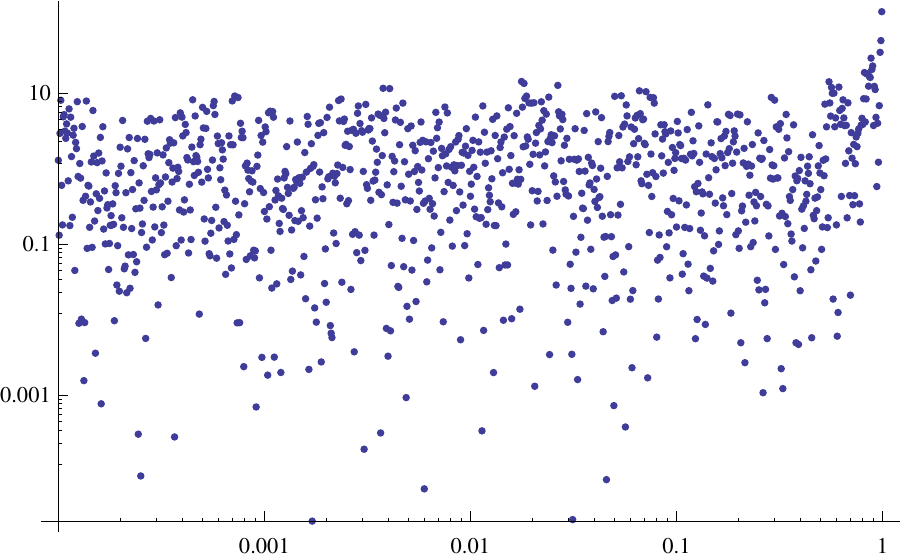}\\
  \includegraphics[width=0.32\textwidth,clip]{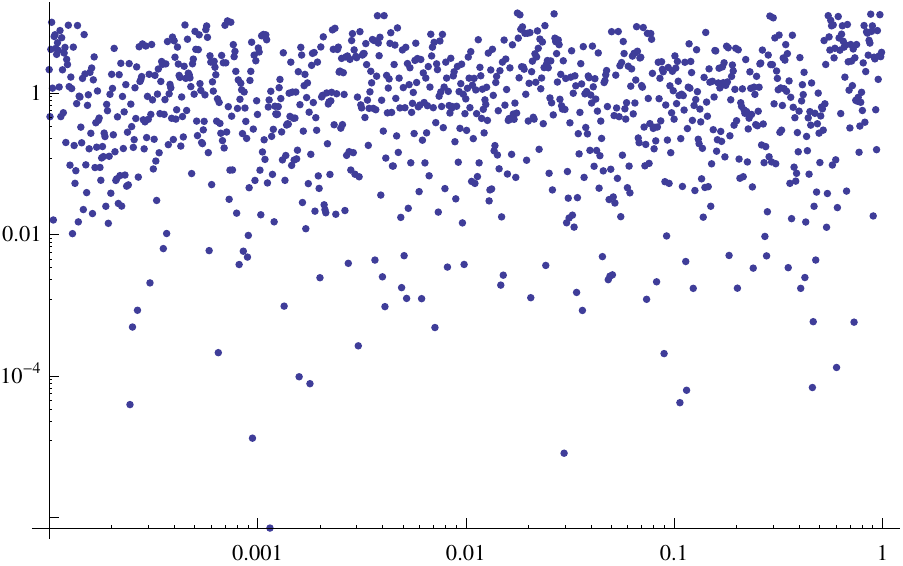}
  \includegraphics[width=0.32\textwidth,clip]{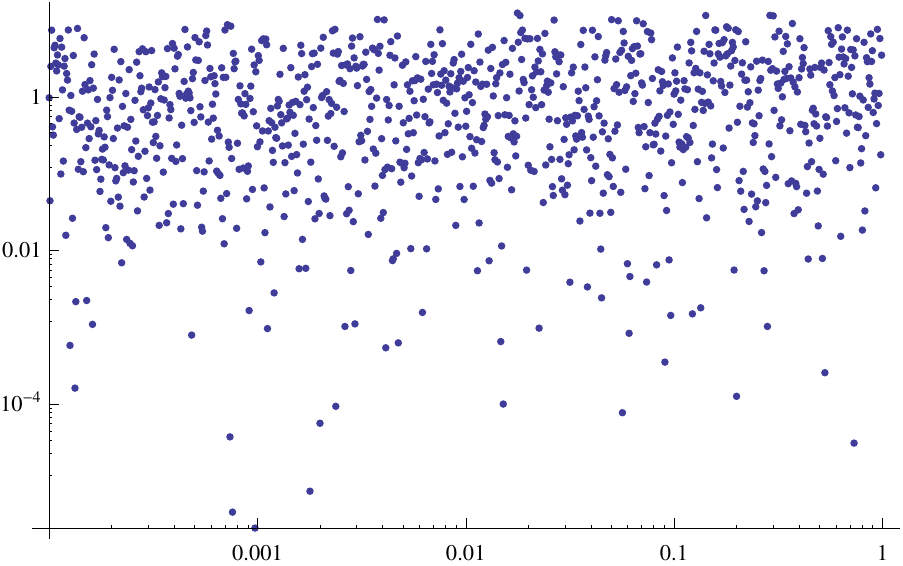}
  \includegraphics[width=0.32\textwidth,clip]{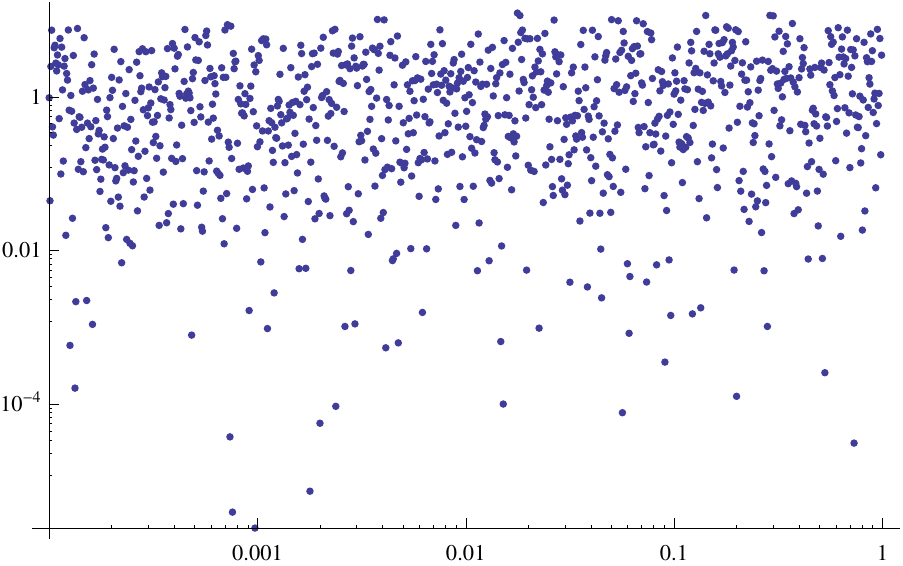}
  \caption{Distribution in $\chi^2$ values versus $x$ for 
fits to a sea quark type of distribution with increasing highest order of 
Chebyshev Polynomials, going from $1^{\rm st}$ (top left) from left to right
for the first row, then from left to right for the second row until the 
$6^{\rm th}$ order. Note that the vertical axis is not the same in all 
plots as the number of points with very large $\chi^2$ decreases with 
the highest order of the polynomial. In this example, 
for four terms onwards there is a 
fairly random distribution. There is 
distinct structure for one term, and for two and even three terms a 
cluster of badly fit points at high $x$.}
  \label{fig:SeaFig33}
  \end{figure}

The exercise is repeated for a sea quark type distribution, i.e.~falling more quickly at 
high $x$ with increasing $x$, and growing far more quickly at small $x$ with decreasing $x$, 
and for which there is no strong sum rule constraint. For this type of 
function the convergence to a very good fit is a little slower. Again one term in the 
polynomial gives a very poor fit and deviations $\sim 10\%$. This time addition of another
term reduces the deviation to 3--4$\%$ at most and 4 polynomials usually results in deviation
$\leq 2\%$ (except at high $x$). For deviations largely guaranteed to be $< 1\%$ 6 terms in the
polynomial are usually required. The results for two examples are shown in 
Fig.~\ref{fig:SeaFig34}. The greater difficulty in obtaining the excellent description
in this case is presumably due both to the lack of the sum rule constraint in the function and 
the much wider variation in values for the PDF than the valence quark case. However, we should 
note that in this case of the sea quark distribution (or gluon distribution) the minimum 1 sigma 
uncertainty in the MSTW2008 input PDFs is $\sim 5\%$. Hence, the deviations with 4 parameters 
are again much smaller than the intrinsic uncertainty in the function and 4 parameters is 
very likely to be more than sufficient. In Fig.~\ref{fig:SeaFig33} we see the $\chi^2$ 
distribution for increasing highest order of Chebyshev polynomials. Again we see that the distribution 
becomes essentially random with 3--4 terms in the polynomial, but it happens a little more 
slowly than in the case of a valence-like distribution. The total $\chi^2$ values suggest 
that there is little evidence for over-fitting
of the sea distribution until we get to at least 6 terms with the Chebyshev polynomial.   

  \begin{figure}[htb!]
  \centering
\includegraphics[width=0.48\textwidth,clip]{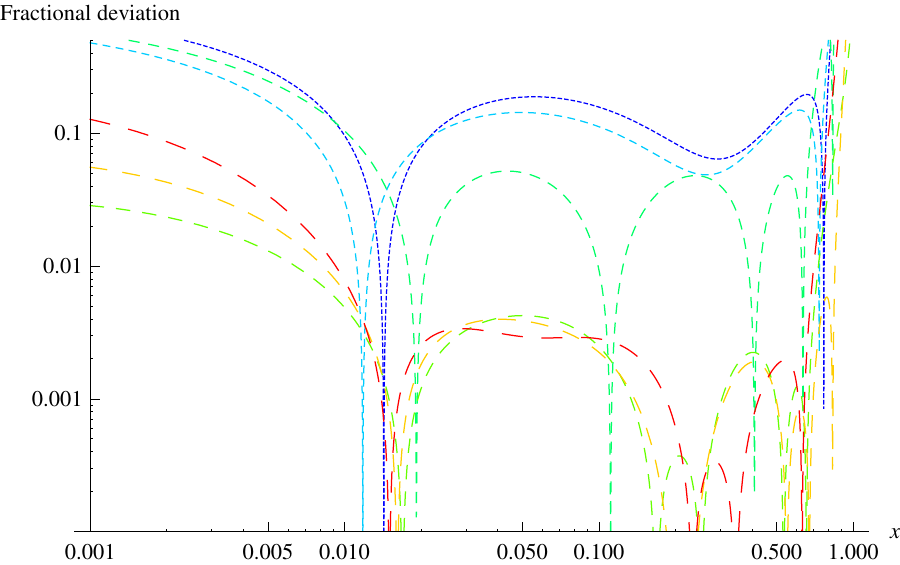}
\includegraphics[width=0.48\textwidth,clip]{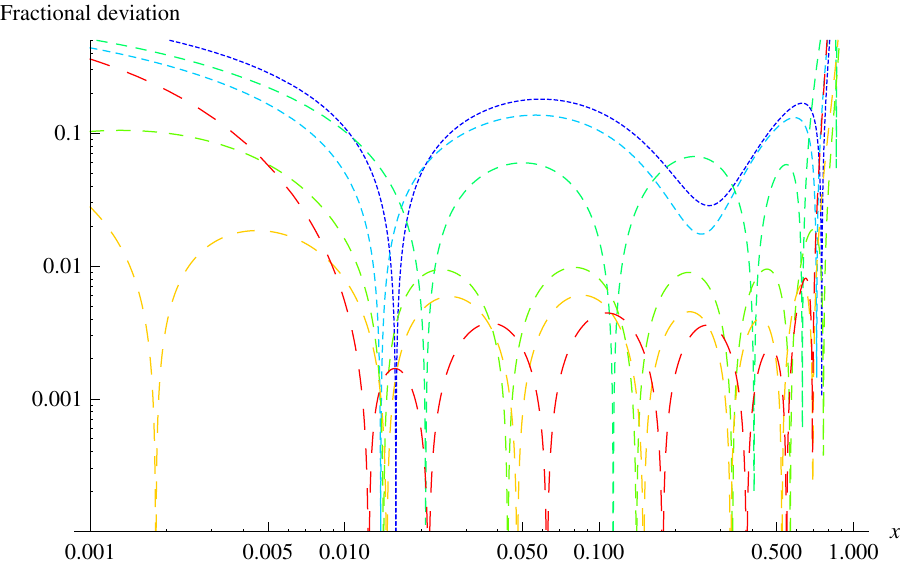}
  \caption{Two examples of the fractional deviation 
between fitted function and true function for 
fits with increasing highest order of Chebyshev Polynomials for 
valence-like distributions with 1000 pseudo-data between $0.01< x < 0.68$.
The dash length decreases as the highest order of the polynomial increases.
The order of the polynomial also increases across the visible spectrum (i.e.~dark blue to red).}
  \label{fig:valrestrict}
  \end{figure}
  \begin{figure}[htb!]
  \centering
  \includegraphics[width=0.24\textwidth,clip]{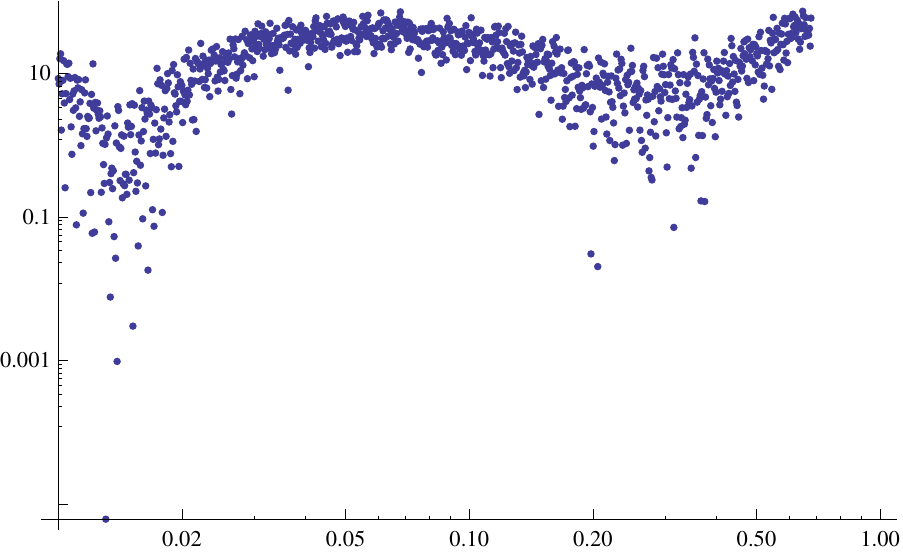}
  \includegraphics[width=0.24\textwidth,clip]{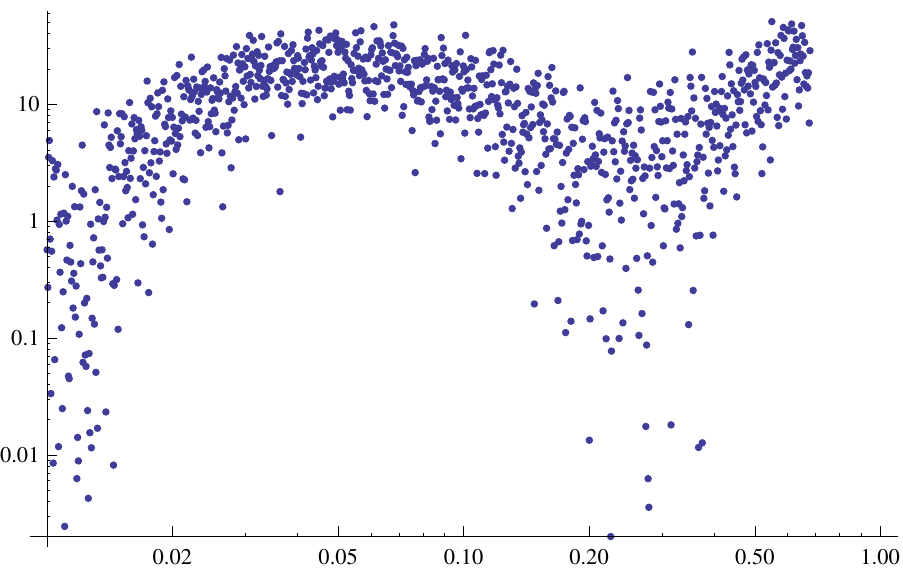}
  \includegraphics[width=0.24\textwidth,clip]{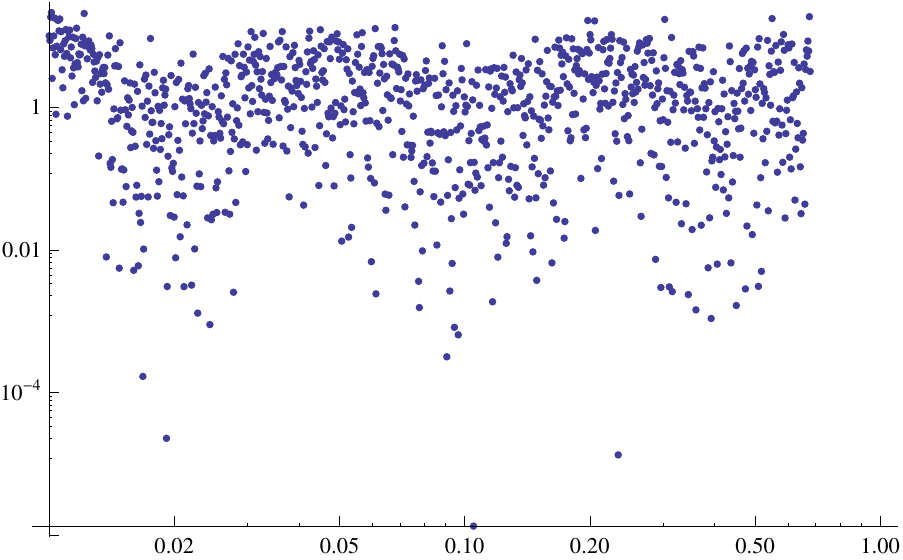}
  \includegraphics[width=0.24\textwidth,clip]{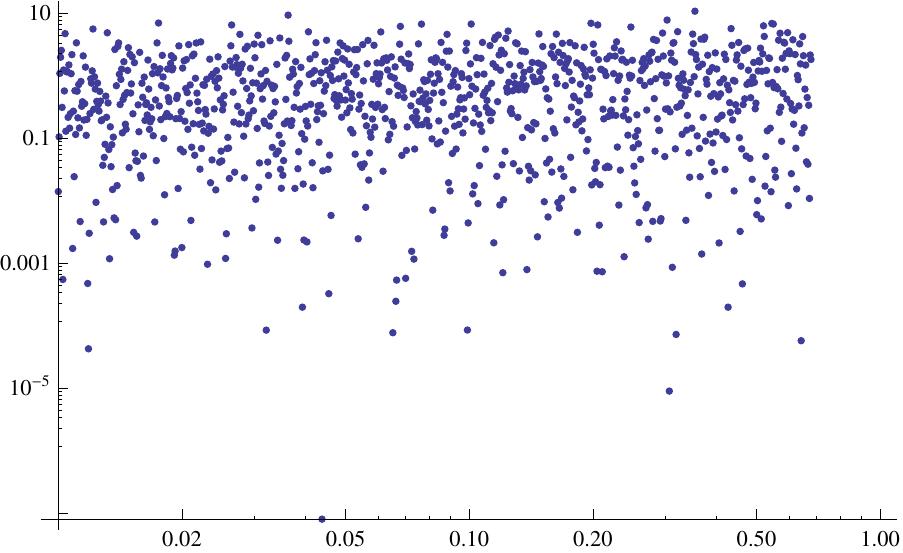}
  \caption{Distribution in $\chi^2$ values versus $x$ for 
fits to a valence quark type of distribution with 1000 pseudo-data between $0.01< x < 0.68$ for increasing highest order of Chebyshev polynomials, going from $1^{\rm st}$ to $4^{\rm th}$ order from left to right. 
Note that the vertical axis is not the same in all 
plots as the number of points with very large $\chi^2$ decreases with 
the highest order of the polynomial. In this example, 
there is structure for one, two and to 
some extent three terms, and only with four terms is there a 
fairly random distribution of $\chi^2$ values.}
  \label{fig:valrestrictchi}
  \end{figure}

We also investigate the case of fitting pseudo-data for a valence-like distribution, but where 
the generated pseudo-data are generated only for $0.01<x<0.68$, rather than for all $x$ down 
to $x=10^{-4}$. This is a more realistic situation, since valence quarks are only constrained 
by data in roughly this region. In this case  two terms in the
polynomial often give quite a very poor fit in some $x$ regions within the fit range with  
deviations $\sim 10\%$. The addition of another
term reduces the deviation to 4--5$\%$ at most and 4 polynomials  usually results in deviation
$\leq 1\%$, except at very high $x$ and much lower $x$ than the range of data. 
The results for two examples are shown in Fig.~\ref{fig:valrestrict}. The addition of more than
4 terms improves the comparison to the true function for the range of $x$ containing data, but
can increase the deviation in the very small-$x$ regime. It can also often lead to over-fitting 
of the data points, which presumably contributes to the variation in the very low-$x$ range. 
The $\chi^2$ distribution as a function of $x$ is shown in Fig.~\ref{fig:valrestrictchi}. As in 
the other examples, a uniform distribution sets in when there are 4 terms in the polynomial. 
Hence, again four terms seems sufficient, but there is more evidence that more than four terms 
corresponds to over-fitting in this case.

  \begin{figure}[htb!]
  \centering
\includegraphics[width=0.6\textwidth,clip]{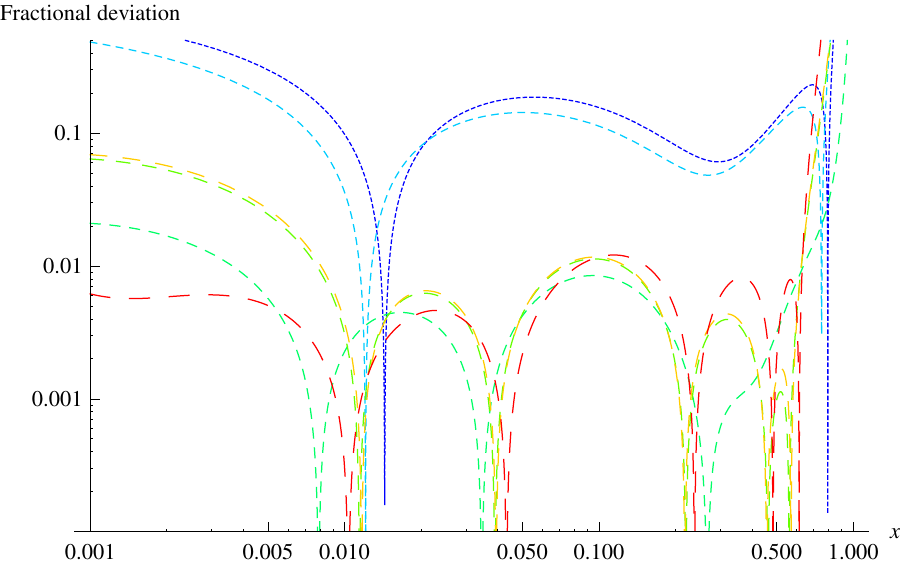}
  \caption{The deviation 
between fitted function and true function for 
fits with increasing highest order of Chebyshev polynomials for 
a valence-like distribution with 100 pseudo-data between $0.01< x < 0.68$.
The dash length decreases as the highest order of the polynomial increases.
The order of the polynomial also increases across the visible spectrum (i.e.~dark blue to red).}
  \label{fig:valrestrict100}
  \end{figure}

Finally we try fitting to the same type of pseudo-data, i.e.~for a valence-like distribution
with points $0.01<x<0.68$, but with only 100 pseudo-data points rather than 1000. The results are
shown in Fig.~\ref{fig:valrestrict100}. In this case the convergence is quicker, with 
deviation from the true function  of $1\%$ or less in the range of $x$ where points exist with 
only three terms in the polynomial. In this case there is then no further improvement with
extra parameters. For 100 points the match to the true function is already as good as the 
representation of the true function by 100 data points allows.

Overall we see that there is some dependence of the quality of the fit using a given number of 
terms in the Chebyshev polynomial on shape of the PDF, whether there is 
a constraint on the function being fit, the $x$ range of the data representing the function
and the number of data points. However, it always seems to be the case that 4 terms are 
sufficient to get an accuracy considerably better than the uncertainty on the MSTW2008 input 
distributions. It also seems to be the case that having many more than 4 terms leads to a distinct 
danger of over-fitting. Also, if there are relatively few data points then it is impossible 
to get a very accurate representation of the true function, and again too many terms in the 
polynomial can lead to over-fitting and instability in extrapolation to ranges of $x$ outside 
the data constraint. This is relevant for PDFs with relatively weak data constraints, such as 
$s - \bar s$, $\bar d -\bar u$ and possibly 
$s + \bar s$.

\section{Impact of Extended Parameterisations on PDF fits}

  \begin{figure}[htb!]
  \centering
\vspace{-6cm}
  \includegraphics[width=0.95\textwidth,clip]{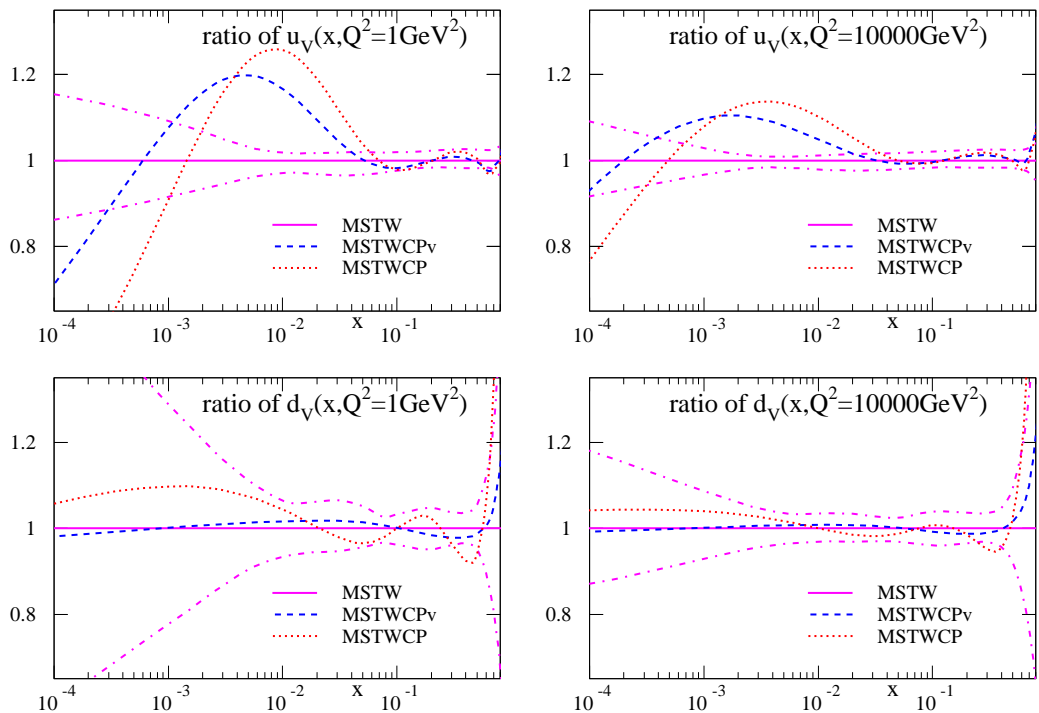}
\vspace{-6cm}
  \caption{The change in the valence PDFs extracted from the MSTW2008 
type fit using 
Chebyshev polynomials for the valence quarks only (MSTW2008CPv) and for valence and sea quarks (MSTW2008CP) compared to the original MSTW2008 PDFs at NLO with their $68\%$ uncertainties given by the dot-dashed lines.}
  \label{fig:Fig38}
  \end{figure}

Having studied the effects of fitting different `extended' input PDF parameterisations
to pseudo-data, we now investigate their effect in the real case of fitting to 
experimental data. The experimental data points are scattered over a wide range of $Q^2$ values, 
so both the evolution and the input distributions are required to be correct. Also, the data points are for structure functions, and other related 
high-energy scattering data, which, in general, not only depend on complicated 
combinations of PDF flavours, but are also related to them via the convolution with 
perturbative coefficient functions for the specific process. We perform the fits at 
next-to-leading order (NLO) in QCD perturbation theory, though MSTW also produce PDFs 
at leading order (LO) and next-to-next-to leading order\footnote{At NNLO it is necessary to make some 
approximations in modelling unknown coefficient functions for some processes.} (NNLO), and we will 
discuss NNLO results, which are very similar, later. 
Here, we perform fits to exactly the same data as used for the MSTW2008 PDF analysis
\cite{Martin:2009iq}, and adopt 
all the same theory decisions, e.g. heavy flavour schemes, nuclear target corrections, 
etc., but now with extended input PDF parameterisations.

  \begin{figure}[htb!]
  \centering
\vspace{-6cm}
  \includegraphics[width=0.95\textwidth,clip]{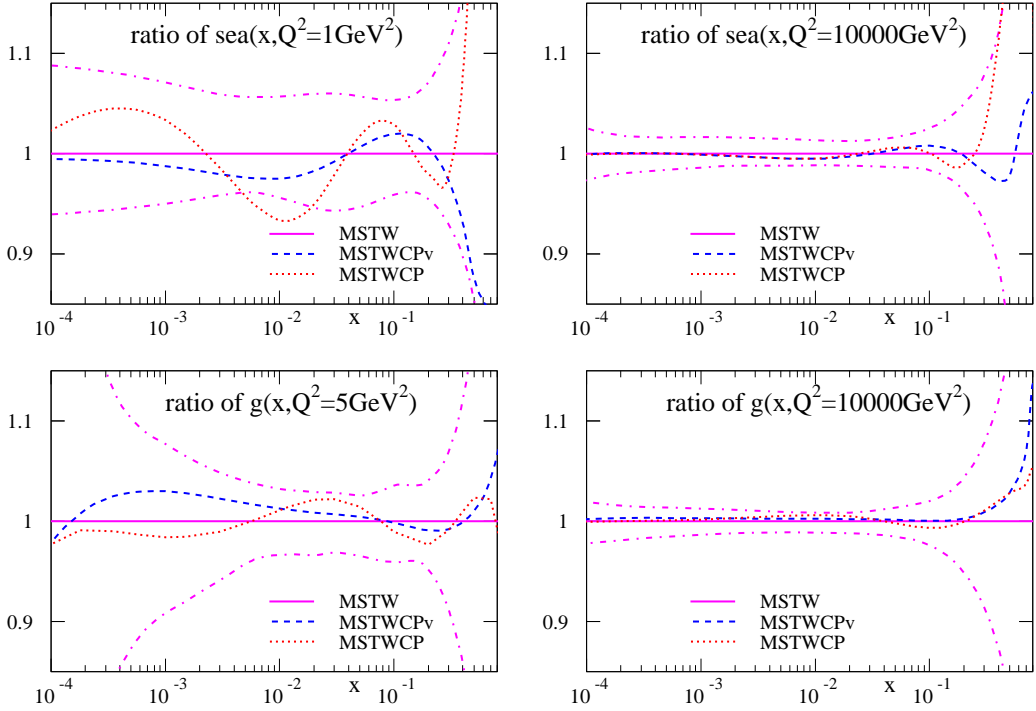}
\vspace{-6cm}
  \caption{The change in the sea and gluon PDFs extracted from the MSTW2008 type fit using 
Chebyshev polynomials for the valence quarks only (MSTW2008CPv) and for valence and sea quarks (MSTW2008CP) compared to the original MSTW2008 PDFs at NLO with their $68\%$ uncertainties shown by the dot-dashed lines. The gluon is shown at $Q^2=5~\GeV^2$ rather than the input scale of $Q^2=1~\GeV^2$ as the fact it 
goes negative at small $x$ at the latter $Q^2$ makes a ratio plot unclear.}
  \label{fig:Fig39}
  \end{figure}

To begin, we apply an extended parameterisation with
Chebyshev polynomials of highest order $n=4$ only for the valence quark PDFs: $u_V$ and $d_V$.  
The resulting improvement to the global fit is 
quite minor; corresponding to $\Delta \chi^2 =-4$ compared to a total of 2543 for 2699 
data points in the MSTW2008 fit. There is a large improvement in the 
description of the BCDMS structure function data \cite{Benvenuti:1989rh}, but a deterioration
in the fit to NMC structure function data \cite{Arneodo:1996qe}. This is very similar 
to the results 
obtained previously by adding just an $ax^2$ term to the valence parameterisations
\cite{Thorne:2010kj},
and, indeed, is similar since Chebyshev polynomials in $(1-2\sqrt{x})$
with highest order $n=4$ 
do add an $x^2$ term, but also a $x^{1.5}$ term. As in this previous study,
the significant change is in $u_V(x)$ for $x \leq 0.03$ at $Q^2=10^4~\GeV^2$. However, it is a larger 
change than previously. The comparison to the MSTW2008 PDFs at $Q_0^2=1~\GeV^2$ and 
at $Q^2=10^4~\GeV^2$ with uncertainty bands is shown in Figs.~\ref{fig:Fig38}
and ~\ref{fig:Fig39}.

A fit is also attempted using $y=(1-2x^{0.25})$ as the argument
of the Chebyshev polynomials, which overlaps with using an extra $x^{0.25}$
term in the parameterisation as tried before. As in the previous study \cite{Thorne:2010kj},
this results in an improvement in the fit of less than one unit in $\chi^2$, and much less 
change in PDFs. Hence, the use of $y=(1-2\sqrt{x})$ receives further justification. 

We then also applied the extended Chebyshev interpolating polynomial 
to the sea distribution. For the sea, the MSTW2008 parameterisation
was exactly the same form as for the valence quark, i.e.~as in (\ref{eq:MSTWparam}),
so the extended parameterisation also has the same form as that for the valence quark PDFs.
The only difference is that 
there is no number sum rule directly constraining one parameter, unlike the valence quark PDFs 
where the normalisation is constrained.  
We apply the extended parameterisation for the sea distribution 
by replacing $(1+\epsilon x^{0.5}+\gamma x)$ by a term including Chebyshev polynomials with highest order $n=4$. 
We tried also extending the parameterisation for the gluon distribution.
For the MSTW2008 gluon distribution
we had the  more flexible parameterisation in (\ref{eq:MSTWgluon})
where, in practice, the second term chooses a negative normalisation. The 
normalisation of the first term is set by the momentum sum rule for the 
PDFs, so in practice there are 7 free parameters.
We replaced the polynomial in the first term by one 
including Chebyshev polynomials with highest order $n=4$, but both the quality 
of the fit and the resulting PDFs were essentially unchanged.
This, together with the fact that the gluon already has 7 free parameters, 
suggests that the extended Chebyshev polynomial parameterisation is not necessary. 
However, using Chebyshev polynomials is a more efficient way of expressing the polynomial 
in the first term of (\ref{eq:MSTWgluon}), so we replaced the previous form 
by an entirely equivalent term using Chebyshev polynomials with highest order 
$n=2$. The resulting global fit for the parameterisation with the extended
sea quark distribution, and the formally modified, but equivalent, gluon 
distribution, has $\Delta \chi^2 =-29$. The improvement is mainly  for 
BCDMS structure function data and E866 Drell--Yan \cite{Webb:2003bj} and Drell--Yan asymmetry 
data \cite{Towell:2001nh}. 
Again there is a deterioration in the description of the NMC structure function 
data, but this time a slight improvement in Tevatron lepton asymmetry 
\cite{Acosta:2005ud,Abazov:2007pm} and $Z$ rapidity data \cite{Abazov:2007jy,Aaltonen:2010zza}. 
Details will be shown in Section 4. 
The PDFs are again shown in Figs.~\ref{fig:Fig38}
and ~\ref{fig:Fig39}, and compared to the MSTW2008 PDFs and their uncertainty. 
The change in $u_V$ is similar to the previous case. There is more change in 
$d_V$ this time, but generally within the $68\%$ uncertainty band. 
The sea quarks at 
input show some differences, sometimes a little outside the $68\%$ confidence
level uncertainty, but this is essentially washed out by evolution for $Q^2=10^4~\GeV^2$.
All other PDFs show changes that are very small compared to the MSTW2008 
uncertainty. 

In all our 
new fits we let $\alpha_S(M_Z^2)$ be a free parameter and in all cases it changes 
by only $0.0002$ or less, i.e.~a change much smaller than the uncertainty
of $+0.0012$ or $-0.0015$ at NLO~\cite{Martin:2009bu}. The set of PDFs, 
with Chebyshev polynomials of highest order $n=4$ applied to the two valence
quarks and sea, and Chebyshev polynomials of highest order $n=2$ to the gluon, 
is denoted by `MSTW2008CP' below. 

As well as studying the central values, we also investigate the uncertainties 
of the new PDFs. The standard MSTW2008 PDFs have 28 free parameters in the best fit,
but 8 are held fixed when determining the uncertainty eigenvectors because there
is too much correlation or anticorrelation between some of the parameters when all 
are left free. This freedom would have resulted in the 8 extra potential eigenvectors having an extremely 
non-quadratic
behaviour in $\Delta \chi^2$; in general behaving quadratically only in the immediate
vicinity of the minimum, then with $\Delta \chi^2$ increasing alarmingly away from this limit.
These are examples of the cases where within a limited range of parameter values the fit quality
changes extremely little with changes of some parameters. However, the actual PDFs tend to remain
very similar as the parameters vary in these cases, it simply being a matter of the redundancy in the 
parameterisation which allows almost identical PDFs with noticeably different parameterisations.

In the MSTW2008 analysis, all PDFs, except the sea (and 
$s-\bar s$ which only has two free parameters), have the small-$x$ and high-$x$ 
powers $\delta$ and $\eta$ free in the eigenvector determination, the small-$x$
power being replaced by the normalisation for the sea. For the gluon the 4 free 
parameters in the eigenvector determination are the two $\eta$ and $\delta$ values
in (\ref{eq:MSTWgluon}). For valence 
and sea quarks we then also let the coefficient of the $x^{0.5}$ term in the polynomial, 
$\epsilon$, be free.  

For the determination of the uncertainties of the MSTW2008CP PDFs, we decide to apply more consistency 
between the PDFs, and for the sea quarks we let the parameter $\delta$ be free in the eigenvector determination, rather than the normalisation.  For all the PDFs, 
other than valence quarks and the light sea, we make the same choices as usual. For the valence and sea
quark PDFs we let the coefficients of the first and third Chebyshev polynomials, $a_1$ 
and $a_3$, be free in the eigenvector determination. Hence we have one more free parameter for each of these PDFs 
and consequently $23$ rather than 20 eigenvectors. Despite having one extra free
parameter the uncertainty on sea quarks for $x\sim 0.001$--$0.01$ at $Q^2=10^4~\GeV^2$ becomes a little 
smaller, but becomes noticeably larger for very low $x$. This is undoubtedly due to 
the exchange of the normalisation for the small-$x$ power as a free parameter. 
The most significant change in the uncertainty, however, is in the same PDF as for the change in the central value, i.e.~the small-$x$ $u_V$ distribution. The uncertainty starts to become larger with decreasing $x$ at a 
higher value of $x$, i.e.~about $x=0.01$.  Indeed, it is markedly larger between $0.001$ and
$0.01$ at $Q^2=10^4~\GeV^2$, 
where there was a rather tight `neck' in the MSTW2008 uncertainty of $u_V$ near 
$x=0.003$, and is significantly larger at very small $x$ values. 
Hence, in the new MSTW2008CP analysis, the increase in uncertainty in $u_V$ is more in line with where the data 
constraint finishes, i.e.~about $x=0.01$. All 23 eigenvectors have reasonably
quadratic behaviour, the worst being comparable to the worst for MSTW2008. 
The worst is in fact the eigenvector comprising largely of the first term in the Chebyshev 
polynomial for $d_V$. This is explained by the fact that this parameter is highly 
correlated with both the small-$x$ power and the third term in the Chebyshev 
polynomial for the same PDF.

\section{Fitting PDFs with Deuteron Corrections Applied}

  \begin{figure}[htb!]
  \centering
\vspace{-1.5cm}
  \includegraphics[width=0.48\textwidth,clip]{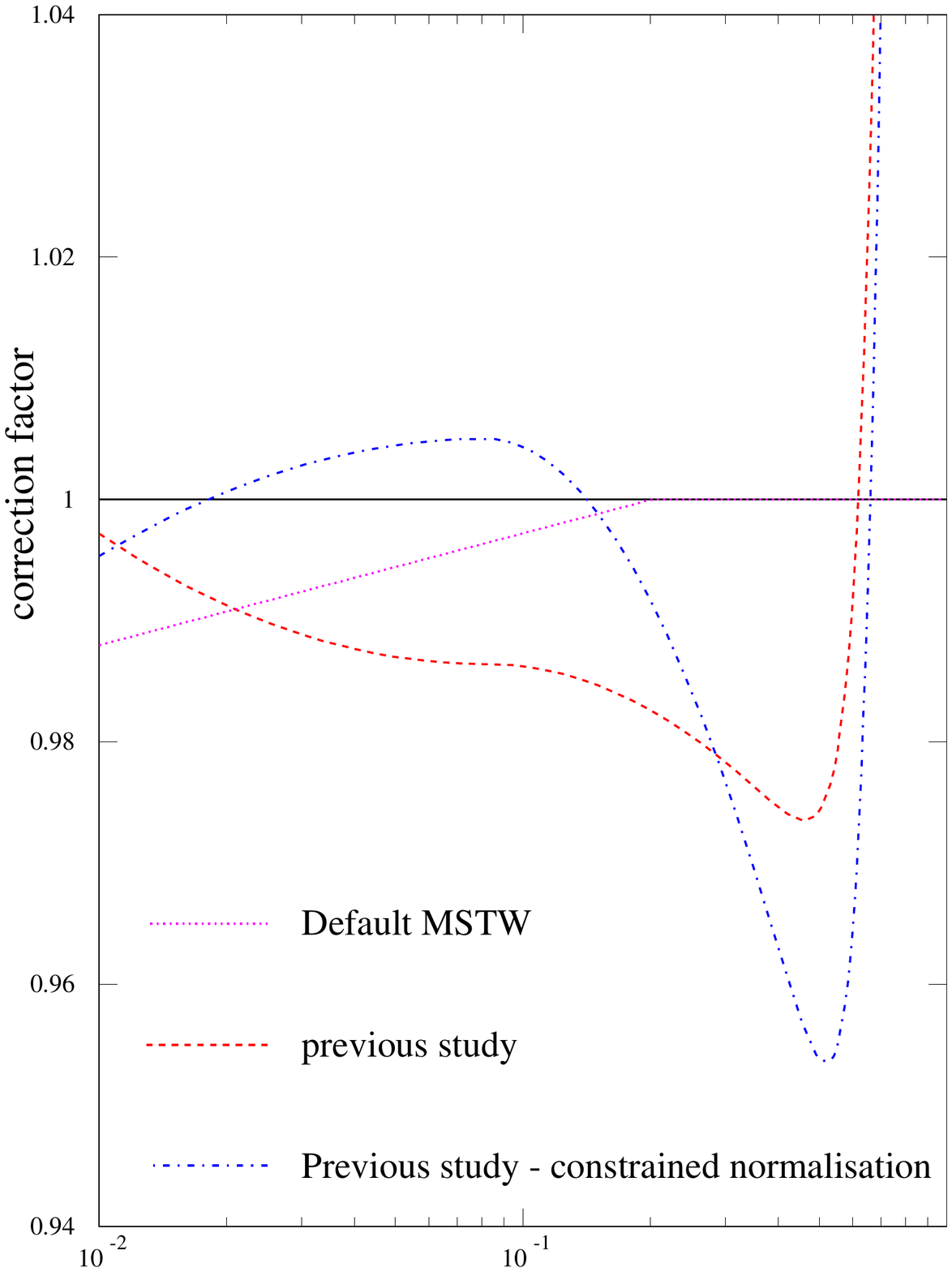}
  \includegraphics[width=0.48\textwidth,clip]{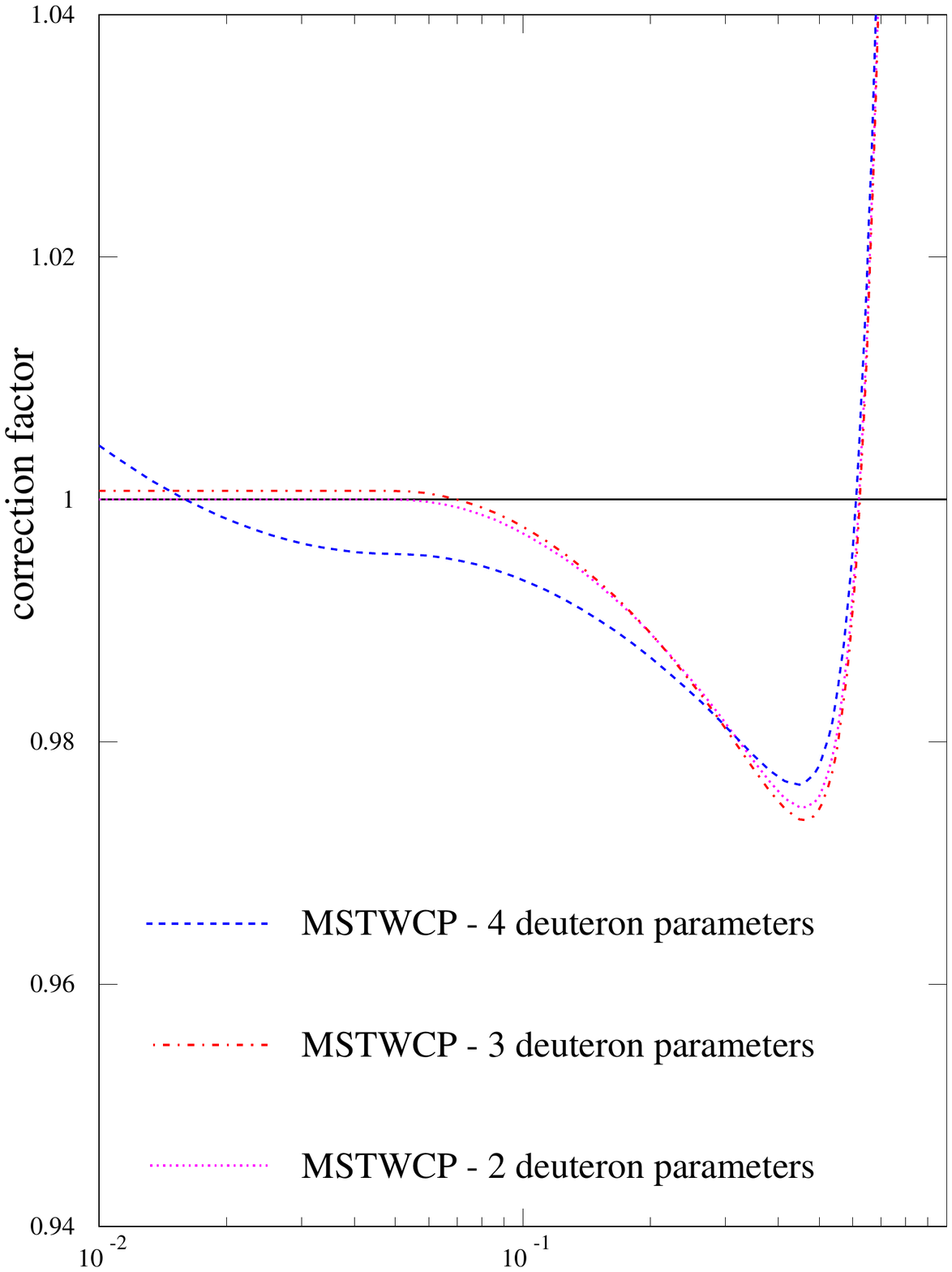}
\vspace{-1.2cm}
 \caption{The deuteron corrections applied to structure functions in 
previous versions of the MSTW2008 type fit (left) and in fits where the 
extended parameterisation for the PDFs is used (right).}
  \label{fig:deutcorrparam}
  \end{figure}

At present, and for at least the short-term future, it is necessary to include the deep-inelastic scattering data on deuteron targets in the global parton analyses, in order  to separate the $u$ and $d$ PDFs at moderate and large $x$ values. Studies of the PDFs obtained with collider data only 
\cite{Ball:2012cx,Watt:2012tq} show much bigger uncertainties for some PDF flavours.
Unfortunately, the deuteron measurements are subject to nuclear  corrections. With the increased precision and variety of data (especially with the advent of the decay lepton charge asymmetry measurements from $W^\pm$ production at the Tevatron and LHC) it is necessary to study the effect of the 
deuteron corrections in some detail. In previous PDF 
determinations we have included the nuclear corrections from \cite{deFlorian:2003qf} for neutrino structure functions taken with either lead or iron targets, 
see the discussion in section 7.3 of \cite{Martin:2009iq}, but
only included a fixed shadowing correction for small $x$ for deuteron structure functions. 
In the PDF analysis of \cite{Alekhin:2012ig}, a specific deuteron correction, 
with associated uncertainty, was applied.
Most of the other groups include no deuteron corrections (the issue does not arise in 
\cite{Aaron:2009aa}), although there have been recent specific investigations (see
e.g. \cite{Accardi:2011fa,Brady:2011hb} 
in the context of the CTEQ-JLab fits, although there are no 
corrections in \cite{Lai:2010vv}) which suggest the effect is not insignificant, 
especially at high $x$. As we will explain below, we base our study in this article on the 
assumption that there need not be deuteron corrections, but allow the fit to choose them 
if required, with some uncertainty determined by the fit quality. Hence, at the very 
least we allow some new degree of PDF uncertainty to be associated with the possibility 
of deuteron corrections. The best fit chooses some small, but significant, corrections at
high $x$.

\subsection{The parameterisation of the deuteron corrections}

In the default MSTW2008 PDF analysis, deuteron structure functions were corrected only for 
shadowing at small values of $x$ \cite{Badelek:1994qg} with a negative correction 
starting a little above $x=0.1$ and becoming as much as $-1.5\%$ near $x=0.01$. 
In \cite{Thorne:2010kj} we presented a much more detailed study. So, first, we briefly summarise the results of this investigation.
Basically, we studied the effect of allowing the deuteron corrections to be described by forms with 4 free parameters, which were allowed to vary 
with no penalty. It was found that the quality of the global fit could improve by 
a very large amount with the optimum deuteron corrections. It was particularly 
clear that the comparison to the Tevatron lepton asymmetry data used in 
the fit improved a great deal, as did the predictions for more recent versions 
of these data. The deuteron corrections required were of the expected general 
form, see e.g. \cite{Melnitchouk:1994rv,Kulagin:2010gd,Accardi:2011fa}, 
with a large positive correction at very high $x$ and a dip for 
$x \sim 0.5$. However, the dip was somewhat larger than expected, the correction 
remained negative near $x=0.1$ where it is likely to be positive due to 
antishadowing effects, and if anything the fit preferred positive corrections 
near  $x=0.03$ rather than the negative shadowing corrections. (Indeed, if the 
shadowing corrections applied in the MSTW2008 analysis were to be simply removed, and no deuteron 
corrections applied, then  the fit would improve slightly.) Hence, the adopted corrections 
seemed unsatisfactory. 

Given that the extended `Chebyshev' parameterisation discussed so far automatically
allows improvement in the global fit, and has by far the most effect on 
valence quarks and the light sea, it seems natural to investigate the 
question of deuteron corrections in this context. The deuteron corrections 
applied previously \cite{Thorne:2010kj} were of the form
\begin{equation}
F^d(x,Q^2) = {\rm c}(x)(F^p(x,Q^2) + F^n(x,Q^2))/2,
\end{equation}
where $F^n(x,Q^2)$ is obtained from $F^p(x,Q^2)$ just by swapping 
up and down quarks, and antiquarks, i.e.~isospin symmetry is assumed. 
The correction factor ${\rm c}(x)$ is taken to be $Q^2$ independent for 
simplicity and is of the form
\begin{eqnarray}
{\rm c}(x)&=& (1+0.01N_{\rm c})(1+0.01{\rm c}_1\ln^2(x_p/x)), \qquad x< x_p,\\
{\rm c}(x)&=& (1+0.01N_{\rm c})(1+0.01{\rm c}_2\ln^2(x/x_p)+0.01{\rm c_3}\ln^{20}(x/x_p)), \qquad x>x_p.
\label{eq:deutcorr}
\end{eqnarray} 
$x_p$ is a ``pivot point'' for which value the normalisation is set to be
$(1+0.01N_{\rm c})$. For $x<x_p$ there is the freedom to increase or decrease smoothly.
The same is true above $x=x_p$, but the very large power is also added to allow the
expected rapid change of the correction as $x \to 1$ due to Fermi motion. In previous studies $x_p$
was chosen to be 0.08 but here we set $x_p=0.05$. If there is 
shadowing at low $x$ and also a dip for high, but not too high, $x$ then $x_p$ 
is where the correction would take its maximum value, expected to be determined by 
antishadowing corrections.  Thus the 4 free parameters describing the deuteron correction, ${\rm c}(x)$, are the ${\rm c}_i$ and $N_{\rm c}$. We do not apply the corrections to the 
E866 data on Drell--Yan asymmetry \cite{Towell:2001nh}, and this could be improved in future. 
However, in the region of the majority of (and most precise) data the correction is very 
small. Very naively the unconstrained 
deuteron correction can simply allow the deuteron structure function 
data to be fit as well as possible while other data sensitive to the separation between 
up and down quarks determine the PDFs. However, there are other constraints, such as sum 
rules, and in practice the many different types of structure function and other data, all
depending on different combinations of flavours, in a global fit, makes the situation
more complicated. In principle, extremely precise collider data will make the fit to 
deuteron data a more-or-less direct fit of deuteron corrections, but this is not yet 
the case with present data.

The deuteron correction \cite{Badelek:1994qg} for the default MSTW2008 fit
is shown in the left of Fig.~\ref{fig:deutcorrparam}. It is negative, 
i.e.~the total correction 
factor is $<1$ below about $x=0.2$, but becomes larger in magnitude as $x$ decreases.
The correction factor for the best fit in our previous study  \cite{Thorne:2010kj}
is also shown. As explained, it is negative everywhere, except at very high $x$, which 
seems unlikely. This gives an improvement in $\chi^2$ compared to our usual global fit 
of $\sim 80$. If the normalisation at $x_p$ was fixed to be 1.005 the correction factor
obtained had the expected type of shape, i.e.~turned below 1 at the lowest $x$ and 
dipped to a minimum near $x=0.6$. However, this resulted in a fit with $\chi^2$ 30 higher
than the free deuteron correction, and as seen the dip is now $-5\%$, which is much 
lower than shadowing models tend to predict. The shape is very different from the correction with
all parameters left free. Fixing the normalisation to 1.0025 and setting ${\rm c}_2$ so that the 
dip is more like $-3\%$, results in a further deterioration of $\Delta\chi^2=5$ to the fit quality. 
This is not particularly significant, but there seemed to be a tension between the best fit and the
expected shape of the correction.

\subsection{Deuteron corrections for `Chebyshev' parameterisations}

We now repeat the exercise using our extended `Chebyshev' parameterisation for the valence quarks
and sea distribution. The quality of the fit is 86 units of $\chi^2$ better than that for the MSTW2008
fit, i.e.~a little better than when we used the standard PDF parameterisation in our previous 
study \cite{Thorne:2010kj}, and 56 units better than the fit with the same parameterisation and fixed deuteron corrections. There are very large 
improvements in the fit to (i) the BCDMS deuteron structure function data, (ii) the E866 Drell--Yan asymmetry data 
and (iii) the Tevatron lepton 
asymmetry data. The last improves from $\chi^2/N_{\rm pts.}=55/32$ in the MSTW2008 fit to $\chi^2/N_{\rm pts.}=27/32$.
 The deuteron correction is shown in the right of 
Fig.~\ref{fig:deutcorrparam}. As with our previous study, when 
all parameters are left free, the normalisation is smaller than 1, but this time by only
$0.5\%$, rather than over $1.5\%$. There is also still a tendency for the correction to 
turn up, rather than down, at lowest $x$. Since this last feature is unexpected, we investigate it by 
fixing ${\rm c}_1=0$, so there is no turn up, but no turn down either. This only changes the fit
quality by $2$--$3$ units (only a couple of data points really preferring the turn-up), and the 
normalisation is now 1.0007. We also try setting the normalisation exactly to unity; the fit quality
and deuteron correction are almost unchanged. The parameters for the different fits are shown in 
Table~\ref{tab:deut}. In all three of these fits, the dip minimises at a value 
of about 0.975 at $x=0.5$. Hence, the deuteron correction is stable as the number of parameters 
held fixed changes, as is the quality of 
the fit. Moreover, with the exception of the slight tendency to 
prefer an upturn near $x=0.01$, the shape is as predicted by standard models
\cite{Melnitchouk:1994rv,Kulagin:2010gd,Accardi:2011fa}.\footnote{We may 
compare our correction with that obtained in a study \cite{Owens:2012bv} 
which appeared after the submission of our paper. Our result lies between the 
smaller (CJ12min) and middle (CJ12mid) estimates of the corrections 
in \cite{Owens:2012bv}; the larger corrections (CJ12max) are disfavoured by the
fit quality. Hence, there is complete compatibility.}

\begin{table}
  \centering
    \begin{tabular}{c|c|c|c|c}
      \hline\hline
      PDF set & $N_{\rm c}$ & ${\rm c}_1$ & ${\rm c}_2$ & ${\rm c}_3\times 10^{6}$ \\
      \hline
    4 parameters  & -0.490  & 0.349  & -0.444  & 0.0340  \\
    3 parameters  & 0.070  & 0.000      & -0.608  & 0.0336  \\
    2 parameters  & 0.000  & 0.000      & -0.573  & 0.0334  \\
      \hline\hline
    \end{tabular}
\caption{The values of the parameters for the deuteron correction in the variety of fits.}
\label{tab:deut}
\end{table}

\subsection{The `MSTW2008CPdeut' fit}

We take the fit with ${\rm c}_1=0$ as the preferred fit since, 
while it produces no small $x$ shadowing, it produces
no enhancement either. It may be that the expected shadowing is not significant in practice until 
$x$ below 0.01. Also, it may be expected that the value in the region $x=0.05$ is larger than the 
1.0007 found, but this is far from certain. 
Using the PDFs from this fit, which we call `MSTW2008CPdeut', 
we find that the prediction for the higher luminosity D{\O} lepton asymmetry data 
\cite{Abazov:2008qv} integrated over
all $p_T$ greater than $25~\GeV$ gives a fit of quality $\chi^2/N_{\rm pts.}=28/12$. 
Although this number seems high, 
this is close to the best that seems possible given the fluctuations in these measurements, and is close to 
the best that we  have obtained even when fitting these data with a very high weight.   
We do not get a good fit to the data split into two different $p_T$ bins, but this seems to be a 
problem found by other PDF fitting groups \cite{Lai:2010vv,Ball:2010gb}.  
The value of $\chi^2$ for each data set is shown for the MSTW2008, MSTW2008CP 
and MSTW2008CPdeut PDFs in Table~\ref{tab:chisquared}.

\begin{figure}[htb!]
  \centering
\vspace{-6cm}
  \includegraphics[width=0.95\textwidth,clip]{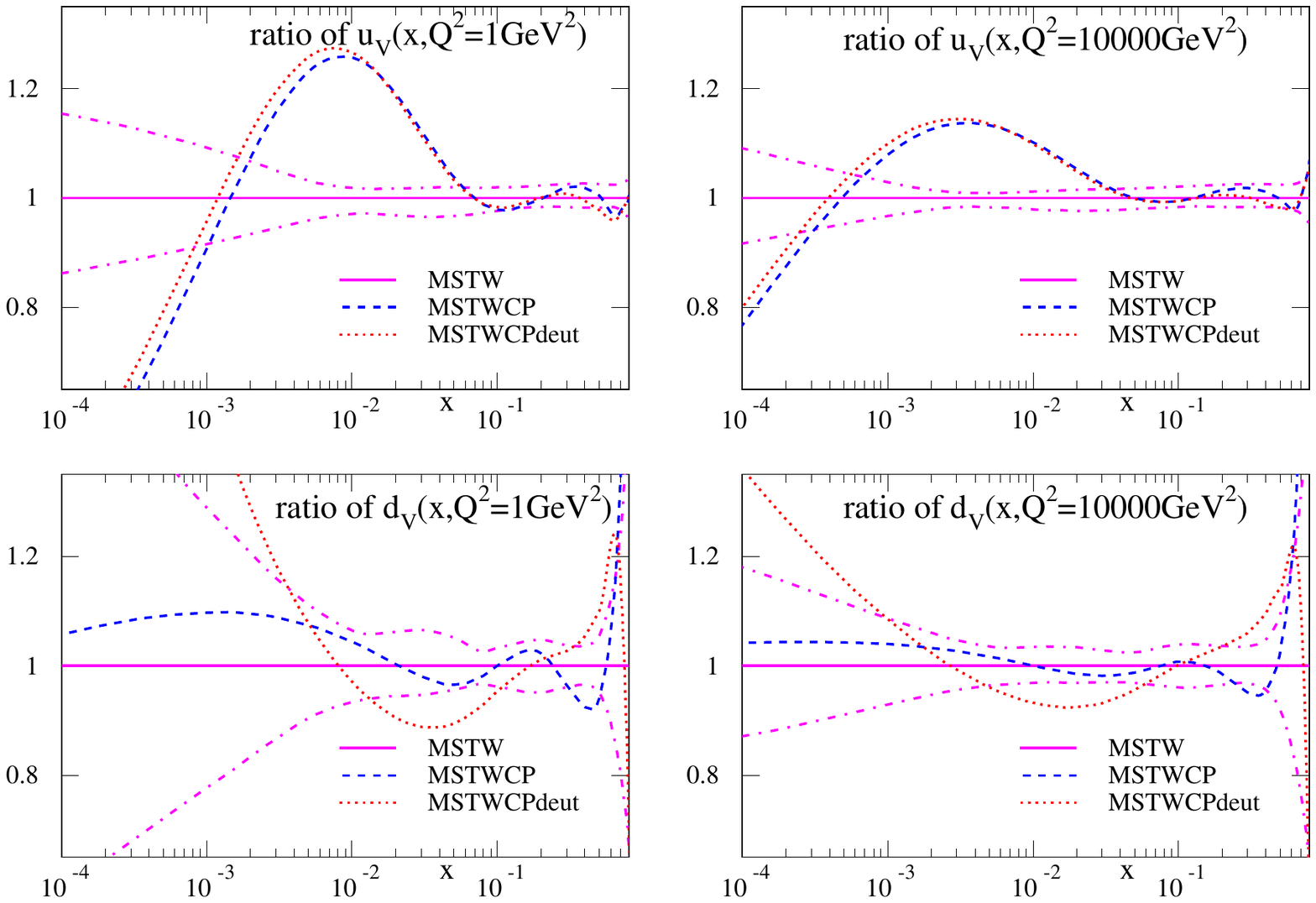}
\vspace{-6cm}
  \caption{The change in the valence quark PDFs extracted from the MSTW2008 type fit using 
Chebyshev polynomials and deuteron corrections (MSTW2008CPdeut) compared to the original MSTW2008 PDFs at NLO with their $68\%$ uncertainties shown using dot-dashed lines.}
  \label{fig:deutPDF}
  \end{figure}

The valence quarks resulting from the fit with deuteron corrections (MSTW2008CPdeut) are shown compared to the corresponding MSTW2008 PDFs 
in Fig.~\ref{fig:deutPDF}. For all PDFs, other than valence quarks, there is negligible change 
compared to those with just the extended parameterisation (MSTW2008CP), i.e.~only the sea distribution 
changes at all significantly compared to MSTW2008 and in a similar way. The $u_V$ distribution also shows little further change.
All the significance is in the change of $d_V$. This increases by a little more than the $68\%$ 
confidence level uncertainty at most near $x=0.6$ and decreases by the $90\%$ confidence level 
uncertainty at $x$ near 0.03, the precise value of $x$ decreasing with increasing $Q^2$. This is as one 
might expect, since the deuteron correction is now larger than the MSTW2008 default at low $x$, 
so the down quark can be smaller, whereas the major increase at high $x$ is in the same position 
as the minimum of the dip in the deuteron correction. Part of the effect is also due to the sum rule -- 
increases at some $x$ values must be compensated by decreases elsewhere. 

Examining the details of the improvement in the fit quality when including deuteron corrections we find that 
a large amount comes from the BCDMS data in the highest $x$ bin at $x=0.75$, the positive deuteron 
correction at this $x$ being preferred. This effect is largest at lowest $Q^2$ and hence lowest $W^2$, so the 
improvement in $\chi^2$ diminishes as the $W^2$ cut is raised. However, the data extends out to high $Q^2$ so 
a cut in $W^2$ of $30~\GeV^2$ is needed to remove the improvement completely. It is also the case that 
the improvement in the fit quality at this $x$ value is essentially all to do with the large deuteron 
correction rather than PDF changes. In general the data for which the fit improves with the free deuteron 
correction is spread over a reasonably wide range of $Q^2$ and $W^2$ and for all $x$ values between $x=0.75$ and
$x=0.01$. 

We perform a fit within the same framework as MSTW2008CPdeut but with cuts in $Q^2$ of $4~\GeV^2$ and 
in $W^2$ of $20~\GeV^2$, i.e.~equivalent to a fit in \cite{Thorne:2011kq}. With these cuts the $\chi^2$ for the
fit with free deuteron corrections and 4 Chebyshev polynomials for the quark input parameterisation is 60 
units better than our standard fit rather than 86 for the standard cuts. It is concentrated in the same data 
sets, i.e.~the BCDMS deuteron structure function data and the Tevatron asymmetry data. However, in the former
the improvement is reduced due to a loss of some of the most sensitive points, including some at $x=0.75$.
For the Tevatron asymmetry data the raised $Q^2$ cut for the standard fit already improves the fit since it
removes some of the constraining data on the up and down quark separation in the region $x\sim 0.05$. 
The form of the deuteron correction with the raised cuts is very similar to that with the full cuts. 
The ratio of the PDFs in the two fits with raised cuts is similar to that with the usual cuts, but 
the largest differences are reduced to about $60$-$70\%$ of the size with larger cuts. There is no real 
evidence that the change between our standard PDFs and those with the extended parameterisation and free 
deuteron corrections is altered by more conservative cuts other than by the fact that the fairly significant 
loss of data reduces the change just by allowing the fits to remaining data to be slightly more self 
consistent even without extended parameterisation or modified deuteron corrections.

  \begin{figure}[htb!]
  \centering
\vspace{-2cm}
  \includegraphics[width=0.6\textwidth,clip]{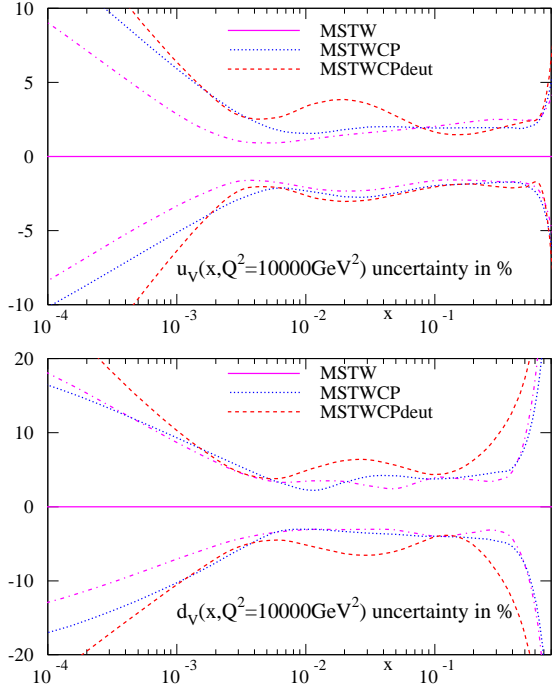}
\vspace{-2cm}
  \caption{The change in the $68\%$ uncertainty bands of the valence quark PDFs extracted from the MSTW2008 type fit using 
Chebyshev polynomials and deuteron corrections (MSTW2008CPdeut) compared to the original MSTW2008 PDFs at NLO.}
  \label{fig:uvdvunc}
  \end{figure}
\begin{table}
  \centering
{\footnotesize
  \begin{tabular}{l|c|c|c}
    \hline \hline
    Data set & MSTW08 & MSTWCP & MSTWCPdeut \\ \hline
    BCDMS $\mu p$ $F_2$              & 182 / 163 & 173 / 163 & 177 / 163 \\ 
    BCDMS $\mu d$ $F_2$              & 190 / 151 & 168 / 151 & 143 / 151 \\ 
    NMC $\mu p$ $F_2$                & 121 / 123 & 123 / 123 & 120 / 123 \\ 
    NMC $\mu d$ $F_2$                & 102 / 123 & 101 / 123 & 103 / 123 \\ 
    NMC $\mu n/\mu p$                & 130 / 148 & 143 / 148 & 143 / 148 \\ 
    E665 $\mu p$ $F_2$               & 57 / 53   & 55 / 53   & 53 / 53 \\ 
    E665 $\mu d$ $F_2$               & 53 / 53   & 58 / 53   & 57 / 53 \\ 
    SLAC $ep$ $F_2$                  & 30 / 37   & 31 / 37   & 31 / 37 \\ 
    SLAC $ed$ $F_2$                  & 30 / 38   & 31 / 38   & 31 / 38 \\ 
    NMC/BCDMS/SLAC $F_L$             & 38 / 31   & 39 / 31   & 39 / 31 \\ \hline 
    E866/NuSea $pp$ DY               & 228 / 184 & 224 / 184 & 221 / 184 \\ 
    E866/NuSea $pd/pp$ DY            & 14 / 15   & 10 / 15   &  7 / 15 \\ \hline 
    NuTeV $\nu N$ $F_2$              & 49 / 53   & 49 / 53   & 54 / 53 \\ 
    CHORUS $\nu N$ $F_2$             & 26 / 42   & 25 / 42   & 26 / 42 \\ 
    NuTeV $\nu N$ $xF_3$             & 40 / 45   & 48 / 45   & 45 / 45 \\ 
    CHORUS $\nu N$ $xF_3$            & 31 / 33   & 34 / 33   & 32 / 33 \\ 
    CCFR $\nu N\to \mu\mu X$         & 66 / 86   & 65 / 86   & 64 / 86 \\ 
    NuTeV $\nu N\to \mu\mu X$        & 39 / 40   & 38 / 40   & 39 / 40 \\ \hline 
    H1 MB 99 $e^+p$ NC               & 9 / 8     & 8 / 8     & 8 / 8 \\ 
    H1 MB 97 $e^+p$ NC               & 42 / 64   & 40 / 64   & 40 / 64 \\ 
    H1 low $Q^2$ 96--97 $e^+p$ NC    & 44 / 80   & 44 / 80   & 44 / 80 \\ 
    H1 high $Q^2$ 98--99 $e^-p$ NC   & 122 / 126 & 121 / 126 & 120 / 126 \\ 
    H1 high $Q^2$ 99--00 $e^+p$ NC   & 131 / 147 & 129 / 147 & 129 / 147 \\ 
    ZEUS SVX 95 $e^+p$ NC            & 35 / 30   & 35 / 30   & 35 / 30 \\ 
    ZEUS 96--97 $e^+p$ NC            & 86 / 144  & 87 / 144  & 87 / 144 \\ 
    ZEUS 98--99 $e^-p$ NC            & 54 / 92   & 53 / 92   & 53 / 92 \\ 
    ZEUS 99--00 $e^+p$ NC            & 63 / 90   & 62 / 90   & 61 / 90 \\ 
    H1 99--00 $e^+p$ CC              & 29 / 28   & 28 / 28   & 31 / 28 \\ 
    ZEUS 99--00 $e^+p$ CC            & 38 / 30   & 38 / 30   & 35 / 30 \\ 
    H1/ZEUS ep $F_2^{\rm charm}$     & 107 / 83  & 108 / 83  & 108 / 83 \\ 
    H1 99--00 $e^+p$ incl.~jets      & 19 / 24   & 19 / 24   & 19 / 24 \\ 
    ZEUS 96--97 $e^+p$ incl.~jets    & 30 / 30   & 29 / 30   & 29 / 30  \\ 
    ZEUS 98--00 $e^\pm p$ incl.~jets & 17 / 30   & 16 / 30   & 16 / 30 \\ \hline 
    D{\O} II $p\bar{p}$ incl.~jets   & 114 / 110 & 117 / 110 & 113 / 110 \\ 
    CDF II $p\bar{p}$ incl.~jets     & 56 / 76   & 57 / 76   & 56 / 76 \\ 
    CDF II $W\to l\nu$ asym.         & 29 / 22   & 26 / 22   & 18 / 22 \\ 
    D{\O} II $W\to l\nu$ asym.       & 25 / 10   & 20 / 10   & 9 / 10 \\ 
    D{\O} II $Z$ rap.                & 19 / 28   & 18 / 28   & 17 / 28 \\ 
    CDF II $Z$ rap.                  & 49 / 29   & 45 / 29   & 52 / 29 \\ \hline 
    All data sets & \textbf{2543 / 2699} & \textbf{2513 / 2699} & \textbf{2457 / 2699} \\
    \hline \hline
  \end{tabular}
}
\caption{The values of $\chi^2 / N_{\rm pts.}$ for the data sets included in the global fits.  The complete references, details of corrections to data, kinematic cuts applied and definitions of $\chi^2$ are contained in~\cite{Martin:2009iq}.}
\label{tab:chisquared}
\end{table}

\subsection{Allowing for uncertainties on deuteron corrections}

We also generate uncertainty eigenvector sets whilst applying
deuteron corrections. Doing this with the deuteron corrections fixed at the position of the
best fit would be straightforward, but would not account for the uncertainty in the deuteron 
corrections themselves. Since our best fit is of roughly the form one would expect for these 
corrections, and since there is no solid basis on which to judge quite how much variation 
in deuteron corrections is allowed, 
we choose to simply let the parameters in the deuteron correction go free with no penalty. This 
is then very similar to our procedure for heavy nuclear corrections, necessary for including neutrino deep-inelastic scattering data in the MSTW2008 analysis
\footnote{We note that an NNPDF study on DIS data only noticed a small change of PDFs relative to uncertainties when nuclear corrections were added to the default fit, in which they are omitted \cite{Ball:2009mk}.}, where we take a set of 
corrections obtained from a global fit to nuclear data \cite{deFlorian:2003qf}, but multiply by a function, similar in 
form to (\ref{eq:deutcorr}), which allows variations away from the default form with no 
penalty. In that case, in  practice, the variations are small, i.e.~our fit is very compatible with 
the determined nuclear corrections and the uncertainty in the nuclear corrections determined 
by the fit quality is a few percent, which seems entirely reasonable. Here we are doing exactly the 
same thing except that we have no starting deuteron correction, other than implicitly zero correction,
to act as a template. Since deuteron corrections are expected to be small, and some groups use zero 
correction (as have we, as default, i.e.~in the MSTW2008 fit and previously, at high $x$), 
using no correction as the template seems reasonable. There is,
however, a complication. There are 4 free parameters in our deuteron correction whereas we have 
3 for our nuclear correction function, and the deuteron structure functions are closely related to 
the down quark distribution. Hence, there are strong correlations between the 4 deuteron correction
parameters, and between them and the parameters for $d_V$. It is impossible to get a stable perturbation
about the best fit in terms of eigenvectors letting all 4 deuteron parameters go free. However, 
we did not even let all parameters go free in the best fit, choosing to fix ${\rm c}_1$ in order to avoid an
unlikely tendency for the correction to grow at the smallest $x$ values. Letting it go free when 
determining eigenvectors would be inconsistent. Hence we let $N_{\rm c}, {\rm c}_2$ and ${\rm c}_3$ go free when 
obtaining the eigenvectors. The freedom in the normalisation lets the correction at low $x$ vary 
and there is not much distance between the lowest $x$ data points and the pivot point at which the 
normalisation is set.  

We attempt to construct 23 eigenvectors by letting the same parameters as before go free. However, 
this results in some severely non-quadratic behaviour. The worst case is the eigenvector comprising 
largely of the first term in the Chebyshev polynomial for $d_V$. As for the fit without 
deuteron corrections this parameter is highly 
correlated with both the small-$x$ power and the third term in the Chebyshev 
polynomial for the same PDF, but the freedom in deuteron corrections make this correlation, and
its effects, worse. This is no longer an acceptable eigenvector. Even fixing one more of the deuteron 
parameters does not help very strongly. Since the main problem is the correlation between the first
Chebyshev polynomial and the small-$x$ power, and higher-order polynomials are less influential at small
$x$, we try instead letting the second and third Chebyshev polynomials have the free parameters for
the valence quarks and sea when finding eigenvectors. This does indeed reduce the correlation between the
parameters for $d_V$. It increases the correlation between the parameters for $u_V$, but this does not
seem to translate into particularly bad behaviour of the eigenvectors, and this change provides 23 
orthogonal eigenvectors with none having worse non-quadratic behaviour than any of the 20 in the 
MSTW 2008 fit. Hence, we have a preliminary set of uncertainty eigenvectors incorporating both the 
extended `Chebyshev' parameterisation and the uncertainties due to deuteron corrections. 

When the fit with deuteron corrections (MSTW2008CPdeut) is compared to the fit using just the extended
parameterisation (MSTW2008CP), the $u_V$ uncertainty increases a little bit more quickly
below $x=0.05$. However the effect for $d_V$ is more dramatic -- it is more than twice as uncertain as that for
MSTW2008 for $x \sim 0.4$,  and also about twice as uncertain for $x=0.01$--$0.05$,
but not near $x=0.1$ where the data constraints are strongest. This expanded uncertainty
means that the MSTW2008 $d_V$ distribution is either within, or just outside the one-sigma uncertainty 
band for $d_V$ obtained in the fit
with deuteron corrections, except for $x < 0.0005$. The uncertainty for the valence quarks for both
the MSTW2008CP and MSTW2008CPdeut
sets is compared to that for the standard MSTW2008 set in Fig.~\ref{fig:uvdvunc}. For other PDFs the change 
is less significant. For
$u_V-d_V$ the uncertainty for MSTW2008CPdeut is more than twice as big as MSTW2008 for $x\sim 0.4$ and
$30$--$50\%$ bigger for $x =0.05$--$0.01$, but only slightly larger near $x=0.1$. The change in the central value 
and uncertainty for $u_V-d_V$ has important implications for the description of the lepton charge asymmetry, 
as we shall discuss in detail in Section 5.

\subsection{Variation of number of parameters in the fits}

So far in the previous two subsections we have considered the results of using 4 Chebyshev
polynomials in the expressions for the valence quarks and sea (and 2 in the gluon) as suggested 
by the results in Section 2. We also consider fits where we reduce the number to 3 (two being equivalent
to the default MSTW fits) and increase to 5 and 6 (and 3 and 4 respectively for the gluon). 
First we consider the improvement in going from MSTW2008 to sets without modified deuteron corrections. A fit with 3 Chebyshev polynomials
only improves the $\chi^2$ by 8 units, i.e.~a further 21 units is achieved with 3 extra parameters going to
the fit with 4 Chebyshev polynomials 
described in Section 3. Much of the change from MSTW2008 in the $u_V$ (and $u_V-d_V$) distribution is already 
achieved with 3 Chebyshev polynomials, but the change in extending to 4 can be larger than the uncertainty 
on the $u_V$ (and $u_V-d_V$) distributions in the MSTW2008CP set. Increasing to 5 Chebyshev polynomials
the fit quality improves by 15 units for 4 extra parameters. The change in some distributions is 
significant, as expected being proportionally greatest in $u_V-d_V$, though this time largely due to 
changes in $d_V$ rather than $u_V$. The change can be a little greater than the uncertainty for a small
range of $x$ near 0.01, but is generally much smaller, especially at higher $x$. When using 6 Chebyshev 
polynomials there is a further improvement in $\chi^2$ of 7 units for 4 parameters. Again the change in 
distributions is most significant for $u_V-d_V$, but is smaller than when increasing from 4 to 5 polynomials,
and is actually such as to reduce the largest changes in the 5 Chebyshev polynomial case. There is no 
particularly obvious case of lack of smoothness in the PDFs with 5 or 6 polynomials, although there is no 
penalty imposed to ensure this. Hence, we conclude that 4 Chebyshev polynomials for valence and sea quarks is 
certainly preferable to 3. It is arguable than 5 may be ideal rather than 4, but that the changes from 
4 onwards are very rarely large compared to the PDF uncertainty.

The case with deuteron corrections included is a little more complicated due to the interplay between 
the $d_V$ distribution and the deuteron correction. As stated above, there is only a little improvement in the 
fit quality with 4 Chebyshev polynomials compared to the default MSTW parameterisation, but the 
deuteron correction is much more sensible in the former case. If we compare the fit quality obtained with a
``sensible'' deuteron correction the improvement in $\chi^2$ is about 20 units when using 4
Chebyshev polynomials. If instead we use 3 we obtain only 7--8 of this. The PDFs make a large fraction of the 
change from the MSTW2008 set to the MSTW2008CPdeut sets, but as in the previous case, the difference can 
be bigger than the uncertainty in the MSTW2008CPdeut PDFs. With 5 Chebyshev polynomials the improvement in 
$\chi^2$ is only 3 for 4 extra parameters. The change in PDFs is proportionally biggest for $u_V-d_V$, and 
in this case is 
always within the uncertainty. With 6 Chebyshev polynomials there is an improvement in $\chi^2$ of 
10 units for 4 parameters. However, the further change in PDFs is  small except for the valence quarks; it is largest for 
$x$ values at most $x=0.02$ at $Q^2=10^4~\GeV^2$ where for $d_V$ it can be comparable to the uncertainty. 
There is no clear sign of lack of smoothness or stability in PDFs. However, for 6 Chebyshev polynomials
our variation away from the default nuclear corrections \cite{deFlorian:2003qf} in the best fit becomes large 
and unrealistic (increasing above 1 and increasing as $x$ decreases near $x=0.01$). Indeed, with a large 
number of 
parameters the position of the best fit with free deuteron corrections becomes more difficult to find. 
Our best fit is found using the Levenberg--Marquardt method 
which combines the advantages of the inverse-Hessian method and the steepest descent method
for minimisation, as described in Section 5.2 of \cite{Martin:2009iq}. It is possible that a 
local rather than global minimum is found, but this can be investigated by starting the fit from different 
places in parameter space.\footnote{There is also in principle some sensitivity to the numerical value 
below which the improvement in $\chi^2$ with successive iterations must fall for the fit to stop. This is normally 
taken as 0.1, but further reductions generally lead to changes very much smaller than the PDF uncertainty,
as has been checked in this study.} 
This is usually not an issue, but difficulty in finding the best possible fit 
has been noticed when jet data is removed from the global fit and $\alpha_S(M_Z^2)$ is left free, as the 
relationship between $\alpha_S(M_Z^2)$ and the gluon then allows many fits with similar quality. A similar problem is found when 5 or 6 Chebyshev polynomials are used
and deuteron corrections are also left free, along with (as usual in the MSTW fit) a parameterisation multiplying the nuclear corrections. (Since much of the nuclear target data is 
for $xF_3(x,Q^2)$ there is a lot of correlation between nuclear corrections and valence quarks.) For 
6 Chebyshev polynomials a rather extreme nuclear correction is ultimately preferred. 

Hence, for the fits with free deuteron corrections we start to lose some stability with enough 
free parameters, and beyond 4 Chebyshev polynomials for valence and sea quarks we do not see changes in
PDFs at all incompatible with the uncertainties, or improvements in $\chi^2$ large when considering the 
extra degrees of freedom. Hence, 4 Chebyshev polynomials seems an optimal choice. For the fits with the fixed
(default) deuteron corrections it is arguable that 5 may be a better choice, but even in this case moving
to 6 moves the biggest change in PDFs back towards the situation with 4. In all cases the changes in 
PDFs when moving from 2 to 4 polynomials for each quark distribution is very much larger than subsequent 
changes. In fact all PDF versions with 4, 5 or 6 Chebyshev polynomials, either with or without free 
deuteron corrections are essentially consistent with the MSTW2008CPdeut set and its uncertainties
(with some systematic differences between the default deuteron correction and free deuteron correction 
cases for $d_V$). Hence,
we conclude that 4 Chebyshev polynomials seems a sensible number at present, and certainly expresses the 
main features of the change away from our standard parameterisation. However, four polynomials does not seem 
to be quite as precise in the real PDF fits as in the fits of a function to pseudo-data in Section 2, presumably
because the relationship between the real data and a specific PDF, especially low-$x$ valence quarks,
is rather less direct than the idealised case of Section 2. We will check in future updates
if a slightly larger number (more than 5 seems very unlikely) might be preferable with the expanded data set.
It is possible that the more direct constraints from new data sets can reduce, rather than expand, the 
optimum number.

\section{Lepton Asymmetry at the LHC and PDF Sensitivity}

\begin{figure}[htb!]
  \centering
\vspace{-1.5cm}
  \includegraphics[width=0.55\textwidth,clip]{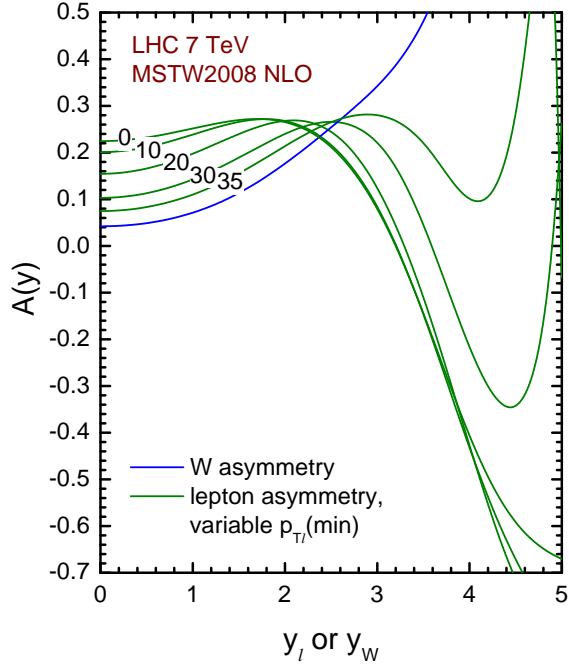}
\vspace{-1.5cm}
  \caption{The dependence of the asymmetry on the lepton minimum $p_T$ cut. The asymmetry is calculated at leading order and zero width using MSTW2008NLO PDFs.}
  \label{fig:varptmin}
  \end{figure}

The measurements of the lepton charge asymmetry, from $W^\pm \to \ell^\pm \nu$
production and decay, at the Tevatron and the LHC, probe novel combinations of 
PDFs. In the next section we shall investigate the effects of the extended 
`Chebyshev' parameterisation, and of the deuteron corrections, on the MSTW 
predictions for the observed asymmetries.  However, first, to gain insight into 
the use of these data in PDF analyses, we explore the predicted behaviour of 
the lepton charge asymmetry at the LHC, based on the LO and zero width expressions for $W$ 
production and decay, using MSTW2008 NLO PDFs. The NLO and NNLO corrections 
\cite{Melnikov:2006kv,Catani:2010en} do 
not change the general picture significantly, though do change the precise values. In 
particular, we explain how the PDFs result in the interesting dependence of the 
asymmetry, shown in Fig.~\ref{fig:varptmin}, on the experimental minimum $p_T$ 
cut applied to the transverse momentum of the decay lepton.

We begin by considering the $W$ charge asymmetry, defined by
\begin{equation}
  A_W(y_W) = \frac{{\rm d}\sigma(W^+)/{\rm d} y_W - {\rm d}\sigma(W^-)/{\rm d} y_W}{{\rm d}\sigma(W^+)/{\rm d} y_W + {\rm d}\sigma(W^-)/{\rm d} y_W},
\end{equation}
where $y_W$ is the rapidity of the $W$ boson.
At leading order, assuming $u$, $d$ quark and antiquark contributions only, taking the CKM matrix to be diagonal, and writing $u=u_V+\bar q$, 
$d=d_V+\bar q$, $\bar u = \bar d  \equiv \bar q$ (which is an approximation, which becomes more accurate at 
very small $x$), we have
\begin{equation}\label{eq:wasymm}
  A_W(y_W)
\approx \frac{u_V(x_1)\bar q(x_2)+ \bar q(x_1) u_V(x_2) - d_V(x_1)\bar q(x_2) -\bar q(x_1)d_V(x_2)}
{u_V(x_1)\bar q(x_2)+ \bar q(x_1) u_V(x_2) + d_V(x_1)\bar q(x_2) +\bar q(x_1)d_V(x_2) + 4\bar q(x_1)
\bar q(x_2)},
\end{equation}
where $x_{1,2}=(M_W/\sqrt{s})\exp(\pm y_W)$.  Contributions from $c,s$ quark scattering can be approximately taken 
into account by $4\bar q(x_1)\bar q(x_2) \to 4(1+\delta ) \bar q(x_1)\bar q(x_2)$ in the denominator. Two important limits are $y_W = 0$ for which 
$x_1 = x_2 = x_0 \equiv M_W/\sqrt{s}$, and $y_W \to y_W^{\rm max} = -\log(x_0)$ for which $x_1 \to 1$ and $x_2 \to x_0^2$. Thus
\begin{equation}\label{eq:wasymmapprox}
 A_W(0)
\approx \frac{u_V(x_0) - d_V(x_0)}
{u_V(x_0) + d_V(x_0) + 2\bar q(x_0)} > 0, \qquad A_W(y_W^{\rm max}) = 1 . 
\end{equation}

In practice, it is usually the lepton charge asymmetry which is measured, defined in a similar way as
\begin{equation} \label{eq:leptonasymm}
  A(y_\ell) = \frac{{\rm d}\sigma(\ell^+)/{\rm d}y_{\ell}-{\rm d}\sigma(\ell^-)
/{\rm d}y_{\ell}}{{\rm d}\sigma(\ell^+)/{\rm d}y_{\ell}+{\rm d}\sigma(\ell^-)/{\rm d}y_{\ell}},
\end{equation}
where $y_{\ell}$ is the (pseudo)rapidity of the charged lepton.\footnote{For massless leptons, the pseudorapidity $\eta_\ell$ is equal to the rapidity $y_\ell$.}  
Defining $\theta^*$ to be the emission angle of the charged lepton relative to the proton beam with 
positive longitudinal momentum in the 
$W$ rest frame, then $\cos^2\theta^* = 1 - 4p_T^2/M_W^2$, where $p_T$ is the lepton transverse momentum. The 
rapidities are related by
\begin{equation}
  y_\ell = y_{W} + y^*, \quad y^* =  \frac{1}{2}\ln\left(\frac{1+\cos\theta^*}{1-\cos\theta^*}\right).
\end{equation}
The leading-order parton momentum fractions are then
\begin{equation}
x_{1,2} = x_0 \exp(\pm  y_W) = x_0  \exp(\pm  y_\ell) \kappa^{\pm 1} ,\quad \kappa = \left( 
\frac{1+\vert\cos\theta^*\vert}{ 1-\vert\cos\theta^*\vert }
\right)^{1/2} > 1,
\end{equation}
i.e.~for a given $p_T$ in $0 \leq p_T \leq M_W/2$, there are two solutions corresponding to positive or negative $\cos\theta^*$, or
equivalently positive or negative $y^*$.

Neglecting overall factors, the analogue of the numerator of \eqref{eq:wasymm} for the lepton asymmetry \eqref{eq:leptonasymm}
can be approximated by
\begin{eqnarray} 
\label{eq:leptonasymmapprox}
&& \left( u_V(x_1^+)\bar q(x_2^+) -\bar q(x_1^+)d_V(x_2^+)  + u_V(x_1^-)\bar q(x_2^-) -\bar q(x_1^-)d_V(x_2^-)          \right)  
   (1-\cos\theta^*)^2  \nonumber \\
&+&   \left( \bar q(x_1^+) u_V(x_2^+) - d_V(x_1^+)\bar q(x_2^+)  + \bar q(x_1^-) u_V(x_2^-) - d_V(x_1^-)\bar q(x_2^-)    \right) 
(1+\cos\theta^*)^2 ,
\end{eqnarray}
where 
\begin{eqnarray}
x_1^+ = x_0 \exp( + y_\ell) \kappa & > & x_1^- = x_0 \exp( +  y_\ell) \kappa^{-1} \nonumber \\
x_2^+ = x_0 \exp( -  y_\ell) \kappa^{-1} & < & x_2^- = x_0 \exp( -  y_\ell) \kappa .
\end{eqnarray}
The explicit $\theta^*$-dependent terms in  \eqref{eq:leptonasymmapprox} originate in the $V\pm A$ structure of the $W$ 
couplings to fermions. Table~\ref{tab:kappa} lists the values of the various quantities that enter in the expression for
the lepton asymmetry as functions of $p_T$.
\begin{table}
  \centering
    \begin{tabular}{c|c|c|c|c|c}
      \hline\hline
      $p_T$ (GeV) & $\vert\cos\theta^*\vert$ & $\vert y^*\vert $ 
&     $(1+\cos\theta^*)^2$ & $(1-\cos\theta^*)^2$ & $\kappa$ \\
      \hline
    0  & 1.000  & $\infty$  & 4.00  & 0.00000 &  $\infty$ \\
    5  & 0.992  & 2.77      & 3.97  & 0.00006 &  16.0 \\
   10  & 0.969  & 2.07      & 3.88  & 0.00099 &  7.91 \\
   15  & 0.928  & 1.64      & 3.72  & 0.00522 &  5.17 \\
   20  & 0.867  & 1.32      & 3.49  & 0.01757 &  3.75 \\
   25  & 0.783  & 1.05      & 3.18  & 0.04705 &  2.87 \\
   30  & 0.666  & 0.80      & 2.77  & 0.11181 &  2.23 \\
   35  & 0.492  & 0.54      & 2.23  & 0.25820 &  1.71 \\
 $M_W/2$ & 0.000 & 0.00    & 0.00  & 1.00000 &  1.00 \\
      \hline\hline
    \end{tabular}
\caption{The dependence of the various $\theta^*$--dependent quantities on the lepton $p_T$.}
\label{tab:kappa}
\end{table}

\begin{figure}[htb!]
  \centering
\vspace{-1.5cm}
  \includegraphics[width=0.55\textwidth,clip]{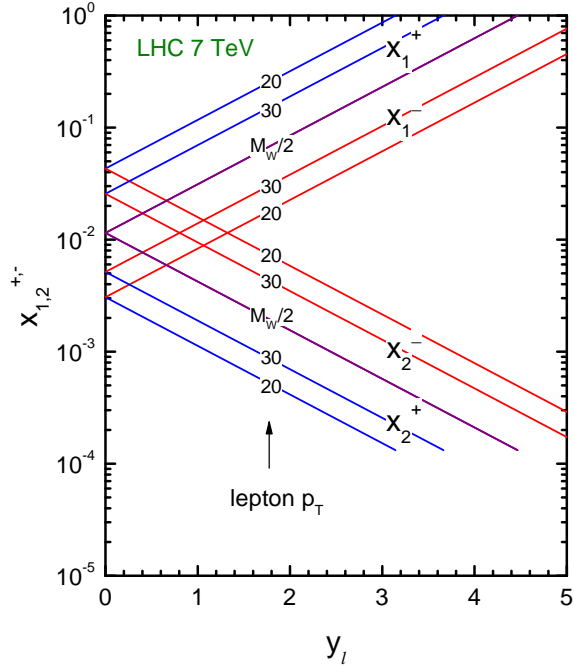}
\vspace{-1.5cm}
  \caption{Dependence of $x_1^\pm$ and $x_2^\pm$ on $y_\ell$, for lepton $p_T = M_W/2$, 30~GeV and 20~GeV, at the 7~TeV LHC. 
For $p_T = M_W/2$ we have $x_1^+=x_1^-$ and $x_2^+=x_2^-$. }
  \label{fig:x_varptmin}
  \end{figure}

Note that whether or not one or both $x$ solutions are physical depends on the values of $y_\ell$ and $\kappa$ (i.e.~$p_T$). 
For the range of lepton $p_T$ accessible to experiment at LHC, there will always be two solutions for sufficiently small $y_\ell$.
Then as $y_\ell$ increases, the {\lq$+$\rq} solution disappears first for $x_1^+ > 1$, i.e.~$y_\ell > -\log(x_0\kappa)$, and both solutions 
disappear for  $x_1^- > 1$, i.e.~$y_\ell > -\log(x_0/\kappa)$. For example, for $p_T = 20$~GeV at $\sqrt{s} = 7$~TeV, 
these limiting values are $y_\ell = 3.14$ and $5.79$ respectively. The $x_1^{\pm}$ and $x_2^{\pm}$ values as a function of $y_\ell$ for 
three different values of lepton $p_T$ are shown in Fig.~\ref{fig:x_varptmin}.

For small or moderate $y_\ell$, the $x_i^+$ and $x_i^-$ contributions in \eqref{eq:leptonasymmapprox} 
give comparable contributions. In particular, for $y_\ell = 0$, 
\begin{eqnarray}
x_1^+ = x_2^- = x_0 \kappa \equiv X \\
x_1^- = x_2^+ = x_0 / \kappa \equiv x
\end{eqnarray}
 with $X/x = \kappa^2 \geq 1$. For small $p_T$ therefore, $X \gg x$ and  as long as $X$ is 
not too close to $1$ we may expect $V(X)\bar q(x) > V(x) \bar q(X)$, where $V(x)$ denotes either 
valence quark, in which case (\ref{eq:leptonasymmapprox}) becomes approximately
\begin{equation} 
\left( u_V(X) - d_V(X) \right)\;  \bar q(x)\; 4 \cos\theta^* ,  
\end{equation}
which, in turn, leads to 
\begin{equation}
 A_\ell(0)
\approx \frac{u_V(X) - d_V(X)}
{u_V(X) + d_V(X) + 2\bar q(X)}  . 
\label{eq:Alep0}
\end{equation}
Since $u_V(X) - d_V(X)$ increases with increasing $X$ at small $X$, this explains, at least qualitatively, 
why the lepton asymmetry grows with decreasing $p_{T{\rm min}}$, see Fig.~\ref{fig:varptmin}. 

As $y_\ell$ increases away from 0, $x_1^+ \to 1$ and the $x_{1,2}^-$ contributions start to dominate.
Furthermore, since in this region $x_1^- \gg x_2^-$, the terms with $V(x_1^-) \bar q(x_2^-)$ are the most important. Thus
\begin{equation} 
\label{eq:leptonasymmapprox2}
A_\ell(y_\ell)\approx \frac{  u_V(x_1^-) (1 -\cos\theta^*)^2 - d_V(x_1^-) (1+\cos\theta^*)^2   }{
u_V(x_1^-) (1 -\cos\theta^*)^2 + d_V(x_1^-) (1+\cos\theta^*)^2  } .
\end{equation}
According to Table~\ref{tab:kappa}, for $p_T \to M_W/2$, $\cos\theta^* \to 0$ and $A_\ell(y_\ell) \to A_W(y_\ell)$ 
because then
 the $(1 \pm \cos\theta^*)^2$ terms in (\ref{eq:leptonasymmapprox2}) are on the same footing 
--- the asymmetry is driven by the $u_V > d_V$ 
inequality which is valid at all $x$, and the lepton asymmetry is always positive.
However, for small or moderate $p_T$, $(1+\cos\theta^*)^2 \gg (1-\cos\theta^*)^2$ 
and so the term proportional to $d_V$ dominates and the asymmetry is negative.


Now $d_V(x)$ decreases faster at large $x$ than $u_V(x)$, and so at some point at large $y_\ell$ the approximation
\begin{equation}
d_V(x_1^-)\bar q(x_2^-) (1+\cos\theta^*)^2  \gg u_V(x_1^-)\bar q(x_2^-) (1-\cos\theta^*)^2
\end{equation}  
breaks down, i.e.~the $V\pm A$ unfavoured forward $u\bar d \to \ell^+ \nu_\ell$ scattering process will eventually dominate. Evidently this will happen
at the $y_\ell$ value for which 
\begin{equation}
u_V(x_1^-) / d_V(x_1^-) \sim (1+\cos\theta^*)^2/ (1-\cos\theta^*)^2 = \kappa^4.
\end{equation} 
The larger the lepton $p_T$ (recall that large $p_T$ means small $\kappa$), the earlier (in terms of increasing $y_\ell$)
 this  will happen, as confirmed by Fig.~\ref{fig:varptmin}.
In principle LHCb data should be sensitive to this, but a very 
significant increase in 
statistics is needed compared to the present data~\cite{Aaij:2012vn}, 
for which the MSTW2008 PDFs give a good prediction. When this is achieved 
then data in bins of different $p_T$ cut will be illuminating.   

In summary, the behaviour of the lepton asymmetry shown in 
Fig.~\ref{fig:varptmin} can now be understood in terms of a fairly complex 
interplay of PDF and $V\pm A$ effects. For small $y_\ell$, the asymmetry is 
sensitive to the combination $u_V-d_V$ at values of $x$ between $M_W/\sqrt{s}$ and $\kappa(p_{T{\rm min}})M_W/\sqrt{s}$, see (\ref{eq:Alep0}), where $p_{T{\rm min}}$ is the minimum observed $p_T$ of the lepton and where values of $\kappa(p_T)$ are shown in Table \ref{tab:kappa}. 
This results in a fairly clear decrease in asymmetry for increasing the minimum lepton $p_T$ 
cut. At high $y_\ell > 3$ the asymmetry is even more sensitive to the minimum 
lepton $p_T$ cut. In this region valence $u$ and $d$ quarks are
 being sampled at very high $x$, see Fig.~\ref{fig:x_varptmin}.

\section{Predictions for the LHC Using 
Modified PDFs}

  \begin{figure}[htb!]
\centering
\vspace{-1cm}
  \includegraphics[width=0.7\textwidth,clip]{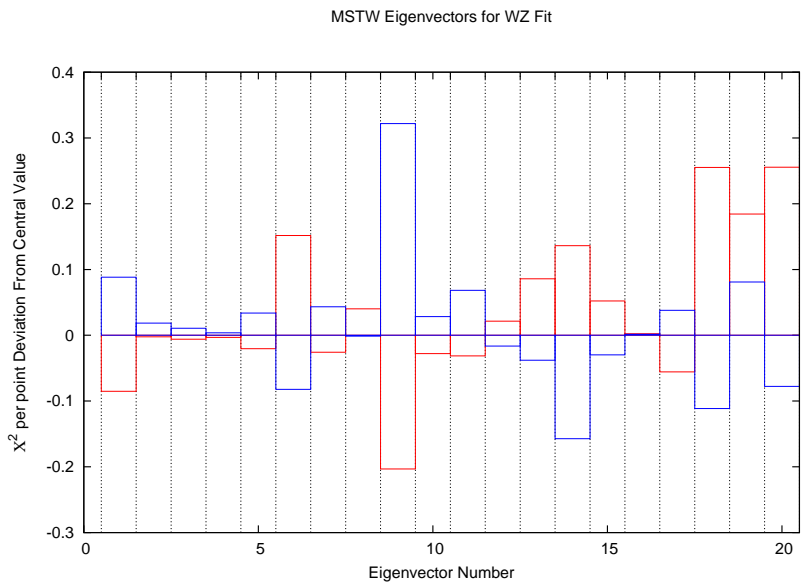}
\vspace{-1cm}
  \caption{The variation in the quality of the fit to ATLAS
vector boson production as a function of rapidity for different MSTW2008 eigenvectors. Eigenvectors 9, 14 and 18 are mainly associated with the gluon, $d_V$ and $u_V$, respectively.}
  \label{fig:WZ-Eig}
  \end{figure}
The MSTW2008 set of PDFs are found to give very good predictions for the vast 
majority of relevant measurements made at the LHC to date, see, for example, \cite{Ball:2012wy}.  However, an 
exception is 
the charged lepton asymmetry arising from $W^\pm$ production, where it is clear that this set of PDFs does not give the optimum description~\cite{Aad:2011dm,Chatrchyan:2012xt,Watt:2012tq,Chatrchyan:2011jz}. In this section we investigate the reason for this deficiency. We will show that the extended `Chebyshev' parameterisation and, to a lesser extent, the improvement in the treatment of the shadowing corrections to the deuteron data, completely remove the problem, without a deterioration in the excellent description of the other data.

\subsection{Preparatory study of LHC $W,Z$ rapidity distributions}

However, first let us start by describing a recent
quantitative study \cite{Watt:2012tq} of the CMS \cite{Chatrchyan:2011jz} and ATLAS \cite{Aad:2011dm}
measurements on the {\it charged lepton asymmetry} with cuts on charged lepton $p_T$ of 
$25~\GeV$ and $20~\GeV$, respectively, with a cut on missing energy of $25~\GeV$ in the latter case. 
In this paper \cite{Watt:2012tq}, 
a {\it reweighting} technique, originally introduced in 
\cite{Ball:2010gb} (modifying an earlier proposal in \cite{Giele:1998gw}) with a slightly different method 
of application, was used to estimate 
the effect these new asymmetry data sets would have on both the central value and  
uncertainty of the MSTW2008 PDF set. The reweighted PDFs were able to turn an initial 
$\chi^2$ of 2-3 per asymmetry data point into just over 1 per point. As is clear from the 
previous section, the 
major PDF sensitivity is in the $u_V-d_V$ distribution. When these new data sets 
were included, $u_V-d_V$ was found to change most for $x=0.02$ for $Q^2=10^4~\GeV^2$, 
increasing by about $5\%$, which is similar to the size of its uncertainty. Due to the sum 
rule, this also resulted in a reduction for $x<0.001$, a region where there is no data 
constraint. The reduction is also about the size of the uncertainty. After the reweighting, the 
uncertainty reduced to about $60$--$70\%$ of its original size near $x=0.02$. The 
data from the higher luminosity CMS measurement \cite{Chatrchyan:2012xt}, obtained 
with a higher minimum lepton $p_T$ cut of $35~\GeV$, were not studied in \cite{Watt:2012tq}, 
but it is clear from the comparison in \cite{Chatrchyan:2012xt} that their description 
using MSTW2008 PDFs is worse than the data with lower $p_T$ cut. 

Here we extend this previous study somewhat. We begin by comparing, at NLO for simplicity, the fit quality 
to the {\it full} ATLAS lepton rapidity data from \cite{Aad:2011dm}, of which the 
asymmetry measurements are only a subset, and where information on the size and shape in rapidity 
of the cross section is lost when taking the ratio of the difference to the sum of $W^+$ and $W^-$ cross sections. 
Comparing the full data set to the MSTW2008 PDFs at NLO 
(without higher order electroweak corrections), 
using APPLgrid \cite{Carli:2010rw} (which uses the MCFM \cite{Campbell:2002tg} code), 
we calculate $\chi^2/N_{\rm pts.} =60/30$, where the 30 data points are 11 each from 
$W^{\pm}$ production and 8 from $Z$ production. As for the asymmetry data, it 
is clear that MSTW2008 PDFs do not provide the optimum fit 
quality, but we note that the PDF 
sets of all other groups (for which APPLgrid can easily be used to calculate
cross section predictions) seem to also obtain $\chi^2/N_{\rm pts.}$ 
noticeably more than 1 per point, the best being the CT10 PDF set, with 
about 1.1 per point.\footnote{We note that the particularly good description by CT10 is probably 
due to the larger strange distribution in their PDF set than in the others. 
A study by 
ATLAS has shown that their data prefer a large strange distribution, in fact 
seemingly one which is the same size as the $\bar u$ and $\bar d$ distributions 
even at low $Q^2$ \cite{Aad:2012sb}. 
We do indeed see some small improvement in fit quality associated 
with the eigenvector most associated with the strange PDF normalisation, but 
rather less than for the three eigenvectors 9, 14, 18 mentioned below. This means that only a 
marginal improvement can be obtained by changing the strange distribution by 
one standard deviation. We have confirmed this by making a more thorough 
study. Moreover, a study by the NNPDF group has reached a 
similar conclusion \cite{Ball:2012cx}.} In order to investigate
the manner in which the description of these data could be improved we look at 
the  quality of the $\chi^2$ using each of the uncertainty eigenvectors. 
This is shown in Fig.~\ref{fig:WZ-Eig}. The 
fit quality to the ATLAS data improves markedly, i.e.~by about $0.2 \times 30= 6$ units, 
in one direction for eigenvector 
9, which is mainly associated with the gluon distribution, so this  
alters the common shape and normalisation of all three $(W^\pm,Z)$ rapidity 
distributions via the dependence of the quarks at high $Q^2$ on the gluon 
due to evolution. 
The value of $\chi^2$  improves by slightly less for eigenvectors 14 and 
18. These are associated  with $d_V$ and $u_V$ respectively; variation in these affects the asymmetry. Due to the fact that we 
underestimate the asymmetry, it is clear that the eigenvector directions, 
leading to the improvement, decrease $d_V$ or increase $u_V$ in the region of 
$x=0.02$. Hence it follows that the fit to these data sets would move the PDFs
towards these directions for $u_V$ and $d_V$ (and also modify the gluon 
distribution to some extent). 

\begin{figure}[htb!]
\centering
\vspace{-1cm}
  \includegraphics[width=0.7\textwidth,clip]{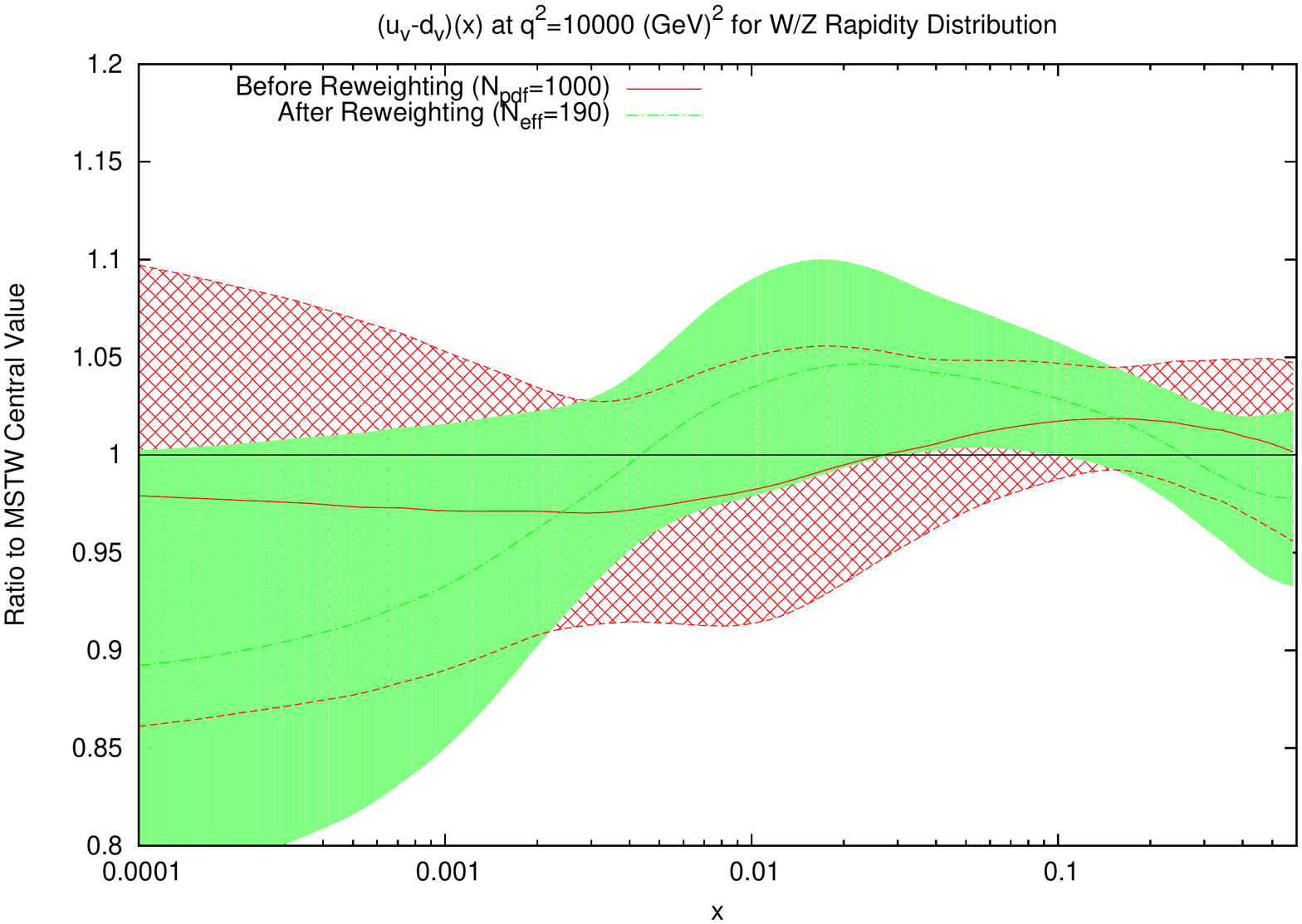}\\
\vspace{-1.5cm}
     \includegraphics[width=0.7\textwidth,clip]{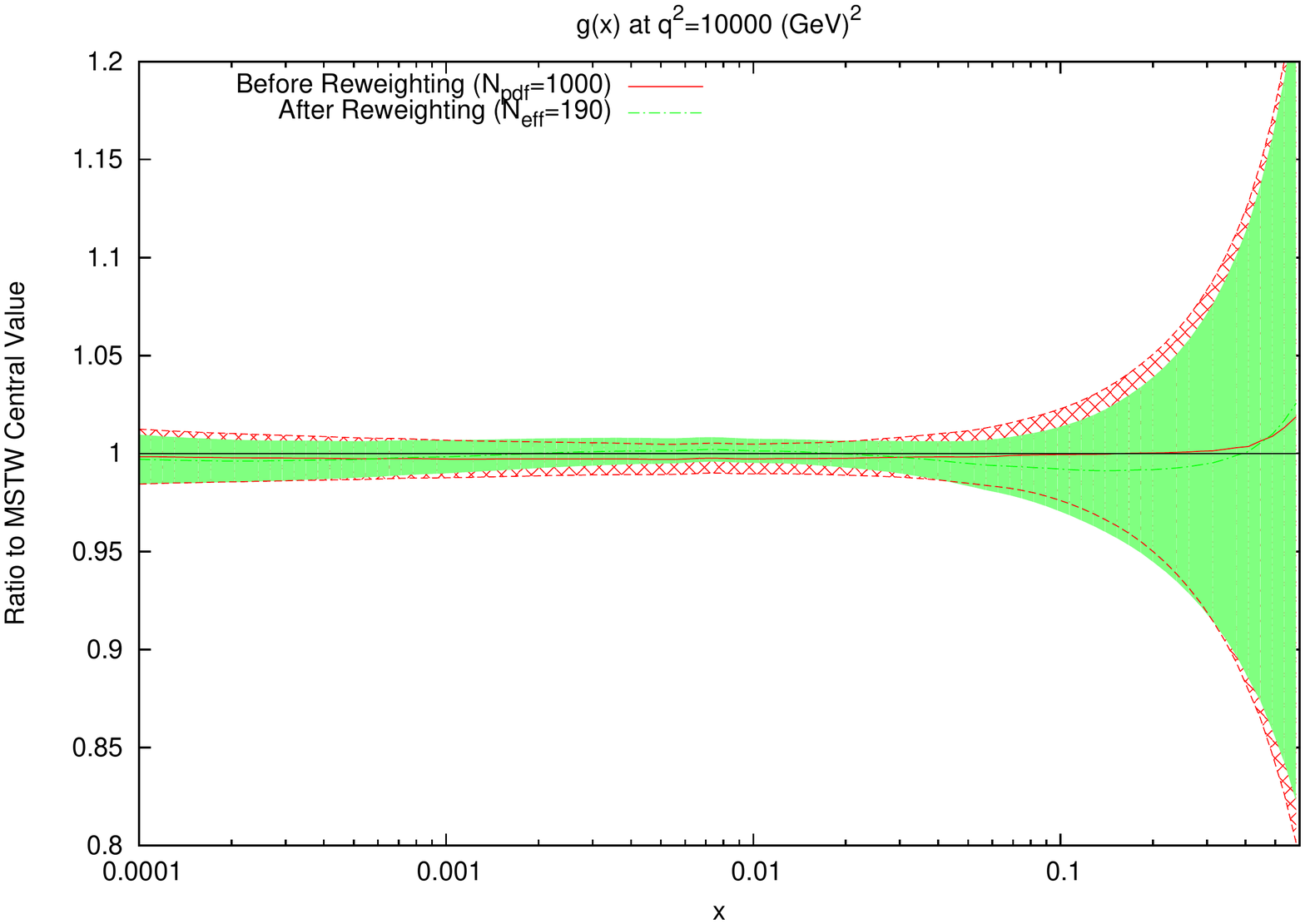}
\vspace{-1cm}
  \caption{The effect of reweighting on the $u_V(x)-d_V(x)$ distribution (top) 
and $g(x)$ distribution (bottom)
at $Q^2=10^4~\GeV^2$. The hatched (red) band shows the average value and standard deviation 
of 1000 randomly generated sets of MSTW2008 PDFs and the continuous (green) band the 
same PDFs after reweighting according to the fit quality for the 
ATLAS data on $W,Z$ rapidity.}
  \label{fig:WZval}
  \end{figure}

This conclusion is verified by reweighting the 
PDFs according to the prescription in \cite{Watt:2012tq}, with asymmetric PDF 
uncertainties, using the full
data on the $W$ and $Z$ rapidity distributions. The result for the difference 
in the valence quarks, $u_V(x)-d_V(x)$ at $Q^2=10^4~\GeV^2$, is shown in the upper of
Fig.~\ref{fig:WZval}, compared to the MSTW2008 PDFs. The study uses 1000
random PDF sets. After reweighting, the effective number of sets $N_{\rm eff}=190$.
The small fraction ($\sim 20\%$) of effective PDFs arising 
in the reweighting procedure shows that there is a significant variation 
in the fit quality for different random sets; some give enhanced quality 
compared to the central MSTW2008 set but many give rather worse quality.
There is a distinct 
tendency for $u_V-d_V$ to increase at $x \approx 0.02$, and correspondingly 
decrease at lower $x$. However, the effect is less pronounced than that seen 
in \cite{Watt:2012tq} when using only the asymmetry data, with the change in 
the average value being less than the uncertainty, even at $x=0.02$. This 
slightly smaller change has two origins. Firstly the small systematic 
uncertainties are treated as entirely uncorrelated in the asymmetry, but some 
history of the correlation persists in the full treatment of the separate
$W^+$ and $W^-$ rapidity distributions. For example, one correlated systematic 
moves the $W^+$ and $W^-$ rapidity distribution up and down in opposite 
directions, independent of rapidity, and clearly contributes to some 
correlation in the asymmetry. Maintaining this information allows for a
slightly easier fit to the asymmetry. Secondly, and more importantly, 
if only asymmetry data are 
used, all PDFs which carry a high weight must improve the comparison for the 
asymmetry. If the full data are used, some higher weight PDFs produce better 
fits due to improvement in overall shape of all distributions with rapidity, 
or improvement in consistency of $W$ and $Z$ data, so not all high weight PDFs
have an increase in $u_V-d_V$ at $x\sim 0.02$. This shows that comparing to
asymmetry data alone can exaggerate their effect and importance. In the lower of
Fig.~\ref{fig:WZval} we also show the effect of the new data on the gluon 
distribution. The change is not large, but is not entirely insignificant, and 
an improvement in the shape of the rapidity distributions does require a 
modification of the gluon distribution. After reweighting the fit quality 
improves to  $\chi^2/N_{\rm pts.}=48/30$.

\subsection{ LHC $W,Z$ rapidity distributions from modified MSTW PDFs}

We now come to an important result: that is, the description of the ATLAS 
$W^\pm,Z$ rapidity data at NLO using the two modified PDF sets extracted earlier in 
this paper. Namely the PDF set based on the Chebyshev polynomial parameterisation (MSTW2008CP), and the set `MSTW2008CPdeut' including the improved deuteron corrections in addition. The change in $u_V$ in MSTW2008CP is actually rather similar to 
that for eigenvector 18, though bigger, and the further change in MSTW2008CPdeut
is similar to eigenvector 14. As one would expect, this does lead to 
an improvement in the comparison to the data. The $\chi^2/N_{\rm pts.}$ improves from $60/30$ to $49/30$ 
for MSTW2008CP, and to 46/30 for MSTW2008CPdeut. This is as good as any other PDF 
set at NLO, except for CT10 which has a much larger strange quark distribution than that of other global analyses (as discussed in the previous footnote).
For the ATLAS asymmetry data the $\chi^2/N_{\rm pts.}$ improves from $30/11$ 
to $15/11$ for MSTW2008CP to $9/11$ for MSTW2008CPdeut.

\begin{figure}[htb!]
  \centering
\vspace{-1.5cm}
  \includegraphics[width=0.7\textwidth,clip]{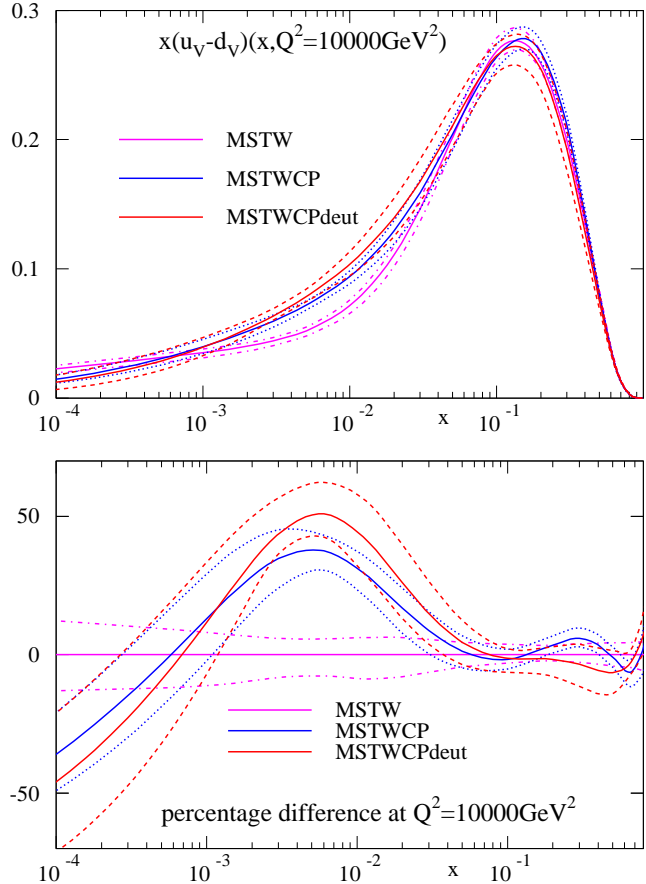}
\vspace{-1.5cm}
  \caption{The combination $u_V-d_V$ obtained in the new MSTW2008CP and MSTW2008CPdeut fits compared with that of the original MSTW2008 analysis. All three analyses fit to exactly the same data sets.}
  \label{fig:uVminusdV}
  \end{figure}

It is particularly informative to study how the combination $u_V-d_V$ is changed in the new fits with the extended parameterisations. Fig. \ref{fig:uVminusdV} compares the values of $u_V-d_V$ obtained in the MSTW2008CP and MSTW2008CPdeut fits with those of the original MSTW2008 analysis.  Indeed, $u_V-d_V$ increases dramatically in the region $x \sim 0.01$--$0.06$ which is probed by the $W^\pm,~Z$ rapidity distributions at the LHC. We emphasise that exactly the same data sets are used in all three analyses.  Although some Tevatron lepton asymmetry data were included in the original MSTW2008 analysis, it is remarkable that the extended parameterisations make such a sizeable change to $u_V-d_V$ for $x \sim 0.01$--$0.06$ and improve the overall fit to the data, as well as giving a good description of the LHC lepton asymmetry data (which were not included in the fits). Moreover, the changes in $u_V,~d_V$ are very small in the region $x > 0.05$ where they are constrained by other types of data in the fits. An exception is the approximately 5$\%$ increase in $d_V$ for $x \sim 0.5$ when the deuteron corrections are included. For $x\lapproxeq 0.01$, where there are no data constraints, and which has little impact on the LHC lepton asymmetry, the changes can be extremely large -- ranging from an increase of up to 50$\%$ at $x=0.005$ to a significant decrease below $x \sim 0.0005$.

The large change in $u_V$ and $d_V$ in the extended `Chebyshev' parameterisations in the 
region $x\lapproxeq 0.01$ unconstrained by data, has an interesting consequence. The small $x$ 
behaviour of the input valence distributions in the original NLO MSTW2008 
parameterisation was controlled by the parametric forms:
\begin{eqnarray}
xu_V \propto x^{\delta_u}~~~~~~{\rm with}~~~~\delta_u=0.29\pm 0.02, \\
xd_V \propto x^{\delta_d}~~~~~\,\,{\rm with}\hspace{0.1cm}~~~~\delta_d=0.97\pm 0.11,
\end{eqnarray}
at $Q_0^2=1~\GeV^2$.  On the other hand with the `Chebyshev' parameterisations we have 
$\delta_u=1.00$ and $\delta_d=0.70$ for the MSTW2008CP fit, and $\delta_u=0.70$ and $\delta_d=0.65$ for the MSTW2008CPdeut fit, where in each case the uncertainties are a little bigger than for the MSTW2008 set. This is much more in line with the Regge 
expectation that the two powers should be the same, particularly in the case of 
MSTW2008CPdeut, where the difference is easily consistent with zero within uncertainties.
The powers are also fairly close to the Regge expectation 
$\delta \sim$ 0.5$--$0.6.   
It seems as though the standard parameterisation for the MSTW2008 valence 
quarks, combined with the constraint from a variety of data, pushes the 
small-$x$ valence quarks in a direction somewhat at odds with the LHC 
asymmetry data. Less constraint from other data, and potentially an equally
restrictive, but different, parameterisation, possibly with the small-$x$ 
behaviour of $d_V$ and $u_V$ tied together more closely, may have provided a 
better prediction. However, the extended parameterisation, introduced here, 
seems to be a preferable approach, describing all data sensitive to $u_V$ and $d_V$ well and 
automatically making the small-$x$ forms of $u_V$ and $d_V$ more similar.

  \begin{figure}[htb!]
  \centering
\vspace{-1.5cm}
  \includegraphics[width=0.6\textwidth,clip]{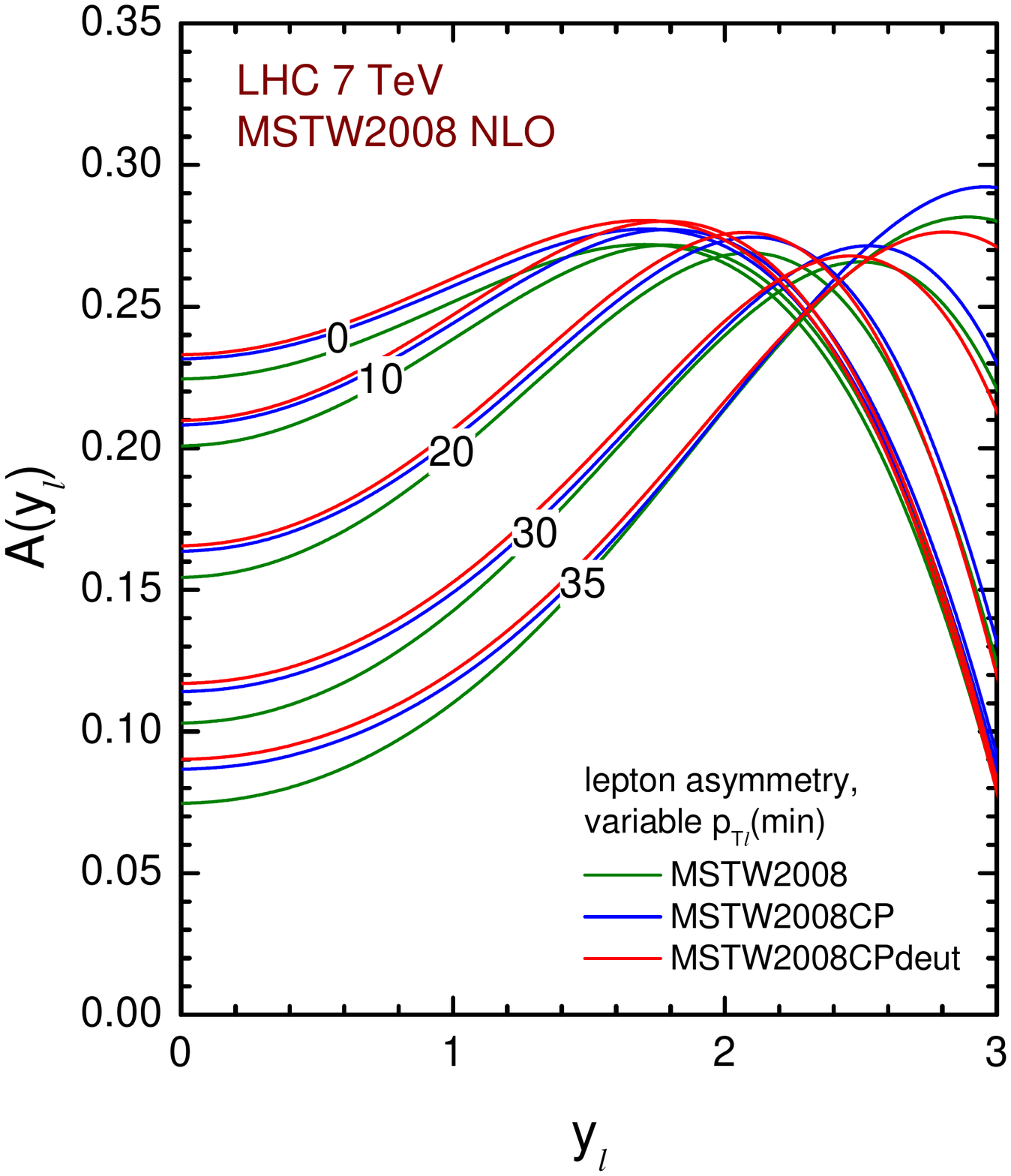}
  \vspace{-1.5cm}
  \caption{The variation in the prediction for lepton asymmetry data 
calculated at LO and zero width using the original and the two modified MSTW2008 PDFs. The sets of curves correspond to different choices of the minimum $p_T$ cut (shown in GeV) applied to the observed charged lepton from the $W$ decay.}
  \label{fig:lasymod}
  \end{figure}

Let us discuss the features of Fig. \ref{fig:uVminusdV}, and the description of the charged lepton asymmetry, in a little more detail.
The change in the $u_V$ and $d_V$ distributions in MSTW2008CP and
MSTW2008CPdeut seems ideal for improving the comparison to the ATLAS
lepton rapidity data, and for removing the shortcomings in the description of these data by the  
MSTW2008 set of PDFs. However, it can be noted that the apparent difficulties
in the MSTW description of the LHC lepton asymmetry data are very much correlated with the minimum 
$p_T$ cut applied to the final state lepton. The comparisons with unpublished data obtained with a $p_T=20~\GeV$ cut seemed
perfectly good \cite{Santaolalla:2012ia,ATLASnote}, but those for data obtained with a higher $p_T=35~\GeV$ cut were much worse \cite{Chatrchyan:2012xt}. From the study 
in the previous section, we see that changes in the minimum $p_T$ cut on the data change
the $x$ range probed in $u_V-d_V$. As $p_{T{\rm min}}$ decreases, the $x$ region expands to include lower and lower values of $x$.  Hence, the change that MSTW2008CP and 
MSTW2008CPdeut make to the lepton asymmetry results are very dependent 
on  the choice of $p_{T{\rm min}}$ of the corresponding data. This is shown explicitly in the results shown in Fig.~\ref{fig:lasymod}
(where, as in the previous section, the curves are made using LO formulae and NLO PDFs). 
We see that the use of MSTW2008CP PDFs increases the lepton asymmetry at low
rapidity, more so for a higher minimum $p_T$ cut. That is, the proportional change 
near $y_\ell=0$ for $p_T=35~\GeV$ is about $15\%$, whereas for $p_T=20~\GeV$ 
it is only about $5\%$. The asymmetry at low rapidity increases slightly further 
when using MSTW2008CPdeut, but by only a small amount compared to the change resulting from using
MSTW2008CP PDFs. Hence, the majority of the change is obtained from an extension in 
the PDF input parameterisation, and only a minor amount when 
flexible deuteron corrections are also included. We note from the figure that for 
the highest $p_T$ cuts the asymmetry for $y_\ell \gapproxeq 2.5$ tends to decrease for 
MSTW2008CPdeut due to the larger value of $d_V$ for $x\sim 0.4$.  

  \begin{figure}[htb!]
  \centering
\vspace{-9cm}
\hspace{1.3cm}  \includegraphics[width=0.8\textwidth,clip]{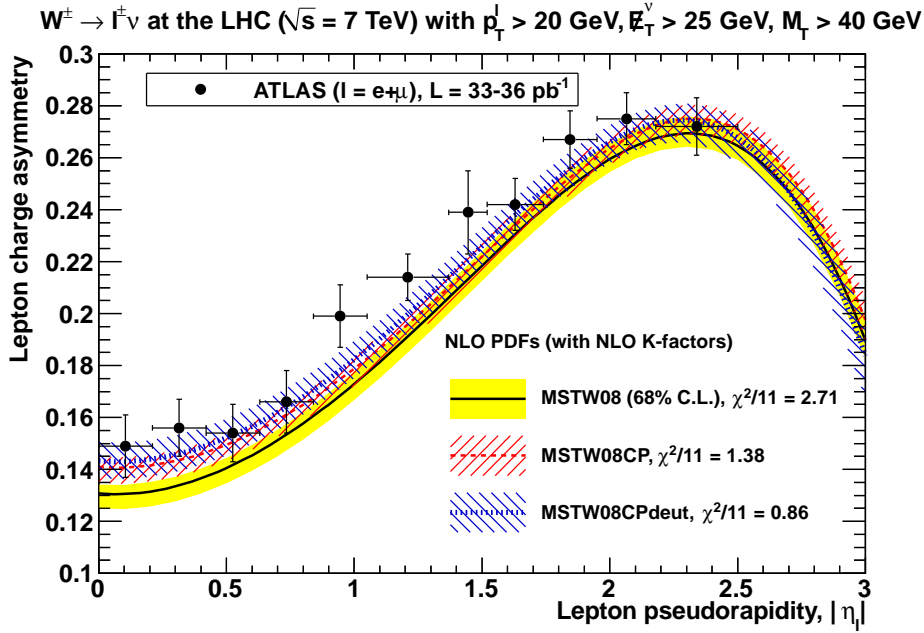}
  \caption{The improvement in the fit quality of the ATLAS
lepton asymmetry data for $p_T > 20$~GeV (and missing transverse energy $\not\mathrel{E}_T^\nu>25$~GeV)~\cite{Aad:2011dm}, in going from the original MSTW2008 $\to$ MSTW2008CP $\to$ MSTW2008CPdeut sets of partons. All three parton sets are obtained by fitting to exactly the same (pre-LHC) data set.}
  \label{fig:ATLAScomp}
  \end{figure}
  \begin{figure}[htb!]
  \centering
\vspace{-9cm}
\hspace{1.3cm}
  \includegraphics[width=0.8\textwidth,clip]{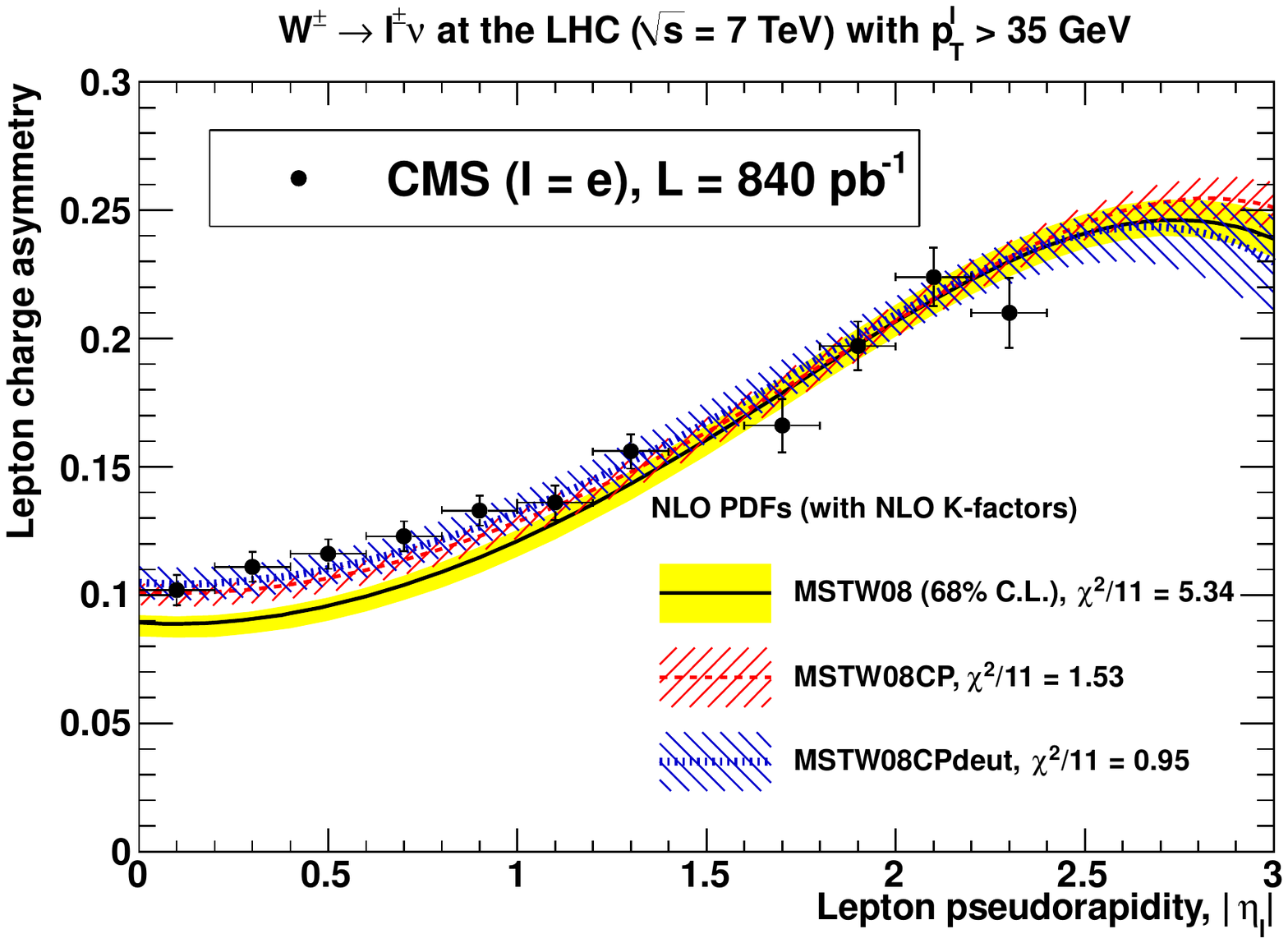}
  \caption{The improvement in the fit quality of the CMS
lepton asymmetry data for $p_{T} > 35$ GeV~\cite{Chatrchyan:2012xt}, in going from the original MSTW2008 $\to$ MSTW2008CP $\to$ MSTW2008CPdeut sets of partons. All three parton sets are obtained by fitting to exactly the same (pre-LHC) data set.}
  \label{fig:CMS35comp}
  \end{figure}

The detailed NLO comparisons to the ATLAS asymmetry data for 
$p_T > 20$~GeV (and missing transverse energy $\not\mathrel{E}_T^\nu>25$~GeV) and CMS electron asymmetry data for 
$p_T > 35 ~\GeV$ (made using NLO $K$-factors \cite{Watt:2012tq} computed 
with \textsc{DYNNLO}~\cite{Catani:2009sm}) 
are shown in Fig.~\ref{fig:ATLAScomp} and Fig.~\ref{fig:CMS35comp}.
The CMS data is most sensitive to the valence quark difference at 
small $x$ and is predicted worst by MSTW2008 PDFs.  
One sees that the initial poor $\chi^2$ per point of 5.3 for MSTW2008 is reduced to
1.5 for MSTW2008CP, and to slightly less than 1 for MSTW2008CPdeut,
where all experimental uncertainties are simply added in quadrature. 
The last could not be a 
much better description of the data. 
The improvement is similar for the ATLAS data, but the standard MSTW2008
PDFs do not give such a poor prediction in this case.
Note that these are all predictions (or more precisely, ``postdictions''); neither these data, nor
 indeed any other LHC data have been used in order to extract the PDFs.
The main reason for the small extra change coming from the MSTW2008CPdeut PDFs is simply due
to the removal of the significant small-$x$ shadowing deuteron correction in the 
default MSTW2008 extraction -- recall that our freely determined deuteron 
correction is extremely small for $x\sim 0.02$.    
The uncertainty band for the prediction for MSTW2008CP is very similar to that for
MSTW2008. The extended parameterisation has changed the average value of 
$u_V-d_V$ far more than it affects the nominal uncertainty. The uncertainty band for MSTW2008CPdeut is a little bigger, reflecting the extra 
uncertainty introduced by having a varying deuteron correction.  

\begin{figure}[htb!]
\centering
\vspace{-1cm}
  \includegraphics[width=0.7\textwidth,clip]{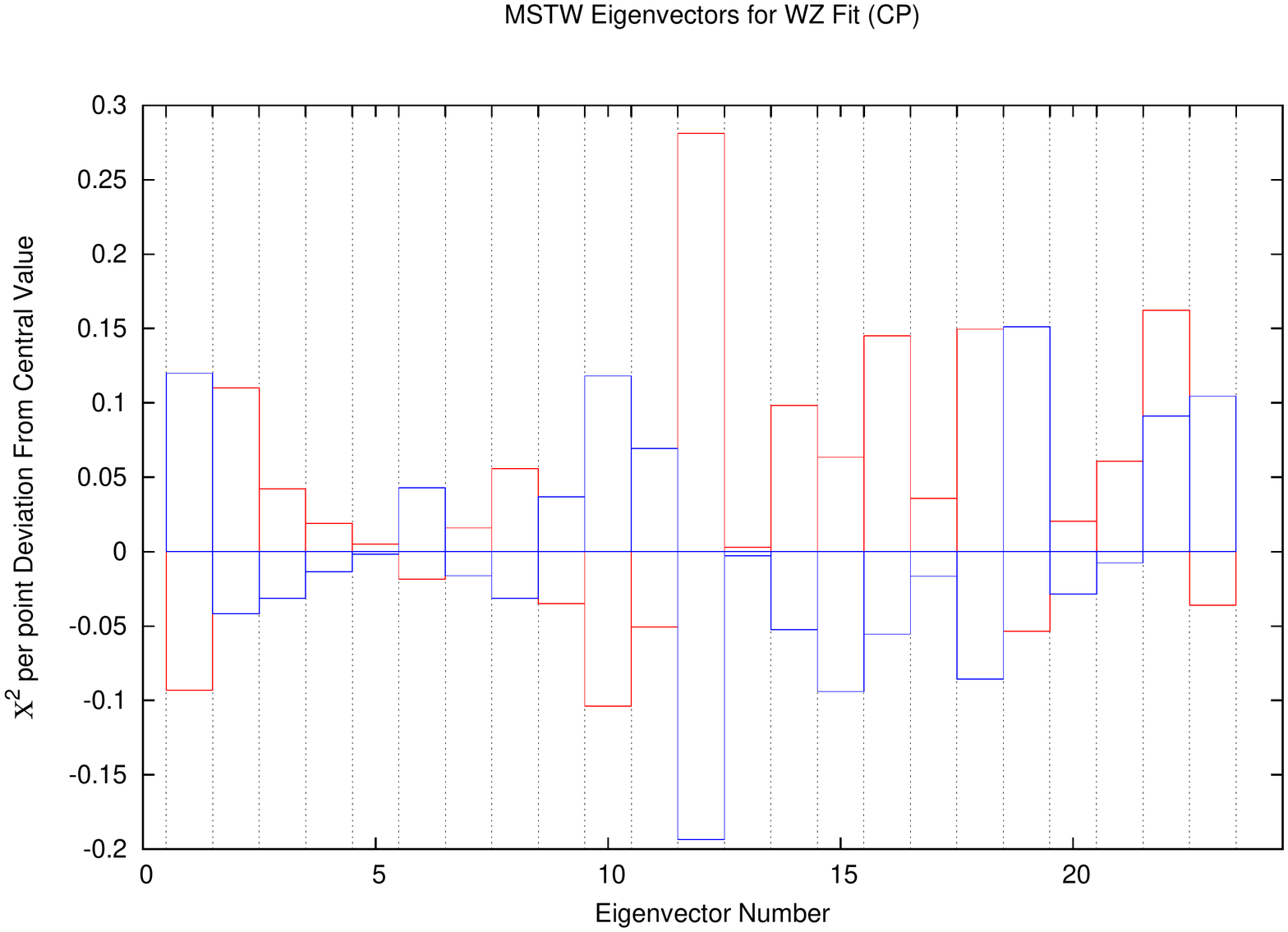}
\vspace{-1cm}
  \caption{The variation in the quality of the fit to ATLAS
vector boson $(W,Z)$ production as a function of rapidity for different MSTW2008CP eigenvectors.}
  \label{fig:WZ-EigCP}
  \end{figure}

\begin{figure}[htb!]
\centering
\vspace{-1cm}
   \includegraphics[width=0.7\textwidth,clip]{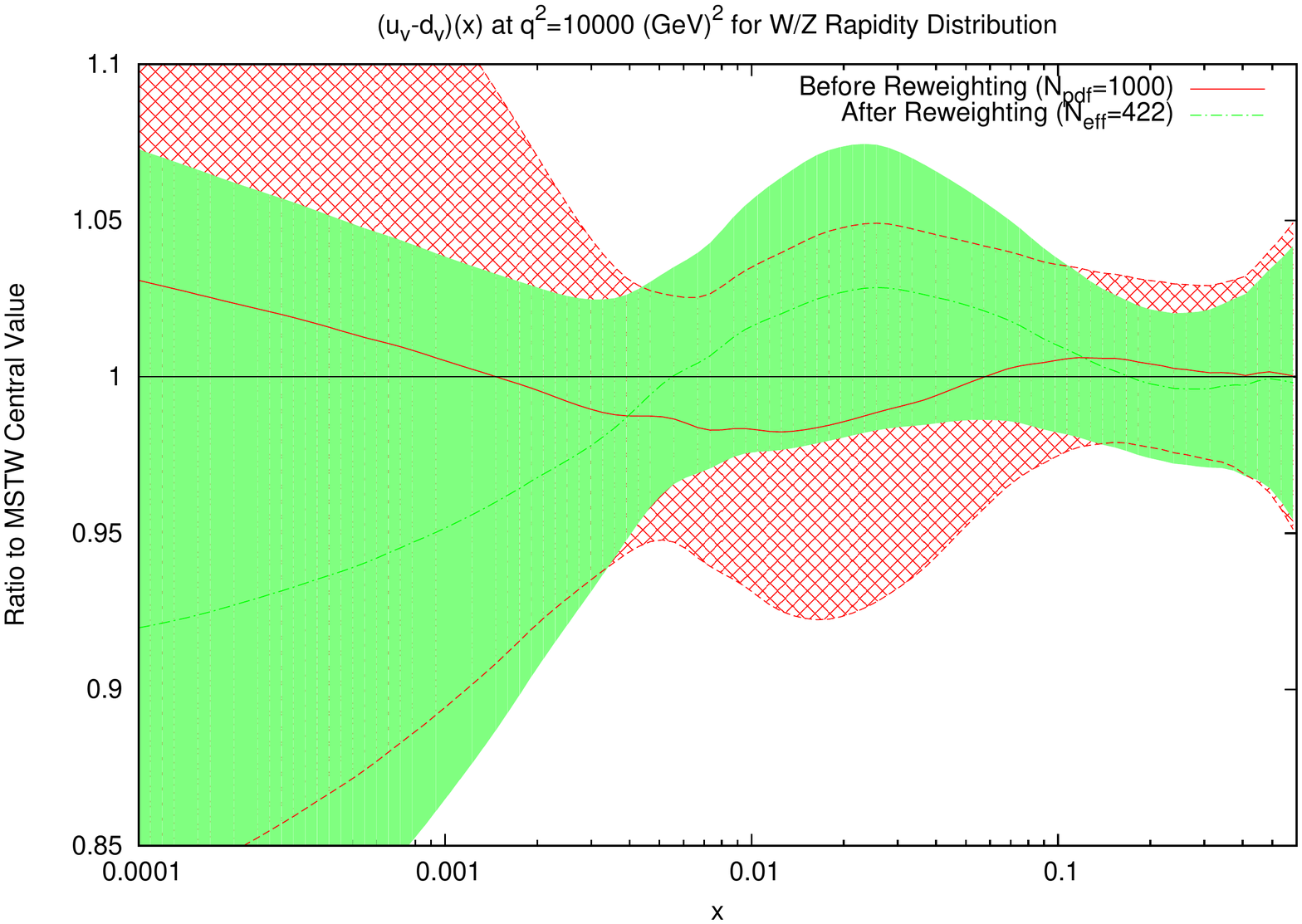}\\
\vspace{-1.5cm}
   \includegraphics[width=0.7\textwidth,clip]{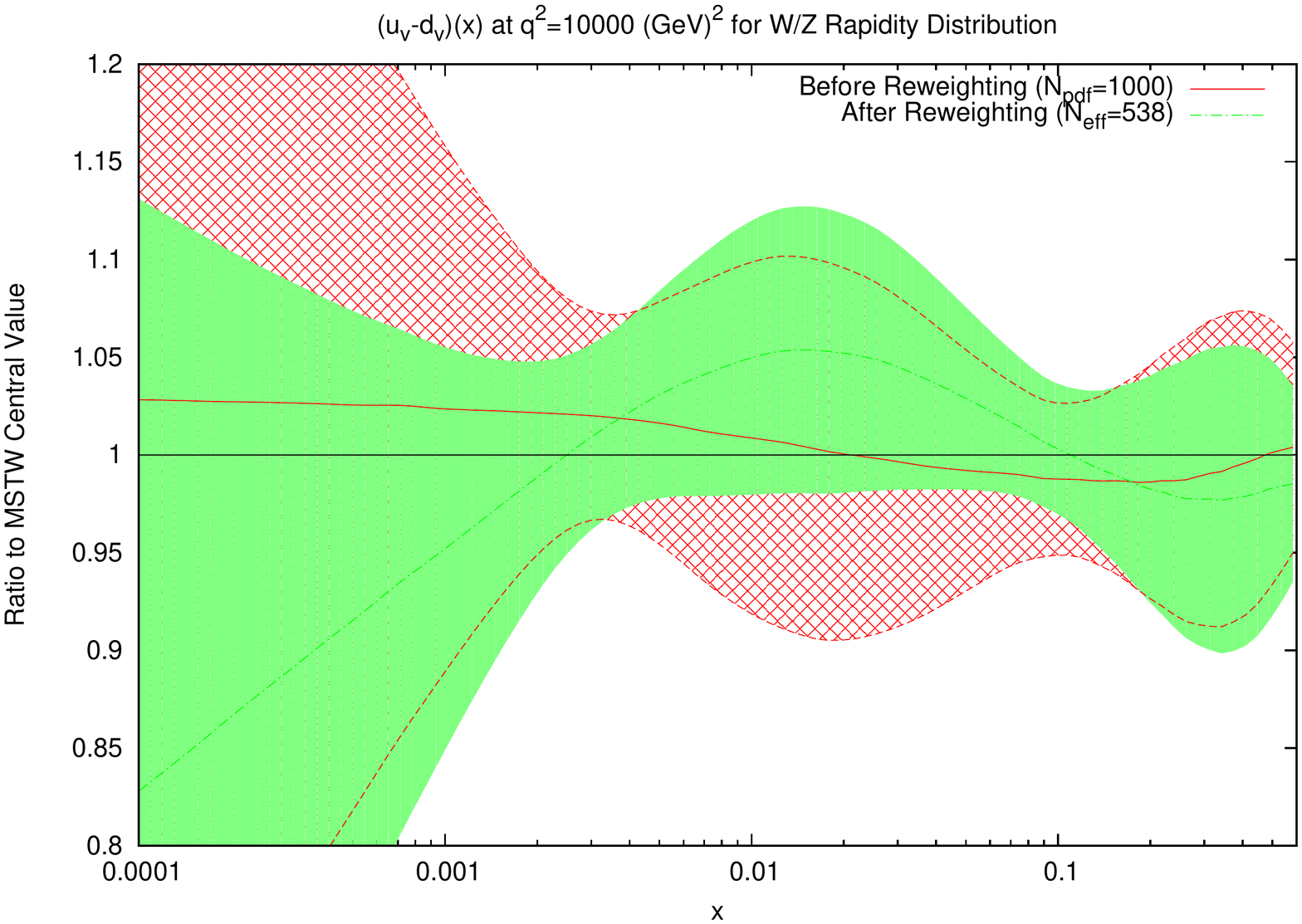}
\vspace{-1cm}
  \caption{The effect of reweighting on the $u_V(x)-d_V(x)$ distribution 
at $Q^2=10^4~\GeV^2$. The hatched (red) band shows the average value and standard deviation
of 1000 randomly generated sets of MSTW2008CP (top) and MSTW2008CPdeut 
(bottom) PDFs and the continuous  
(green) band the 
same PDFs after reweighting according to the fit quality for the 
ATLAS data on $W,Z$ rapidity.}
  \label{fig:WZvalCPdeut}
  \end{figure}

We also examine the effect of the $W$ and $Z$ rapidity 
data on the MSTW2008CP and 
MSTW2008CPdeut PDFs by looking at the eigenvector sensitivity and using the {\it reweighting 
procedure}. The change in $\chi^2$ for each of the eigenvectors of the MSTW2008CP set is shown in 
Fig.~\ref{fig:WZ-EigCP}. The dominant eigenvector, number 12, is still mainly 
to do with the gluon, but some variation of the strange quark is mixed in. The situation is 
very similar for the MSTW2008CPdeut set. Hence, we still obtain a small effect on 
the gluon distribution similar to that for MSTW2008 shown in the lower of Fig.~\ref{fig:WZval}.  
However, even with the modified sets there are still some small changes required for 
the $u_V-d_V$ distribution, as shown in Fig.~\ref{fig:WZvalCPdeut}. 
After reweighting the fit quality improves to  $\chi^2/N_{\rm pts.}=39.5/30$
for MSTW2008CP and $\chi^2/N_{\rm pts.}=38.5/30$ for MSTW2008CPdeut.
Note however, that the 
effective number of PDFs is far greater than in the case for MSTW2008, showing the increased 
compatibility of the data and the PDFs. Even though some of the eigenvectors that show an 
improved fit to the data are those with a larger strange quark at small $x$, the amount of 
weight the random PDFs with larger strange fraction carry is not such to show a clear 
increase in the strange fraction after reweighting. Essentially the direct constraint from 
the dimuon data is overwhelming any pull from the ATLAS data. 

We have also calculated the $\chi^2$ for the fits with 3, 5 and 6 Chebyshev 
polynomials. For 3 polynomials the $\chi^2/N_{\rm pts.}$ is a little above 
$54/30$ both with and without free deuteron corrections, noticeably worse 
than for 4 polynomials, where the corresponding numbers were $46/30$ and 
$49/30$ respectively. For 5 and 6 polynomials the $\chi^2$ is always in the 
low 40s. Hence this is an improvement on the results with 4 polynomials, but 
is not as good as the result obtained from reweighting using MSTW2008CP and
MSTW2008CPdeut, where the effective number of PDFs remains large. Hence,
the additional improvement in the prediction using 5 or 6 polynomials can also 
be comfortably obtained by minor variations in the PDFs with 4 polynomials, 
variations which are consistent with the size of the uncertainties on these 
PDF sets. This adds additional weight to the conclusion that the use of 5 or 6 
polynomials adds little value to the use of 4.         

Hence the MSTW2008CP and MSTW2008CPdeut PDFs result in a big change in the 
high-$p_T$ cut, low-rapidity lepton asymmetry at the LHC. However, as we have 
shown, this quantity is extremely specifically sensitive to 
$u_V -d_V$ for $x \sim M_W/\sqrt{s}$, the PDF combination that changes by far the most.

\subsection{Predictions for other LHC data using modified MSTW PDFs} 

\begin{figure}[htb!]
  \centering
\vspace{-2cm}
  \includegraphics[width=0.6\textwidth,clip]{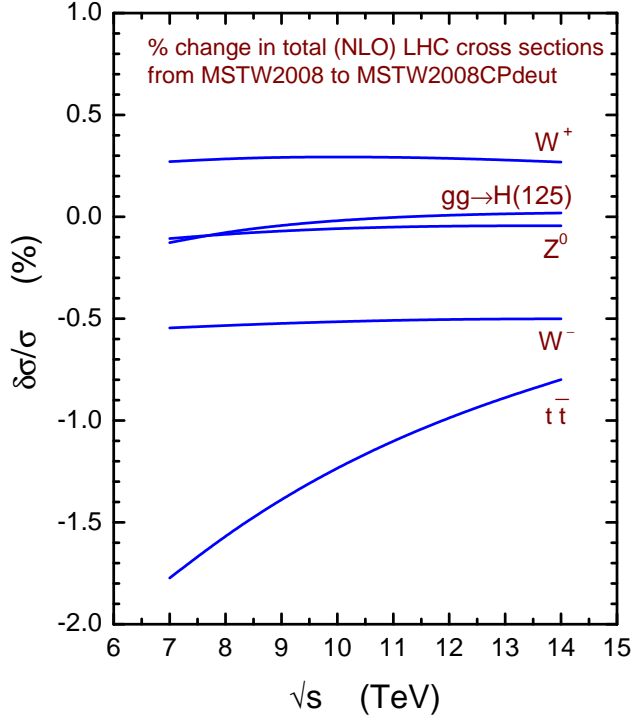}
\vspace{-2cm}
  \caption{The variation in the prediction for various cross sections as a function of energy for the MSTW2008CPdeut PDFs compared to the values obtained from the original MSTW2008 PDFs.}
  \label{fig:sigmaCPdeut}
  \end{figure}

\begin{figure}[htb!]
\centering
\vspace{-1cm}
   \includegraphics[width=0.495\textwidth,clip]{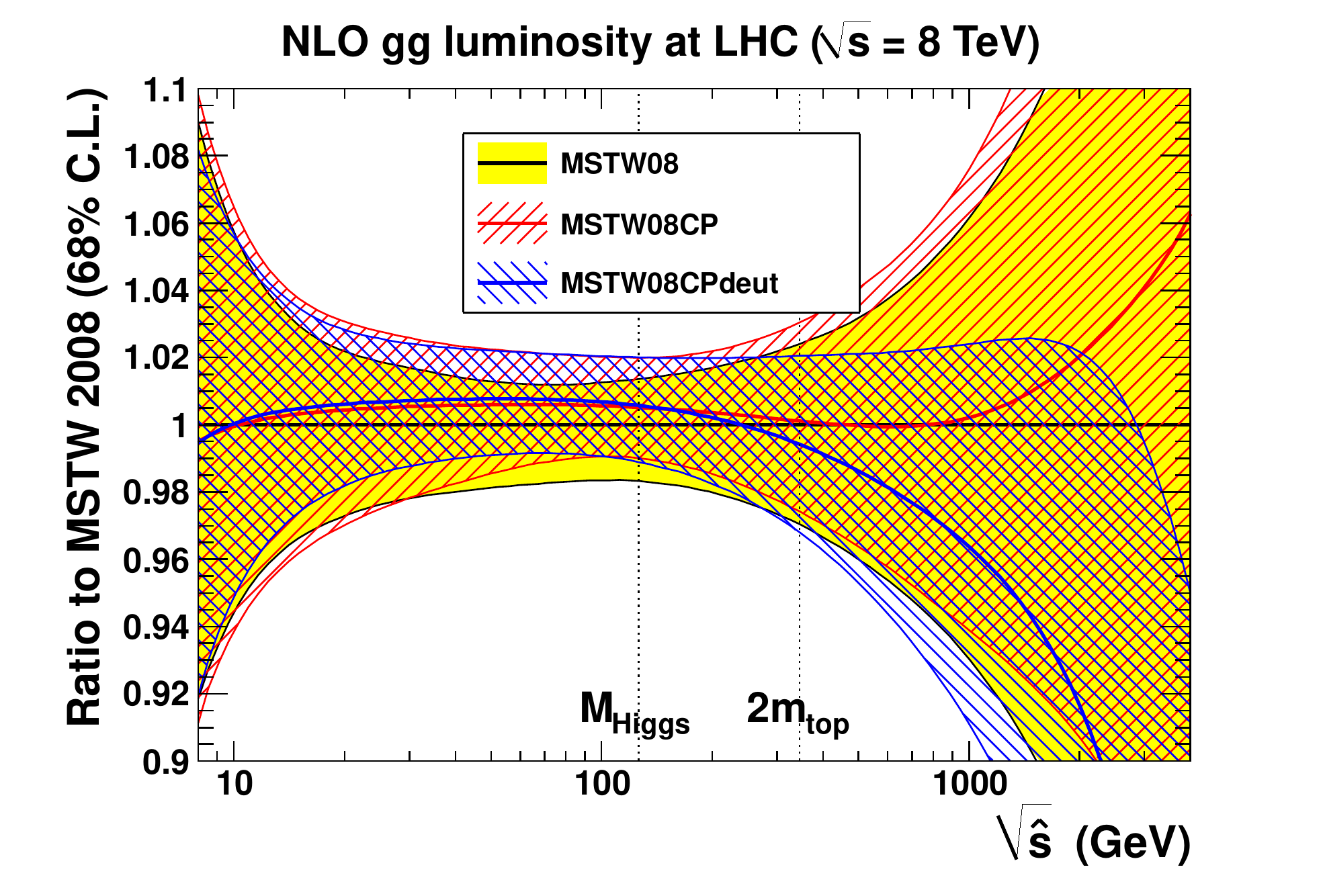}
   \includegraphics[width=0.495\textwidth,clip]{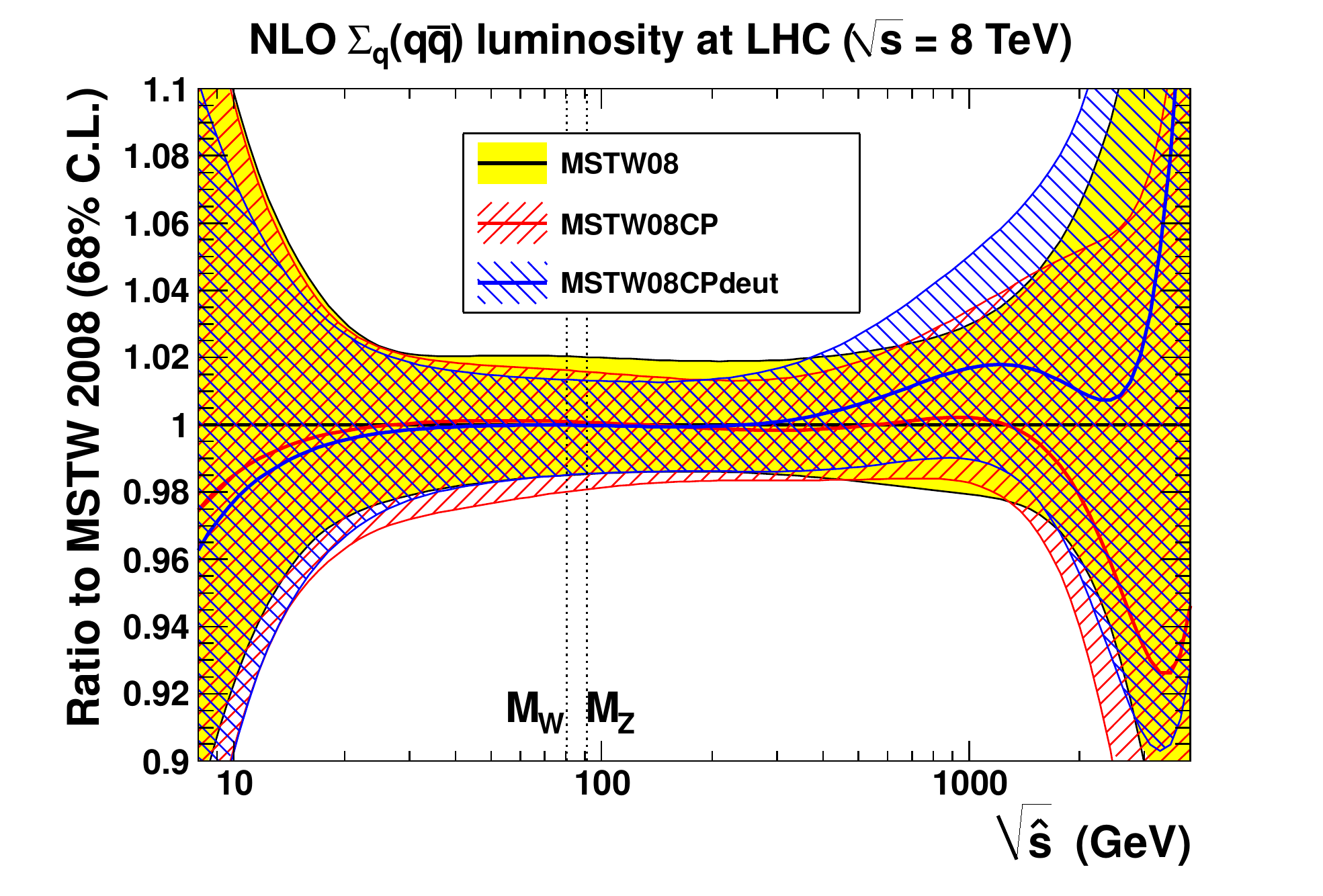}\\
\vspace{-0.1cm}
   \includegraphics[width=0.495\textwidth,clip]{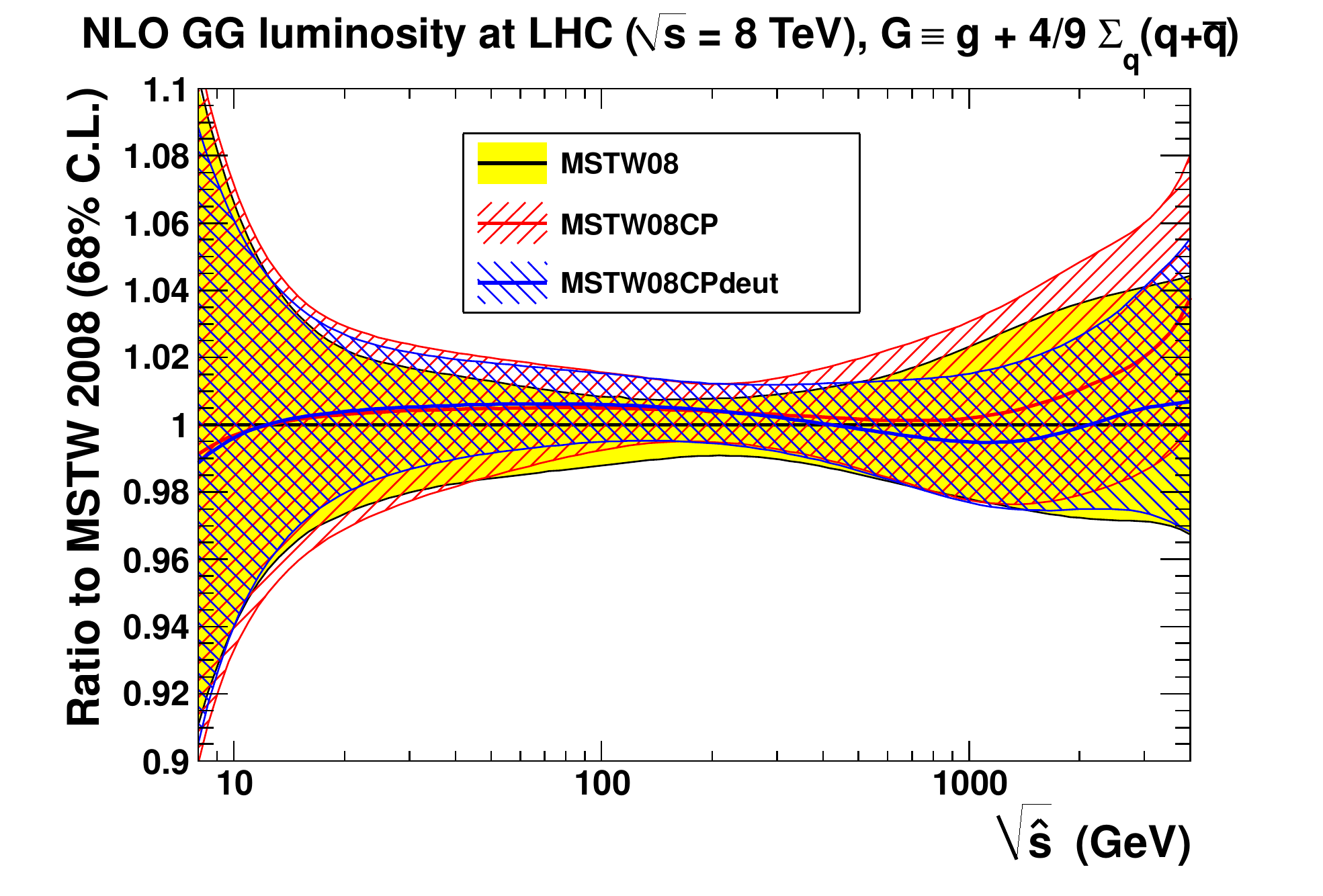}
\vspace{-0.2cm}
  \caption{The parton luminosities for the MSTW2008, the MSTW2008CP and 
the MSTW2008CPdeut PDFs for production of a final state of 
particular invariant mass $\sqrt{\hat s}$ at the LHC for $\sqrt{s} = 8~\TeV$.
The top left plot is for the gluon--gluon luminosity, the top right plot for 
the quark--antiquark luminosity (summed over flavours) and the lower plot for 
$g+4/9 \sum_q (q + \bar q)$, which is 
relevant for some quantities that can be initiated by both gluons and quarks,
e.g. inclusive jets.}
  \label{fig:luminosities}
  \end{figure}

What about the predictions for other LHC observables? Apart from $u_V,d_V$ at low $x$, other 
PDFs have changed little, especially the gluon distribution which 
hardly changes at all from MSTW2008 compared to the size of its uncertainties. Similarly, in 
the new fits $\alpha_S(M_Z^2)$ is left free, but only experiences a 
tiny change. Hence, we would expect little variation in most cross-section 
predictions, as compared to those of the MSTW2008 PDFs.  This is verified 
in Table \ref{tab:sigma} where we show the 
percentage variation in predictions for various standard cross sections 
compared to those using MSTW2008 PDFs (the Higgs boson predictions are for 
$M_H=125~\GeV$). 
We see that there is extreme stability in the total cross-section predictions. 
All the changes are inside the uncertainties, 
and in most cases by very much less than the uncertainty.
Even $\sigma(W^+)/\sigma(W^-)$ changes by barely more than $1\%$, reflecting the 
fact that the largest change in the asymmetry is for a small region of phase space, i.e.~high-$p_T$ and 
low-$y_\ell$, and, moreover, is rather small compared to the individual 
cross sections. Hence, the excellent agreement of predictions using 
MSTW2008 PDFs with the measurements of 
$W$ and $Z$ total cross sections by ATLAS  \cite{Aad:2011dm} and CMS \cite{CMS:2011aa}
is not altered. The change is also displayed in Fig.~\ref{fig:sigmaCPdeut},
for a continuous range of LHC collider energy. This illustrates the same results. 
There is a reasonable change of $\sim 1.5\%$ in the $\bar t t$ cross section 
for the lowest LHC energies, but this is where the gluon distribution is probed at 
relatively high $x$, where the uncertainties are largest. The PDF uncertainty on 
$\sigma_{t \bar t}$  for $7~\TeV$ is $2.9\%$, or $3.9\%$ when the $\alpha_S(M_Z^2)$
uncertainty is also included. As smaller $x$ values are probed at higher energy 
the uncertainty reduces, e.g. to $3.1\%$ at $14~\TeV$, 
and the difference between the MSTW2008 and 
MSTW2008CPdeut predictions also reduces to less than $ 1\%$.   
We also show in Fig.~\ref{fig:luminosities} the parton luminosities for 
the MSTW2008CP and MSTW2008CPdeut sets compared to those of MSTW2008.
These are consistent with the cross section results, i.e.~there is little 
change in the luminosities except a tendency for the quark--antiquark 
luminosity to be slightly higher at the largest $\sqrt{\hat s}$ and the 
gluon--gluon luminosity to be slightly lower in the same region for the 
MSTW2008CPdeut set. 
This is because of the increase of the down quark at high $x$ 
in this set, and a corresponding small decrease in the gluon luminosity 
coming from the fit to inclusive jet data. However, these changes are  
comfortably smaller than the PDF uncertainty.

\begin{table}
\begin{center}
\begin{tabular}{|l|l|l|l|}
\hline
& {CP} & {CPdeut} & unc. \\
\hline
$\!\! W\,\, {\rm Tevatron}\,\,(1.96~\TeV)$                      & +0.6   &  +0.1 & 1.8 \\   
$\!\! Z \,\,{\rm Tevatron}\,\,(1.96~\TeV)$                      & +0.8   &  +0.7 & 1.9 \\    
$\!\! W^+ \,\,{\rm LHC}\,\, (7~\TeV)$        & +0.7   &  +0.3 & 2.2\\    
$\!\! W^- \,\,{\rm LHC}\,\, (7~\TeV)$        & $-0.7$  &  $-0.4$ & 2.2\\    
$\!\! Z \,\,{\rm LHC}\,\, (7~\TeV)$          & +0.0   &  $-0.1$ & 2.2\\    
$\!\! W^+ \,\,{\rm LHC}\,\, (14~\TeV)$       & +0.6   &  +0.3 & 2.4\\    
$\!\! W^- \,\,{\rm LHC}\,\, (14~\TeV)$       & $-0.6$  &  $-0.5$ & 2.4 \\    
$\!\! Z \,\,{\rm LHC}\,\, (14~\TeV)$         & +0.1  &  $-0.1$ & 2.4\\    
$\!\! {\rm Higgs} \,\,{\rm Tevatron}$            & $-0.5$  &  $-1.8$ & 5.1\\
$\!\!{\rm Higgs} \,\,{\rm LHC}\,\,(7~\TeV)$  & +0.2  &  $-0.1$ & 3.3 \\
$\!\!{\rm Higgs} \,\,{\rm LHC}\,\,(14~\TeV)$ & +0.1  &  +0.1 & 3.1\\ 
$\!\! t\bar t \,\,{\rm Tevatron}$            & $+0.5$  &  $-0.6$ & 3.2\\
$\!\! t\bar t\,\,{\rm LHC}\,\,(7~\TeV)$  & $-0.4$  &  $-1.8$ & 3.9 \\
$\!\! t\bar t\,\,{\rm LHC}\,\,(14~\TeV)$ & $-0.2$  &  $-0.8$ & 3.1\\ 

\hline
    \end{tabular}
\end{center}
\caption{The percentage change of various cross sections due to the modifications of the MSTW2008 PDFs. CP denotes the fit with the Chebyshev polynomial input parameterisation, and CPdeut denotes the fit with the deuteron corrections included in addition. To demonstrate the small changes in the cross sections, we also
show, in the final column, the symmetrized PDF $+\,\,\alpha_S(M_Z^2)$ 
percentage uncertainties for the MSTW2008 PDFs.}
\label{tab:sigma}
\end{table}

  \begin{figure}[htb!]
\centering
\vspace{-1cm}
  \includegraphics[width=0.7\textwidth,clip]{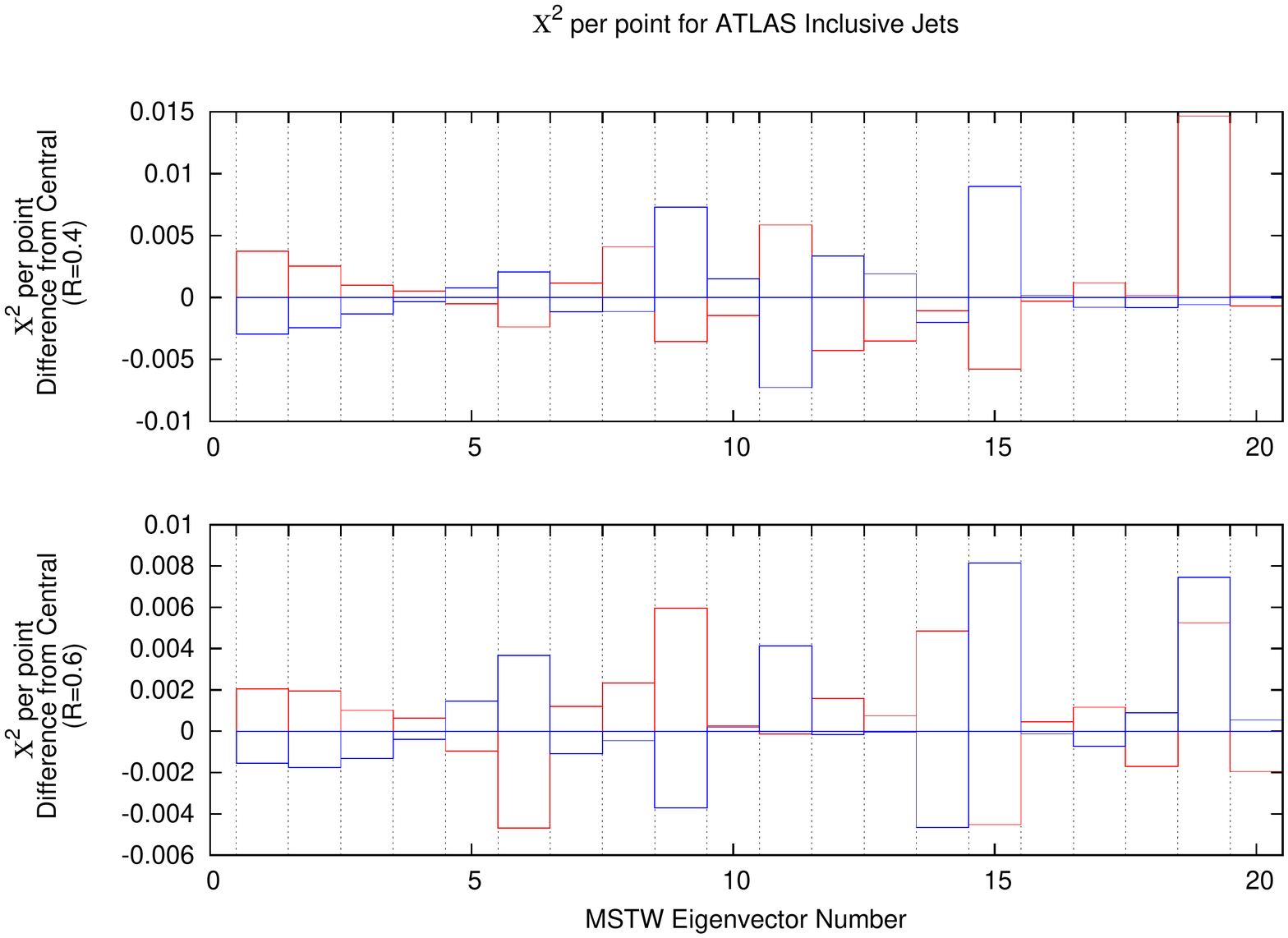}\\
\vspace{-1.2cm}
  \includegraphics[width=0.7\textwidth,clip]{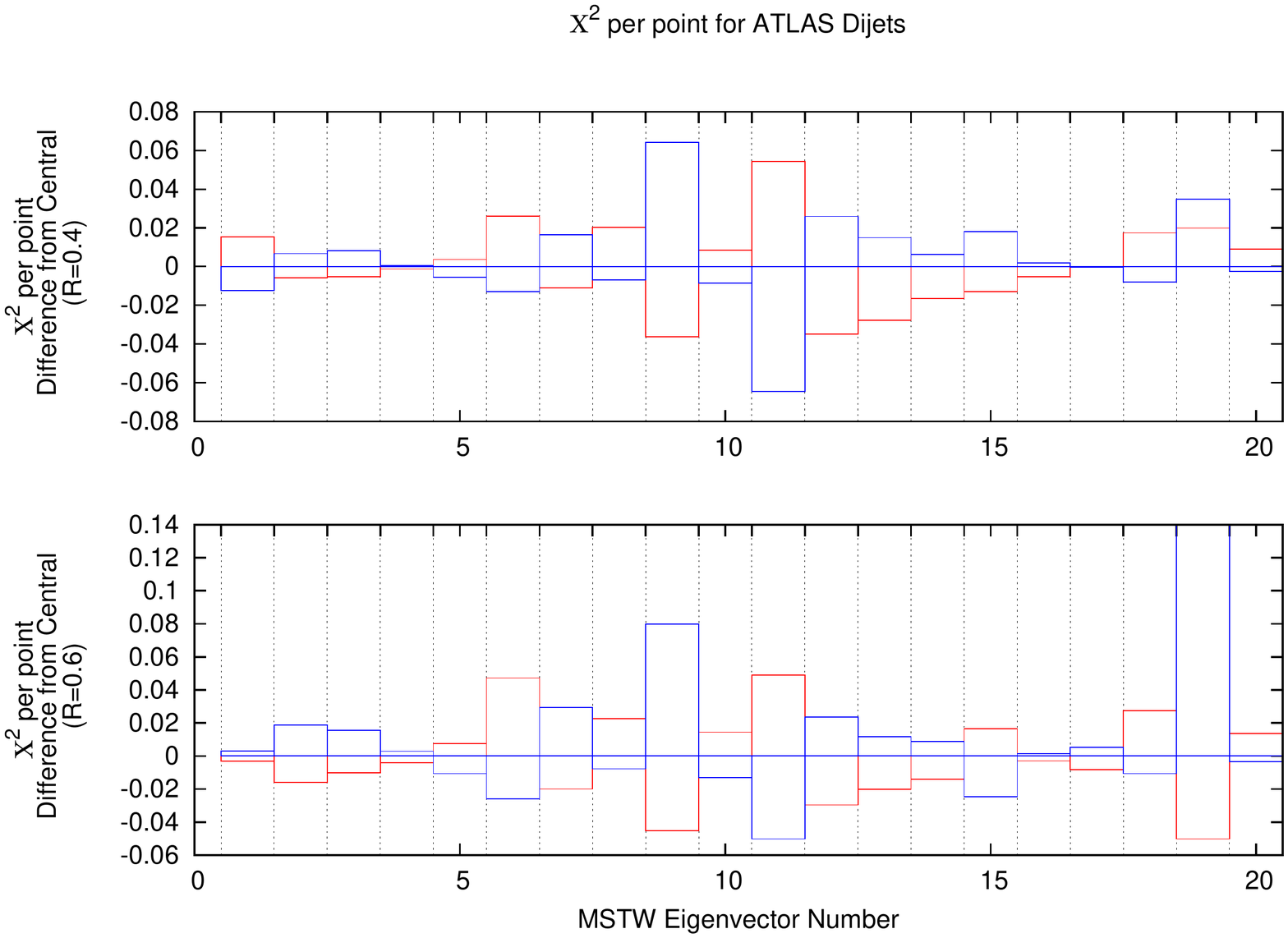}
\vspace{-1cm}
  \caption{The variation in the quality of the fit to ATLAS
inclusive jet production (top) and dijet production (bottom) 
as a function of rapidity for different MSTW2008 eigenvectors.}
  \label{fig:inclusive-Eig}
  \end{figure}

  \begin{figure}[htb!]
\centering
\vspace{-1cm}
  \includegraphics[width=0.7\textwidth,clip]{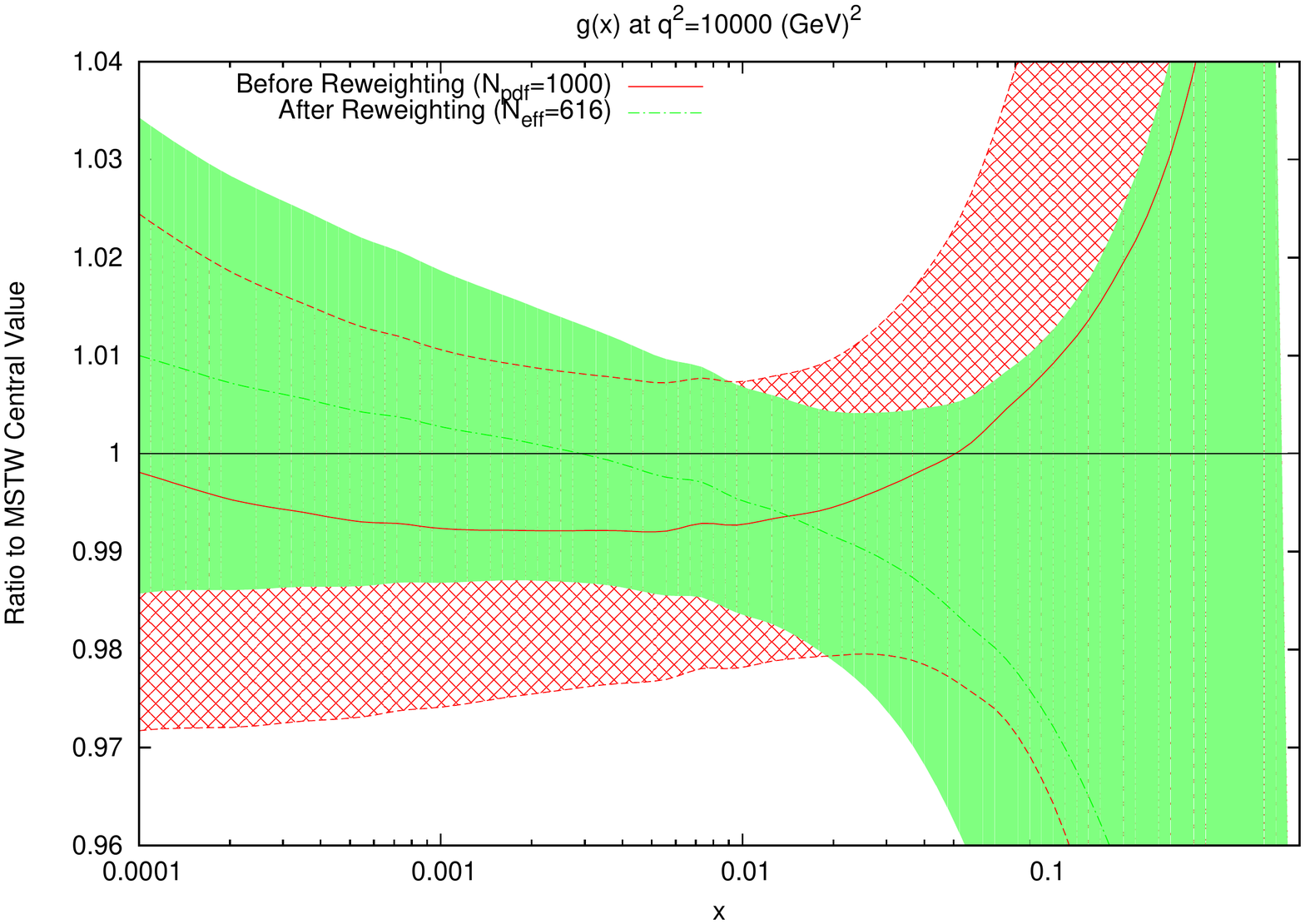}\\
\vspace{-1.5cm}
    \includegraphics[width=0.7\textwidth,clip]{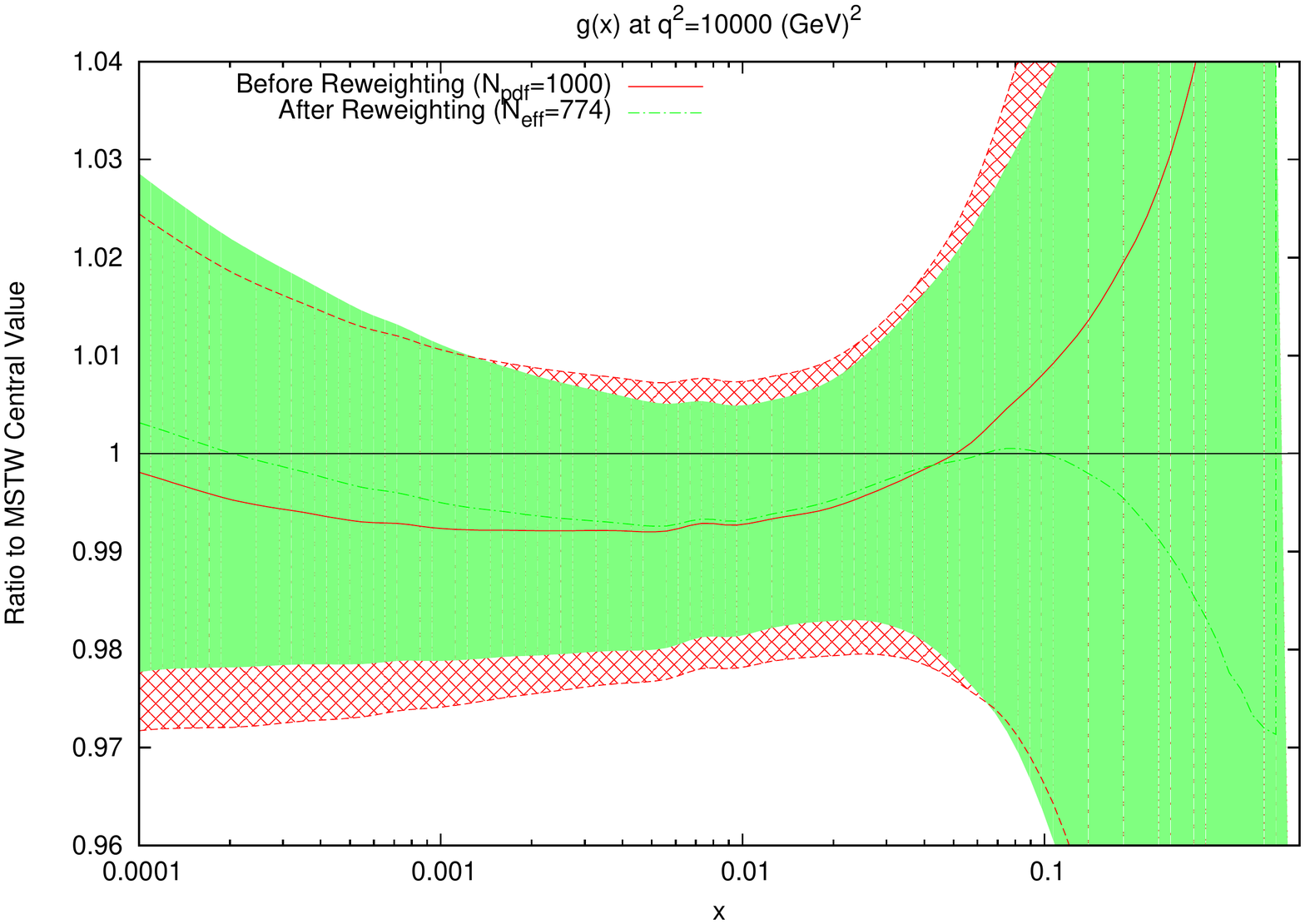}
\vspace{-1cm}
  \caption{The effect of reweighting on the $g(x)$ distribution 
at $Q^2=10^4~\GeV^2$. The hatched (red) band shows the average value and standard deviation 
of 1000 randomly generated sets of MSTW2008 PDFs and the continuous (green) band the 
same PDFs after reweighting according to the fit quality for the 
ATLAS data on inclusive jet production using $R=0.4$ (top) and $R=0.6$ (bottom).}
  \label{fig:inclusive04}
  \end{figure}

Finally, we consider the description of LHC jet data. 
The $\chi^2$ is defined as 
\begin{equation}
  \chi^2 =\sum_{i=1}^{N_{\rm pts.}} \left(\frac{\hat{D}_{i}-T_{i}}{\sigma_{i}^{\rm uncorr.}}\right)^2 + \sum_{k=1}^{N_{\rm corr.}}r_{k}^2,
\label{eq:chi2def}
\end{equation}
where $\hat{D}_{i} \equiv D_{i} - \sum_{k=1}^{N_{\rm corr.}}r_{k}\,
\sigma_{k,i}^{\rm corr.}$ are the data points allowed to shift by the 
systematic errors in order to give the best fit,  
$\sigma_{k,i}^{\rm corr.}$ is an 
absolute correlated uncertainty and normalisation is treated as the other correlated 
uncertainties. The same definition is used for the comparison to $W,Z$ data, 
but the definition is less important in that case. The $\chi^2$ per 
point for the ATLAS inclusive jet data \cite{Aad:2011fc}
is $\chi^2/N_{\rm pts.}=70/90$ for jet radius $R=0.4$ and $\chi^2/N_{\rm pts.}=71/90$ for $R=0.6$. 
The variation as a function of the MSTW 
eigenvectors for the two different choices of the jet radius 
is shown in the upper of Fig.~\ref{fig:inclusive-Eig}. 
(The precise value of $\chi^2$ 
depends on the detailed manner in which the correlated systematic 
uncertainties are treated.) The calculation is made at NLO using APPLgrid (which uses the NLOjet++ 
\cite{Nagy:2001fj,Nagy:2003tz} code) 
using a renormalisation and factorisation scale choice of $p_T^{\rm jet, max}$ in each rapidity bin, 
though extremely similar results are obtained using FastNLO \cite{Kluge:2006xs}. The $\chi^2$ 
for the MSTW2008 
central set is very good, at least as good as the PDF sets of other groups, though the data 
do not discriminate strongly. Similarly the variation in $\chi^2$ between
MSTW2008 eigenvectors is very small: at the absolute most about 2 units in 
$\chi^2$ for the 90 data points -- though this variation is rather smaller than 
the few units difference observed between different PDF groups. Hence, there is 
even less possibility of these data improving the knowledge of a single set of 
PDFs than there is for them to discriminate between different sets. As one might 
expect, the $\chi^2$ using MSTW2008CP and MSTW2008CPdeut is hardly changed from that for 
MSTW2008.  In the lower of Fig.~\ref{fig:inclusive-Eig} we see that exactly the same
picture holds for the ATLAS dijet data (which corresponds to a different analysis 
of basically the same data set). Here, the scale choice is multiplied by a factor of $\exp(0.3y^{\star})$, where 
$y^{\star}$ is half the rapidity separation, 
to avoid instabilities at high rapidity.
In this case the $\chi^2$ values are larger
 but 
similar between the different PDF groups, and at the very most there is $10\%$ variation in 
$\chi^2$ value between different eigenvectors, and in most cases much less. 
In Fig.~\ref{fig:inclusive04} we see the effect of reweighting when 
using the MSTW2008 PDFs. The small variation between the random PDF sets is illustrated by the 
very high effective number $N_{\rm eff}$ of PDFs, and in both cases the change in the average 
value is well within uncertainties, and the decrease in the standard deviation is minimal. 
With reweighting the fit quality improves by only one or two units in $\chi^2$. 
Similarly the MSTW2008CP and MSTW2008CPdeut sets give a $\chi^2$ within a couple 
of units of MSTW2008 for both values of $R$. 
Future data based on higher luminosity, and with a better determination of 
systematic uncertainties, will be more constraining and discriminating.

\section{PDFs at NNLO}

We have repeated the study of extending the parameterisation in the valence and
sea quarks, and of investigating deuteron corrections, using NNLO PDFs 
extracted from an NNLO fit to data. We do not go into details here as all the 
conclusions are qualitatively similar. The input PDFs at NNLO are not exactly 
the same shape as at NLO due to higher-order corrections, mainly in the coefficient 
functions. However, as we will discuss, the relative change in the PDFs
from the extended parameterisation and deuteron corrections is altered 
rather little by this.   

To be a little more precise, the extended parameterisation leads to an NNLO 
fit with $\Delta \chi^2 = -37$ compared to the standard NNLO fit, in comparison to 
$-29$ for the NLO fit. 
The changes in PDFs 
are completely analogous, i.e.~the $u_V(x)$ distribution increases for 
$x \sim 0.01$ and decreases at very small $x$, but there is little change 
in other PDFs. Despite the slightly larger fit improvement at NNLO, the 
change in $u_V(x)$ is slightly smaller than at NLO. When deuteron 
corrections are allowed to be free they choose a form very similar to 
the NLO case, though with the normalisation marginally smaller than 1 and 
the precise value changing slightly if the pivot point $x_p$ is varied.
If $x_p=0.03$ is used instead of $x_p=0.05$ the fit is only three units worse, 
but the deuteron correction is very similar to that at NLO, with 
normalisation marginally above 1.  
The fit has $\Delta \chi^2=-82$ compared to our default fit, similar to the 
value of $-86$ at NLO. As at NLO
the only additional change in PDFs is another slight change in $u_V(x)$ and
a similar change in $d_V(x)$ to that at NLO. Again the changes are  
qualitatively the same to those at NLO, but slightly smaller. 
The changes to predicted cross
sections (such as those shown in Table \ref{tab:sigma} and Fig. 
\ref{fig:sigmaCPdeut}) using the modified
NNLO PDFs are very similar to those at NLO, 
perhaps a little smaller in general, 
and as at NLO are much smaller than the PDF uncertainties.

The changes in the modified NNLO PDFs automatically significantly improve the deficiencies that MSTW2008 
PDFs have with the LHC asymmetry. The effect from the PDFs is a little 
less than at NLO due to the 
smaller change. However, this is well within uncertainties, and is, for example,
no longer true for the $x_p=0.03$ fit mentioned above. (Similarly the agreement of 
the small-$x$ powers $\delta$ for the up and down valence distribution are 
in better agreement for $x_p=0.03$ -- the values of $\delta$ being very sensitive 
to small changes in detail.) We also note that the NNLO 
cross section corrections themselves automatically improve the 
description of data very slightly. 
As at NLO, the best fit to the total $W,Z$ rapidity 
data can still be improved, and at NNLO the fit to the general 
shape in rapidity is a little worse than at NLO. While the modified PDFs
essentially cure the problem with the asymmetry, detailed changes in the 
gluon and sea are required for the best possible fit as shown by the 
reweighting exercise at NLO described above.

\section{Conclusions}

In this paper we have performed global PDF analyses in which the standard 
MSTW input parameterisations of the PDFs have been made more flexible by 
replacing the usual $(1+\epsilon x^{0.5} + \gamma x)$ factors in the valence, sea 
and gluon distributions by Chebyshev 
polynomial forms $(1+\sum a_iT_i(y))$, with $y=1-2\sqrt{x}$.  A Chebyshev 
form has the advantage that the parameters $a_i$ are well-behaved and, 
compared to the coefficients in our standard parameterisation, are rather 
small, with moduli usually $\leq 1$. We demonstrated that about four Chebyshev polynomials are 
sufficient for high precision and used this number in the valence and sea distributions. 
However, the gluon distribution, which already had seven parameters, did not 
require extra free parameters. Hence, only two Chebyshev polynomials were used in this case, 
which is equivalent to the usual $(1+\epsilon x^{0.5} + \gamma x)$ factor.

To explore the effects of using these more 
flexible input forms we fit to exactly the same data set as was used for the 
MSTW2008 analysis \cite{Martin:2009iq}.  The resulting parton set was called 
MSTW2008CP. We found some improvement in the fit to the data, but the only 
significant PDF change was in the valence up-quark distribution, $u_V$, at 
small $x$. Although left free in the fit, the change in $\alpha_S(M_Z^2)$ 
is tiny.

The use of Chebyshev forms allows a more consistent determination of the 
uncertainties of the PDFs. In our best fit we have 6 more free parameters 
than in the original MSTW2008 fit. In the determination of the uncertainties 
we allow one extra parameter to be free for the $u_V,~d_V$ and sea quark 
PDFs when evaluating the uncertainty eigenvectors.  Hence, in the 
determination of the uncertainties of the MSTW2008CP PDFs we have 23 
eigenvectors, rather than the 20 eigenvectors of the MSTW2008 analysis. 
Despite having extra eigenvectors, the uncertainties were found to be similar 
to those of the MSTW2008 partons, but tend to be larger, particularly for 
valence quarks, at small $x$. The most significant change is in $u_V$ which 
now has a more realistic uncertainty, without the artificial `neck' for 
$x\sim 0.003$ at $Q^2=10^4~\GeV^2$.

We also performed a detailed investigation of the nonperturbative corrections 
to be applied when fitting to the data obtained from deuteron targets. It 
will be important to continue to include these deuteron data in global PDF 
analyses for the relatively short term, in order to separate the $u_V$ and $d_V$ 
PDFs. The deuteron correction factor was parametrised in terms of 4 
variables, and various MSTW2008CP global fits were performed allowing some, 
or all, of these variables to be free. We found that large improvements could 
be obtained in the description of the deuteron data, and also in the 
Tevatron charged lepton asymmetry data, as compared to the MSTW2008 analysis. 
The MSTW2008 fit had a fixed deuteron correction imposed, and then only at 
small $x$.  Using the results from the present study, we have adopted the 
best, and most realistic, deuteron correction.  The corresponding parton 
set is called MSTW2008CPdeut.  The most significant change, in comparison to 
MSTW2008CP, is in the $d_V$ PDF, and in the uncertainties of both valence PDFs.
Again the change in $\alpha_S(M_Z^2)$ is insignificant.

In summary, the main changes to the MSTW2008 PDFs obtained in the CP and 
CPdeut `Chebyshev' fits are in $u_V$ and $d_V$ for $x \lapproxeq 0.03$ at high $Q^2\sim 10^4~\GeV^2$,
or slightly higher $x$ at low $Q^2$: a region where there are few or weak 
constraints on the valence PDFs from the data used in these fits. There is also an 
approximately 5$\%$ increase in $d_V$ for $x \sim 0.5$ in the CPdeut fit.
 
We have drawn attention to one type of measurement that is particularly 
sensitive to $u_V$ and $d_V$ in the small $x$ region. That is the 
decay charged lepton asymmetry from $W^\pm$ production, which probes a little 
more deeply into the small $x$ region as the collider energy is increased. 
For a 7 TeV collider, the probed region is $0.01 \lapproxeq x \lapproxeq 0.05$.
The PDF combination sampled by the asymmetry is ($u_V-d_V$) at low lepton 
rapidities. However, we showed that the description of the asymmetry data has 
a more intricate dependence on the PDFs at larger rapidities, and also 
depends sensitively on the experimental minimum $p_T$ cut applied to the 
observed lepton. In Fig. \ref{fig:uVminusdV} we plotted ($u_V-d_V$) at the 
scale $Q^2 = 10^4 ~\GeV^2$ approximately relevant for $W,Z$ production, 
obtained from the original MSTW2008 analysis together with the behaviour 
coming from the CP and CPdeut PDF sets; all three sets were fitted to exactly 
the same (pre-LHC) data.

In Figs. \ref{fig:ATLAScomp} and \ref{fig:CMS35comp} we showed the predictions for the ATLAS 
and CMS asymmetry 
measurements obtained from the three sets of PDFs.  It is remarkable that the 
`Chebyshev' sets, and, in particular the MSTW2008CPdeut set, give such 
excellent descriptions of these data, which were not well predicted at low 
rapidities by the original MSTW2008 set. Note that the improvement is due to 
using a more flexible and more physically suitable (`Chebyshev') 
parameterisation of the input PDFs, and to a lesser extent, to taking more 
care with the deuteron corrections. We emphasise again that the main 
changes to the MSTW2008 PDFs are in the valence distributions in the small 
$x$ region which is barely probed by the existing data. Lepton asymmetry at 
higher LHC energies will sample this region more and at smaller $x$ values. 
It is not surprising, therefore, that we found that the predictions of the 
original MSTW2008 PDFs are essentially unchanged for all other observables.
In all the cases that we investigated, the changes in the cross sections, compared to
those of 
MSTW2008, were much smaller than the PDF uncertainties. This was even the case 
for the total $\sigma(W^+)/\sigma(W^-)$ ratio, since each cross section only 
obtains a small contribution from the high-$p_T$ cut, low-$y_\ell$ region, and 
moreover the change in the PDFs does not even change the individual
differential cross sections by very much, but in a manner which is 
maximised in the asymmetry measurement. Hence, for the overwhelming majority 
of processes at the LHC the MSTW2008 PDFs give essentially the same result 
as those resulting from the investigations in this article. 

From the results of this paper it is clear that a full update of MSTW2008 
PDFs would benefit from an extended parameterisation similar to the form 
presented, and a modification of the deuteron corrections of some sort,  
together with some account of the uncertainty associated with these. Moreover, it would 
require other features, most particularly an inclusion of new data, 
including the LHC data considered in this article. 
However, from the preliminary studies already undertaken of including new
data in \cite{Thorne:2010kj,Watt:2012tq} and in this article, it is 
clear that the effects of individual improvements considered so far give no signs of very 
significant changes (and no obvious signs of accumulation of changes from 
individual improvements) other than the form
of the valence quarks at small $x$. However, the precise results of a full update await a 
much more extensive study than the more specific one which is the focus of this article.

\section*{Acknowledgements}

We would like to thank Stefano Forte and Jon Pumplin for discussion on some
of the issues in this article. The work of R.S.T. is
supported partly by the London Centre for Terauniverse Studies (LCTS),
using funding
from the European Research Council via the Advanced Investigator Grant 267352.

\end{document}